\documentclass[twocolumn,tighten,numberedappendix]{aastex631}
\usepackage{amsmath}

\defcitealias{Beloborodov23}{B23}

\def\simlt{\lower.5ex\hbox{$\; \buildrel < \over \sim \;$}}
\def\simgt{\lower.5ex\hbox{$\; \buildrel > \over \sim \;$}}

\def\beq{\begin{equation}}
\def\eeq{\end{equation}}
\def\ba{\begin{eqnarray}}
\def\ea{\end{eqnarray}}
\def\bB{\boldsymbol{B}}
\def\bE{\boldsymbol{E}}

\def\bv{\boldsymbol{v}}
\def\bOm{\boldsymbol{\Omega}}
\def\bk{\boldsymbol{k}}

\def\Eq{Equation}
\def\Eqs{Equations}

\def\bb{\boldsymbol{\beta}}

\def\E{{\cal E}}
\def\RLC{R_{\rm LC}}

\def\tobs{t_{\rm obs}}
\def\EFRB{{\cal E}_{\rm FRB}}

\def\Bbg{B_{\rm bg}}
\def\bBbg{\boldsymbol{B}_{\rm bg}}

\def\bk{\boldsymbol{k}}
\def\bv{\boldsymbol{v}}
\def\bE{\boldsymbol{E}}
\def\bB{\boldsymbol{B}}

\def\Ep{E_{\rm p}}
\def\Bp{B_{\rm p}}
\def\Up{U_{\rm p}}
\def\bEp{\bE_{\rm p}}
\def\bBp{\bB_{\rm p}}
\def\Bpf{{B'_{\rm p}}}
\def\bBpf{\bB'_{\rm p}}

\def\eF{\epsilon_0}
\def\D{{\cal D}}

\def\omf{{\omega'}}
\def\omFf{\omega'_{\rm F}}
\def\omAf{\omega'_{\rm A}}

\def\kk{k'}
\def\bkk{\bk'}

\def\kF{k_{\rm F}}

\def\kkF{{\kk_{\rm F}}}
\def\kkA{{\kk_{\rm A}}}

\def\bkkF{\bkk_{\rm F}}
\def\bkkA{{\bkk_{\rm A}}}

\def\bkF{\bk_{\rm F}}
\def\bkA{\bk_{\rm A}}
\def\omF{\omega_{\rm F}}
\def\omA{\omega_{\rm A}}

\def\Om{\varOmega}
\def\bOm{\boldsymbol{\Om}}
\def\Omf{\Om'}

\def\OmF{\Om_{\rm F}}
\def\bOmF{\bOm_{\rm F}}
\def\OmFf{\Om'_{\rm F}}
\def\bOmFf{\bOm'_{\rm F}}

\def\Uf{U'}
\def\qf{q'}
\def\tf{t'}

\def\UF{U_{\rm F}}

\def\Rout{R_{\rm out}}
\def\Rin{R_{\rm in}}
\def\Gout{\Gamma_{\rm out}}

\def\Enp{\E_{\rm p}}
\def\Lp{L_{\rm p}}

\def\n{n}
\def\nF{n_{\rm F}}

\def\nuFRB{\nu_{\rm FRB}}
\def\Kf{{\cal K}'}
\def\Uf{U'}

\def\Upf{U'_{\rm p}}
\def\Ubg{U_{\rm bg}}

\def\thf{\theta'}
\def\phif{\varphi'}

\def\bn{\boldsymbol{n}}

\def\Qseed{Q_{\rm seed}}

\def\nseed{n^{\rm seed}}
\def\Useed{U^{\rm seed}}
\def\als{\alpha_{\rm s}}

\def\Omseed{\Omega_{\rm s}}
\def\Rseed{R_{\rm s}}
\def\Lseed{L^{\rm seed}}

\def\A{\cal A}

\def\eFout{\eF^{\rm out}}

\def\nupeak{\nu_{\rm peak}}

\def\xiadv{\xi_{\rm adv}}

\def\Gmax{\Gamma_{\rm max}}

\def\xisat{\xi_{\rm sat}}

\def\R{{\cal R}}

\def\tauatt{\tau_{\rm att}}

\def\nseedF{\nseed_{\rm F}}

\def\tauseed{\tau_{\rm seed}}

\def\ompeak{\omega_{\rm peak}}

\def\omsat{\omega_{\rm sat}}

\def\tausat{\tau_{\rm sat}}

\def\bsat{b_{\rm sat}}

\def\ssat{s_{\rm sat}}

\def\dd{\text{d}}

\def\Lpeak{L_{\rm peak}}

\def\xip{\xi_{\rm p}}

\def\cnst{{\rm const}}

\def\N{{\cal N}}

\def\fFA{f_{\rm F+A}}
\def\fAA{f_{\rm A+A}}
\def\WFA{W_{\rm F+A}}
\def\WAA{W_{\rm A+A}}

\def\nx{\Om_{{\rm F}x}}
\def\ny{\Om_{{\rm F}y}}
\def\nz{\Om_{{\rm F}z}}
\def\bn{\bOmF}

\def\aFA{g}
\def\aAA{h}

\newbox\grsign \setbox\grsign=\hbox{$>$} \newdimen\grdimen \grdimen=\ht\grsign
\newbox\simlessbox \newbox\simgreatbox \newbox\simpropbox
\setbox\simgreatbox=\hbox{\raise.5ex\hbox{$>$}\llap
     {\lower.5ex\hbox{$\sim$}}}\ht1=\grdimen\dp1=0pt
\setbox\simlessbox=\hbox{\raise.5ex\hbox{$<$}\llap
     {\lower.5ex\hbox{$\sim$}}}\ht2=\grdimen\dp2=0pt
\setbox\simpropbox=\hbox{\raise.5ex\hbox{$\propto$}\llap
     {\lower.5ex\hbox{$\sim$}}}\ht2=\grdimen\dp2=0pt
\def\simgt{\mathrel{\copy\simgreatbox}}
\def\simlt{\mathrel{\copy\simlessbox}}

\begin{document}

\title{A mechanism for fast radio bursts and their narrow spectra}

\author[0000-0001-5660-3175]{Andrei M. Beloborodov}
\affiliation{Physics Department and Columbia Astrophysics Laboratory, Columbia University, 538 West 120th Street, New York, NY 10027, USA}
\affiliation{Max Planck Institute for Astrophysics, Karl-Schwarzschild-Str. 1, D-85741, Garching, Germany}

\author[0000-0002-9217-6964]{Ivan Demidov}
\thanks{A.M.B. and I.D. contributed equally to this work.}
\affiliation{Physics Department, Ben-Gurion University, PO Box 653, Beer-Sheva 84105, Israel}

\begin{abstract}
We find an emission mechanism for fast radio bursts (FRBs) capable of producing broad and narrow spectra. It operates in magnetar explosions, which launch an electromagnetic pulse from the inner magnetosphere at radii $r_0\sim 10^{7}$\,cm. Plasma inside the pulse accelerates outward with Lorentz factor $\Gamma(r)\approx r/r_0$. Before exiting the magnetosphere at radius $r\sim 10^{10}$\,cm, the pulse passes through ambient kHz waves that populate the magnetospheres of active magnetars. We show that the waves experience stimulated scattering inside the pulse, with the boost factor $\Gamma^2$ converting them into radial GHz radiation. This process exponentially amplifies tiny GHz seeds to extreme luminosities consistent with observed FRBs. The produced GHz burst rides inside the explosion pulse, which helps it escape to distant observers. The FRB duration and energy are controlled by the pulse duration and energy. In a saturated emission regime, the predicted burst contains an ultrastrong microsecond component. 
\end{abstract}

\vspace*{-1cm}

 \keywords{
X-ray transient sources (1852);
Neutron stars (1108);
Magnetars (992);
Radiative processes (2055);
Radio bursts (1339);
Plasma astrophysics (1261)
}


\section{Introduction}

The origin of FRBs is a fascinating puzzle of radio astronomy (e.g.  \citealt{Petroff2022, Lyubarsky2021}). So far, one source of (weak) FRBs has been identified with a known magnetar, SGR~1935+2154, which is located in our Galaxy. It is plausible that the numerous bright cosmological FRBs with typical energies ${\cal E}\sim 10^{38}-10^{40}$\,erg are also produced by magnetars.

It has been argued that the escape of FRBs through the plasma surrounding magnetars requires an explosion \citep{Beloborodov2024}. Numerical simulations demonstrate that explosions do  occur in the perturbed magnetospheres of magnetars \citep{Parfrey2013,Mahlmann2023,Yuan2020,Chatterjee2026}. They generate a powerful outgoing electromagnetic pulse with a typical duration of $\sim 1$\,ms. The pulse is a macroscopic disturbance described by magnetohydrodynamics (MHD) in the extreme relativistic limit, where the plasma mass is strongly dominated by the electromagnetic field. The pulse is typically launched at radii $r_0\sim 10^7$\,cm; then, it propagates outward with the speed of light $c$ as a shell of growing radius $r\approx ct$ and constant width $\delta r\sim r_0$. The plasma Lorentz factor inside the shell grows as $\Gamma\approx r/r_0$.

How such explosions can produce GHz bursts is a topic of debate. It was first proposed that FRBs can be emitted by synchrotron maser at the wind termination shock (where the magnetar wind inflates a nebula) when the shock becomes boosted by an explosion \citep{Lyubarsky2014}. It was then realized that explosions can launch new shocks inside the wind itself and emit FRBs at smaller $r\sim 10^{13}-10^{14}$\,cm, which helps explain frequent repeaters \citep{Beloborodov2017}. The shock may also convert small-scale wind fluctuations to radio waves \citep{Thompson2023}. Another scenario envisions production of GHz waves when the explosion pulse hits the current sheet that warps around the rotating magnetar outside its light cylinder $\RLC\sim 10^{10}$\,cm (\citealt{Lyubarsky2020, Mahlmann2022}). For all the models, a key challenge is the observed temporal and spectral structure of FRBs. An outstanding puzzle is the narrow spectral peak $\Delta\nu/\nu<1$ observed in some bursts, in particular in repeaters
\citep{Kumar2024}.

In this paper, we find that magnetar explosions naturally generate powerful GHz bursts at radii $r\sim 10^9-10^{10}{\rm \,cm}<\RLC$. The emission process develops as the explosion pulse propagates through the sea of kHz waves that fill the outer magnetospheres of active magnetars with a small energy density $U_{\rm kHz}$, typically $\sim 10^{-14}$ of the pulse energy density $\Up$. We show that stimulated scattering  of kHz waves exponentially amplifies tiny seed GHz waves inside the pulse until they reach extreme luminosities and consume the ambient kHz waves. The pulse effectively sweeps the kHz waves, giving them a radial direction and compressing them into a thin shell while changing their frequency $\nu_0$ to $\nuFRB\sim \Gamma^2\nu_0$, which falls in the GHz band. Remarkably, GHz waves grow strongest in a narrow frequency band, which results in a narrow FRB spectrum. The emission mechanism is robust and calculated from first principles.


\section{Basic picture}
\label{picture}

\begin{figure}
  \centering
  \includegraphics[width=0.97\columnwidth]{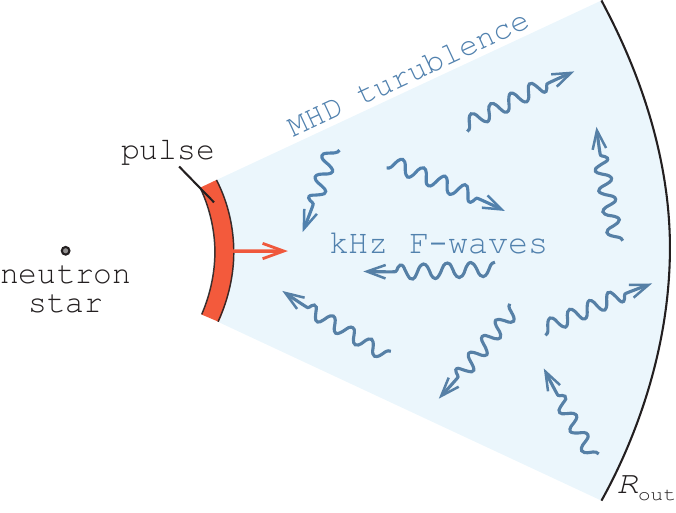}
  \caption{An explosion pulse (orange) propagates outward through the turbulent magnetosphere and interacts with kHz waves crossing the pulse. At $r\gg r_0$, the pulse is a shell of radius $r$ far exceeding its width $cT\sim r_0$. The pulse will exit the magnetosphere at radius $\Rout\approx \RLC$ at time $t=\Rout/c$.}
\label{fig:Fig1}
\end{figure}

Figure~\ref{fig:Fig1} schematically shows the explosion pulse moving outward and interacting with ambient waves. The magnetospheres of active magnetars are generally populated with waves of two types: Alfv\'enic (A) and fast magnetosonic (F). Their characteristic frequencies are in the kHz range, as discussed in section~\ref{ambient}. The F-waves in a magnetically dominated plasma are practically identical to electromagnetic waves in vacuum; they freely propagate in any direction and can cross the pulse.

The pulse exposed to ambient kHz waves inevitably becomes an exponential amplifier of seed GHz radiation, and we will show below that the generated GHz luminosity scales with the pulse power. The emission process is best viewed in the local rest frame of the plasma $\Kf$; hereafter we call it ``fluid frame.'' Plasma inside the pulse moves outward with a high Lorentz factor $\Gamma\approx r/r_0$ (section~\ref{pulse}), reaching $\Gamma\sim 10^3$ at the light cylinder. In frame $\Kf$, the ambient kHz F-waves with frequency $\nu_0$ are Doppler boosted to $\nu'\sim \Gamma\nu_0$ and form a ``head-on'' MHz beam. A key point is that the beam acts as a pump F-wave that can exponentially amplify a pair of new seed waves $F+A$. This process occurs with conservation of energy and momentum in frame $\Kf$: $F\rightarrow F+A$. Like the pump $F$ wave, the new waves $F$ and $A$ have MHz frequencies in frame $\Kf$, but propagate in different directions. When viewed in the lab frame, the new F-waves have GHz frequencies and are beamed radially outward.

The new waves grow from a seed amplitude, which may be provided  e.g. by radio pulsations observed in some galactic magnetars \citep{Camilo2006}. Any weak seeds (practically arbitrarily small) turn out sufficient to ignite the dramatic exponential amplification. A similar process is well known in nonlinear optics as stimulated Raman scattering, where a pump wave generates a scattered wave plus a molecular vibration. Stimulated Raman scattering is also well studied in laser plasma \citep{Kruer2003}, where the pump wave (laser beam) amplifies another (seed) electromagnetic wave plus a Langmuir wave: $F\rightarrow F+L$. The process $F\rightarrow F+A$ differs in that it involves an Alfv\'en wave instead of a Langmuir wave. This ``3-wave'' process is known to occur in magnetically dominated plasmas with a well defined rate \citep{Thompson1998, Lyubarsky2019, GolbraikhLyubarsky2023, Solanki2026a}. 

\begin{figure*}
  \centering
  \includegraphics[width=0.97\textwidth]{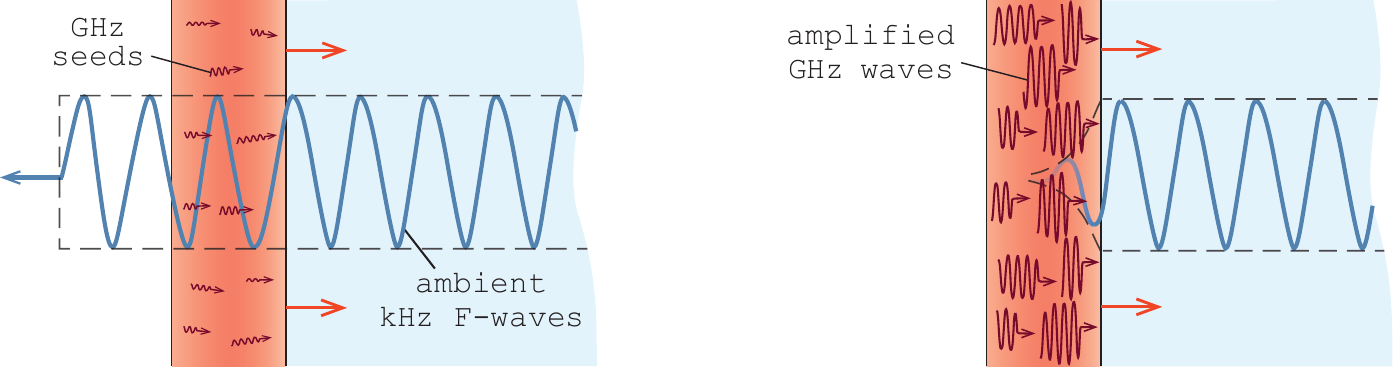}
  \caption{Schematic picture of the stimulated conversion of ambient kHz F waves into GHz radio waves. The conversion occurs inside the explosion pulse as it propagates from a radius $\Rin$ to $\Rout\sim 10^{10}$\,cm. The ambient kHz wave passing through the pulse feeds GHz seeds (which propagate outward with the pulse). Left: initial unsaturated stage. The GHz radiation is still weak, the stimulated scattering is slow and weakly affects the kHz wave that freely passes through the pulse. Right: later saturated stage near $\Rout$. The GHz radiation is enormously amplified, and the kHz wave is depleted inside the pulse.}
\label{fig:Fig2}
\end{figure*}

The process $F\rightarrow F+A$ has enough time to develop many $e$-foldings of growth because the amplified waves $F+A$ stay inside the pulse. When viewed in frame $\Kf$, the strongest amplification happens for F-waves propagating perpendicular to the pump beam, and so also perpendicular to the pulse propagation direction. When viewed in the static lab frame, the amplified F-waves are beamed radially within an angle $\psi\approx\Gamma^{-1}$ and propagate outward together with the explosion pulse (Figure~\ref{fig:Fig2}). Their frequency in the lab frame is $\Gamma\nu'\approx \Gamma^2\nu_0$, in the GHz band. As a result, the pulse becomes loaded with extremely strong GHz radiation before leaving the magnetosphere at $\RLC$.

The gain factor $\Gamma^2$ is similar to that for inverse Compton scattering, which was discussed previously as a possible FRB mechanism \citep{Zhang2022}. Unlike usual Compton scattering (a spontaneous process), here we deal with a stimulated process, which has a huge rate because the waves involved have low frequencies and high occupation numbers. The rate is insensitive to the fact that the MHD fluid is composed of individual particles, and does not depend on the particle mass or charge, so the amplification of GHz radiation is a pure MHD phenomenon. Its rate grows exponentially because it is proportional to the intensity of the amplified waves. Although it is intuitive to think of stimulated scattering in the quantum picture, the process is classical, as expected in the limit of high occupation numbers. The exponential growth is a standard parametric instability, enabled by the nonlinearity of MHD.

The ambient kHz F-waves crossing the explosion pulse amplify the GHz seeds throughout its interior until the accumulated GHz radiation becomes extremely bright. Eventually, it can make the pulse so aggressive in stimulated scattering that it depletes the kHz waves in a certain frequency band. This saturated regime yields FRBs with a special spectral and temporal structure. The kHz waves in the saturated band become unable to cross the pulse, as they convert into GHz radiation as soon as they enter the pulse. Effectively, the pulse develops a thin ``wind shield'' sweeping the ambient waves in a certain frequency band. Just outside this band, kHz waves continue to flow through the pulse and exponentially amplify GHz radiation in its interior. This qualitative picture will be rigorously developed with detailed calculations below.

The energy of the resulting FRB can be roughly estimated by using the quantum description and thinking of the kHz waves as photons. Stimulated scattering converts the kHz photons to GHz  photons. Since the energy of each scattered photon is boosted by $\Gamma^2$, the energy of the produced FRB is related to energy $\Delta\E$ consumed from the ambient kHz waves as follows
\beq
  \EFRB\approx\Gamma^2\Delta\E.
\eeq
Here, the scattering process is viewed in the lab frame, so it is not conservative: the scattered photons gain energy from the moving MHD explosion pulse, which carries energy $\Enp\gg\EFRB$ and acts like an infinitely heavy wall.

The energy density of ambient kHz F-waves may be parameterized as $U_0=\eF\Bbg^2/8\pi$, where $\Bbg$ is the magnetic field of the background (unperturbed) magnetosphere. The kHz wave energy swept up by the pulse traversing distance $\Delta r\sim r$
in the saturated regime is 
\beq
\label{eq:dEn}
  \Delta\E\approx 4\pi r^2   \Delta r\, U_0 = \frac{1}{2}\,\eF \Bbg^2  r^2 \Delta r. 
\eeq 
The saturated regime is reached if the growth of GHz waves undergoes many e-foldings, $q\Delta r/c\gg 10$, where $q$ is the growth rate of parametric instability. The standard theory of 3-wave interactions yields $q\sim\eF\nu_0$ within a numerical factor $\sim 1$ \citep{GolbraikhLyubarsky2023}. This gives a rough estimate:
\beq
\label{eq:EFRB_sat}
  \EFRB\sim \frac{1}{2}\,\Gamma^2 \Bbg^2  r^2 \frac{c}{\nu_0} \left(\frac{q r}{c}\right),
\eeq
where $c/q\sim (c/\eF\nu_0)$ is the distance corresponding to one e-folding of FRB amplification. Using $\nu_{\rm FRB}\approx\Gamma^2\nu_0$ and the relation $\Bp=2\Gamma^2\Bbg$ between $\Bbg$ and the pulse field $\Bp$ (see section~\ref{pulse}), we find
\beq
\label{eq:EFRB_sat}
  \EFRB\sim 10^{-2}\, \left(\frac{q r}{c}\right) \frac{\Lp}{\nu_{\rm FRB}},
\eeq
where $\Lp=cr^2\Bp^2$ is the isotropic equivalent of the explosion power and $r\sim \Rout$. The saturation parameter $qr/c\sim \eF\nu_0 \Rout/c$ can exceed several hundred, which implies $\EFRB\gtrsim 10^{39}$\,erg for strong explosions.

Next sections describe the outlined mechanism in detail, beginning with the origin of kHz waves and dynamics of the explosion pulse, and then presenting the scattering process and the resulting FRB spectrum.


\section{Ambient kHz waves}
\label{ambient}

Waves with frequencies considered in this paper are MHD modes propagating in the magnetized plasma. The magnetic energy around neutron stars far exceeds the rest mass and kinetic energy of the plasma. At the same time, plasma density is sufficiently high to satisfy two conditions:
(1) plasma frequency exceeds the wave frequency, and 
(2) there is no ``charge starvation,'' i.e. the density
is sufficient to support electric currents required by the MHD waves. These conditions are further discussed in section~\ref{pulse}.

In the magnetically dominated plasma, MHD waves can propagate in two eigen modes: Alfv\'en (denoted as A) and fast magnetosonic (denoted as F). Each mode is characterized by its wavevector $\bk$, frequency $\omega(\bk)$, and linear polarization. A-waves have electric fields $\bE_{\rm wave}$ oscillating in the $(\bk,\bB)$ plane, and F-waves have $\bE_{\rm wave}$  perpendicular to this plane. Both modes have group speeds $v_{\rm gr}=c$. A-modes are ducted along the magnetic field lines ($\boldsymbol{v}_{\rm gr}\parallel\bB$), and F-modes freely propagate across the magnetic field lines like usual electromagnetic waves in vacuum (radio waves).

Powerful explosions in magnetars are often preceded by low-level activity, which is associated with small quakes of the neutron star and its magnetosphere. The quakes occur in the kHz frequency band and generate MHD waves in the magnetosphere. In particular, the spectrum of Alfv\'en waves transmitted from the oscillating neutron star is shaped by the transmission coefficient ${\cal T}(\nu)$ \citep{Blaes1989,Bransgrove2020}. ${\cal T}$ is suppressed at low frequencies $\nu\lesssim 1$\,kHz and approaches a nearly constant value for $\nu\gtrsim 10$\,kHz. When the intrinsic spectrum of crustal oscillations peaks at any frequency below 10\,kHz, the excited magnetospheric waves will peak at $\nu_0\sim 3-10$\,kHz. 

The transmitted Alfv\'en waves fill the entire magnetosphere as they bounce along closed magnetic field lines. Their amplitude relative to the local background field $\Bbg$ increases with radius as $\delta B/\Bbg\propto  r^{3/2}$, where $\Bbg$ is approximated as a dipole field.\footnote{The dipole field component dominates at large $r$, so the outer magnetosphere is usually described as dipole. Near the light cylinder radius $\RLC$, it strongly deviates from the dipole configuration and transitions to the spindown wind.} This increase follows from the conservation of the wave energy flux ducted along $\bBbg$, and is caused by the decline of $\Bbg\propto r^{-3}$. The turbulence energy density normalized to the energy density of the background magnetosphere scales as
\beq
\label{eq:dB}
\frac{(\delta B)^2}{\Bbg^2}\propto r^3.
\eeq
It peaks in the outermost magnetosphere, at $r\sim \RLC$.

Waves with large $\delta B/\Bbg$ can develop a turbulent cascade to small scales \citep{Thompson1998}. More precisely, the cascade development is controlled by the parameter
\begin{equation}\label{chi}
  \chi=\frac{k_\perp}{k_\parallel}\frac{\delta B}{B_{\rm bg}},
\end{equation}
where $k_\parallel$ and $k_\perp$ are the components of the wavevector parallel and perpendicular to the local $\bBbg$. Since the frequency of Alfv\'en waves satisfies $\omega=ck_\parallel$, one can think of the parameter $\chi$ as the ratio of the oscillation time $\omega^{-1}$ to the nonlinear shearing time of the wave packet $t_{\rm NL}\sim (k_\perp \delta v_\perp)^{-1}$, where $\delta v_\perp \sim c\,\delta B/B_{\rm bg}$ is the fluid velocity in the wave. The quick development of turbulence is assisted by the growth of $k_\perp$ due to linear shearing of wave packets as they propagate along the closed magnetic field lines of different lengths \citep{Bransgrove2020,Chen2022}.

During the development of Alfv\'enic turbulence, nonlinear interactions also excite F-modes \citep{Thompson1998,Cho2005,Takamoto2016}. As a result, 10--20\% of the turbulence energy converts into F-waves. Their spectrum peaks at a frequency comparable to the ``driving'' frequency $\nu_0$ at which Alfv\'enic turbulence is injected by the quakes.

In addition, starquakes can directly emit kHz F-modes (see Figure~1 in \cite{Beloborodov2023} and \cite{Qu2026}). As they expand through the magnetosphere, the relative amplitude of F-waves $\delta B/\Bbg$ grows as $r^2$, i.e. they grow even faster than A-waves and typically become strongly nonlinear before reaching $r>10^9$\,cm. The nonlinearity of kHz F-waves results in formation of shocks \citep{Chen2022a, Beloborodov2023} or the waves become damped through excitation of a sea of Alfv\'en waves \citep{GolbraikhLyubarsky2023, Solanki2026b}. These additional nonlinear processes may significantly contribute to the MHD turbulence in the outer magnetosphere.

In view of all these processes, one expects that the pre-explosion magnetospheres of active magnetars are filled with a population of A and F kHz waves at radii $r$ up to $\RLC$, with wavevectors $\bk$ in all directions. We are particularly interested here in F-waves because they can propagate across the magnetic field lines and will enter the explosion pulse. Their  energy density $U_{\rm kHz}$ may be parameterized as
\beq
   U_{\rm kHz}\equiv U_0=\eF\Ubg, \qquad \eF\ll 1.
\eeq
Note that $\eF$ grows with radius $r$ (cf.~\Eq~\ref{eq:dB}) while $\Ubg\propto r^{-6}$ steeply decreases with $r$. We are mainly interested in radii approaching $\RLC\sim 10^{10}$\,cm because here (i) the explosion develops a high $\Gamma\sim 10^3$  and (ii) the ambient kHz waves have a large $\eF$, which enables the fast exponential growth of GHz waves calculated in the next sections. Note that  $U_{\rm kHz}$ is smaller than the energy density of the explosion pulse $\Up$ (\Eq~(\ref{eq:Up1}) below) by the factor of $\eF/8\Gamma^4$, i.e. typically by $\sim 14$ orders of magnitude.

The exact spectrum and angular distribution of the kHz F-waves in the turbulent outer magnetosphere are not essential for the FRB mechanism proposed in this paper. As a concrete example, our numerical calculations will use an isotropic wave distribution with a spectral density of the form
\beq
\label{eq:bg}
 U_\omega=\frac{U_0}{K\omega_0}\times \left\{\begin{array}{lr}
\vspace*{2mm}
(\omega/\omega_0)^{\alpha_1}  
 & \omega<\omega_0 \\
(\omega/\omega_0)^{-\alpha_2} 
 & \omega>\omega_0
    \end{array}\right.
\eeq
where $K=(\alpha_1+1)^{-1}+(\alpha_2-1)^{-1}$. In numerical examples we will use $\alpha_1=\alpha_2=3$. Their values weakly affect the results; any $U_\omega$ with a smooth peak at some $\omega_0$ will give approximately the same FRB production, as described below.


\section{Explosion pulse}
\label{pulse}

Explosions result from a magnetospheric instability developing at some characteristic radius $r_0$ \citep{Parfrey2013, Mahlmann2023,Chatterjee2026}. They launch a compressive pulse with duration $T\sim r_0/c$ and energy 
\beq
  \Enp\sim \frac{\mu^2}{cTr_0^2}\sim 10^{45}\,\mu_{33}^2\left(\frac{r_0}{10^7{\rm\, cm}}\right)^{-3} \,{\rm erg},
\eeq 
where $\mu$ is the magnetic dipole moment of the magnetar, normalized here to a typical $\mu=10^{33}$\,G\,cm$^3$. Both $\Enp$ and the corresponding explosion power $\Lp\sim\Enp/T$ can vary by orders of magnitude. The explosion energy is usually related to observed gamma-ray output $\E_\gamma$. Magnetar flares detected so far have reached $\E_\gamma\sim 10^{47}$\,erg \cite{Kaspi2017}. This energy is emitted in a fraction of a second, which implies explosion power exceeding $10^{47}$\,erg\,s$^{-1}$.

One can think of the outgoing compressive MHD pulse as a superposition of a normal vacuum electromagnetic wave $\bEp$, $\bBp$ with the static background magnetic field $\bBbg$ \citep{Lyubarsky2020, Beloborodov2023}. At radius $r_0$, the pulse has amplitude $\Bp=\Ep\sim\Bbg$. The pulse propagates with speed $c$ to $r\gg r_0$, and its amplitude decreases as $r^{-1}$ while its thickness, profile, and energy stay constant.\footnote{This description holds for a compressive pulse with $\bBp\cdot\bBbg>0$. Pulses with a rarefaction part $\bBp\cdot\bBbg<0$ are different, as they launch a monster shock at the trough of the wave \citep{Beloborodov2023}. We here consider pulses with $\bBp\cdot\bBbg>0$.}  
The constant power  $\Lp=cr^2\Bp^2$ (isotropic equivalent) corresponds to $\Bp\propto r^{-1}$.

We will approximate the background field $\bBbg$ as dipole at radii $r_0\ll r<\RLC$. The pulse has a toroidal electric field $\bEp\perp\bBbg$, and its magnetic field satisfies $\bBp\cdot\bBbg=\Bp\Bbg\cos\alpha>0$.  The exact value of $\cos\alpha$ is not essential for the model described below and we set $\cos\alpha=1$ for simplicity. Since $\Bbg$ decreases as $r^{-3}$, one finds 
\beq
 \frac{\Bbg}{\Bp}\approx \frac{r_0^2}{r^2}\ll 1\qquad {\rm at} \quad r\gg r_0.
 \eeq

As usual in MHD, plasma drifts with velocity $\bb=\bv/c=\bE\times\bB/B^2$, where $\bB=\bBp+\bBbg$ and $\bE=\bEp$. Expansion in the small parameter $\Bbg/\Bp$ yields 
\begin{equation}
\label{eq:Gamma}
   \beta\equiv\! \frac{v}{c} \approx 1 \! - \! \frac{\Bbg}{\Bp}, \qquad
   \Gamma \approx \sqrt{\frac{\Bp}{2\Bbg}} \approx \frac{r}{r_0},
\end{equation}
where $\Gamma=(1-\beta^2)^{-1/2}$. The linear growth of $\Gamma(r)$ is confirmed by full MHD simulations of magnetar explosions \citep{Chatterjee2026}. Hereafter all quantities measured in the plasma rest frame $\Kf$ inside the pulse will be denoted with a prime. Note that 
\begin{equation}
    \Ep'=0, \qquad \Bp'=\frac{\Bp}{\Gamma}, \qquad \Upf=\frac{\Bpf^2}{8\pi}.
\end{equation}
The energy density of the pulse in the lab frame is given by 
\beq
\label{eq:Up}
 \Up=\frac{\Bp^2}{4\pi}=\frac{\Gamma^2\Bpf^2}{4\pi}=2\Gamma^2\Upf.
\eeq 
It exceeds the energy density of the outer magnetosphere at $r\gg r_0$ by many orders of magnitude:
\beq
\label{eq:Up1}
   \Up = 8\Gamma^4\Ubg, \qquad \Ubg=\frac{\Bbg^2}{8\pi}.
\eeq

Note that the magnetospheric plasma inside the pulse drifts with respect to it with a small relative speed $v_r-c\approx -c/2\Gamma^2$, which does not allow the plasma to cross the pulse width $cT$ on the timescale $r/c$. Thus, the plasma becomes trapped inside the pulse. 

A minimum amount of plasma in the pulse is inherited from the pre-explosion background magnetosphere, which has a density distribution $n_{\rm bg}(r)\approx \N_{\rm bg}/r^3$ with a typical $\N_{\rm bg}\sim 10^{37}$ \citep{Beloborodov2020}. The pulse traps $\sim \N_{\rm bg}$ particles in a shell of a growing volume $\Delta V\propto r^2$ and advects them with the growing Lorentz factor $\Gamma\propto r$. When a magnetospheric plasma layer initially residing at a radius $r_1$ is swept by the explosion pulse and pushed to a bulk Lorentz factor $\Gamma_1$, its proper density is increased to $n_e'(r_1)\sim \Gamma_1 n_{\rm bg}(r_1)$. Then, the density of the swept-up layer decreases as $n_e'\propto (r^2\Gamma)^{-1}\propto r^{-3}$, which gives 
\beq
\label{eq:n_e}
    n_e'\sim \frac{\Gamma_1\N_{\rm bg}}{r^3}.
\eeq 
Layers swept-up later (at larger $r_1$) are denser because they have larger $\Gamma_1$. 

Most of the pulse volume is occupied by the plasma carried from small radii $r_1\gtrsim  r_0$. These deep layers may be loaded with additional $e^\pm$ pairs created during the early, hot phase of the explosion. The density of hot $e^\pm$ deviates from annihilation balance when the explosion temperature drops to $\sim 20$\,keV due to adiabatic cooling, leaving an $e^\pm$ ``freeze-out.'' The plasma proper density at freeze-out radius $R_\pm$ is $n_e'(R_\pm)\sim (\Gamma_\pm/\sigma_{\rm T} R_\pm)$ where $\Gamma_\pm \sim 10$ \citep{Beloborodov2017}. At $r>R_{\pm}$, the proper density decreases as $r^{-3}$, which gives a rough estimate
\beq
  n_e' \sim \frac{\N_\pm}{r^3}, \qquad \N_\pm\sim \frac{r_0^2\Gamma_\pm^3}{\sigma_{\rm T}}\sim 10^{41}.
\eeq

In all layers of the explosion pulse, the density behaves as $n_e'=\N/r^3$ at large $r$, and we assume $\N\gtrsim 10^{39}$.
This density is sufficient to sustain the MHD regime for waves considered in this paper. Indeed, the plasma frequency is
\beq
  \omega_{\rm p}'=\left(\frac{4\pi e^2 n_e'}{m_e}\right)^{1/2}\approx \frac{2\times 10^9\,\N_{39}^{1/2}}{r_{10}^{3/2}}\,{\rm rad\,s}^{-1}.
\eeq
It is well above the MHz frequencies $\omega'$ (measured in the fluid frame) of waves participating in FRB production. Density $n_e'$ is also sufficient to avoid charge starvation. In particular, A-waves with wavevector $\bkA'$ and amplitude $E_{\rm A}'$ need electric current $j'=(c/4\pi) k_{\rm A,\perp}'E_{\rm A}'$. They become charge starved if $n_e'$ falls below the critical density defined by
\beq
   n_{\rm st}' \! \equiv \frac{j'}{ec}=\frac{k_{\rm A\perp}'E_{\rm A}'}{4\pi e}
  \!  \sim \frac{\omega'}{4\pi e c} \sqrt{\frac{L}{cr^2\Gamma^2}}
   \sim \frac{\nu_0}{2e c r} \sqrt{\frac{L}{c}},
\eeq
where $L$ is the isotropic equivalent of the wave power inside the explosion pulse measured in the lab frame, $\omega'=2\pi\nu'\sim 2\pi\Gamma\nu_0$ is the characteristic wave frequency in the fluid frame, and $\nu_0\sim 10$\,kHz is the pump wave frequency in the lab frame. Then, one finds
\beq
  \frac{n_e'}{n_{\rm st}'}\sim  \frac{2e c \N}{\nu_0 r^2}\sqrt{\frac{c}{L}}\approx 5\,\N_{39} r_{10}^{-2}L_{42}^{-1/2}>1.
\eeq

\begin{figure}
  \centering
\includegraphics[width=0.95\columnwidth]{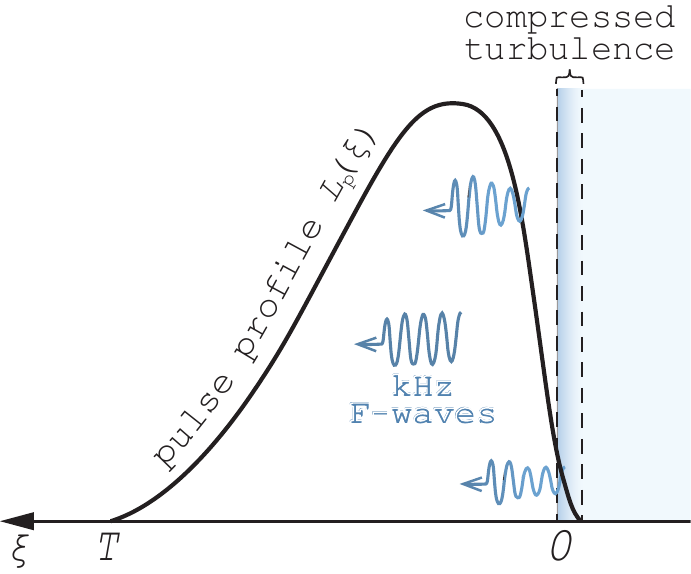}
  \caption{Schematic pulse profile $\Lp(\xi)$ in coordinate $\xi=t-r/c$. The explosion power $\Lp$ is expected to steeply  rise at the leading edge at $\xi\approx 0$ and gradually drop toward the end of the pulse at $\xi=T$. The outer magnetosphere swept by the pulse is compressed into a thin layer $\delta\xi_{\rm swept}\ll T$ (\Eq~\ref{eq:swept}). The pulse is static in coordinate $\xi$ while the ambient F-waves flow to larger $\xi$ and cross the pulse.}
\label{fig:pulse}
\end{figure}

The outer magnetosphere, which is swept by the pulse at $r\gg r_0$, becomes compressed into a thin shell surfing at the leading edge of the explosion (Figure~\ref{fig:pulse}). The pulse structure is convenient to view in coordinate $\xi=t-r/c$ where $\xi=0$ at the leading edge of the pulse and $\xi=T\sim 1$\,ms at its end. The pulse profile $\Lp(\xi)=cr^2\Bp^2$ is static in coordinate $\xi$. The outer magnetosphere swept up by the explosion resides at $\xi\ll T$; its turbulent magnetic field lines are compressed into a layer of thickness
\beq
\label{eq:swept}
  \delta\xi_{\rm swept}\approx \frac{(1-\beta)r}{c}\approx \frac{r}{2\Gamma^2 c}\approx 1.7\times 10^{-4}\,\Gamma_3^{-2}\,{\rm ms}.
\eeq
In contrast to the trapped and compressed A-waves, the magnetospheric F-waves initially do not feel the explosion. They freely propagate across the magnetic field lines and fill the entire pulse $0<\xi<T$ until they become consumed by stimulated scattering. Hereafter we focus on the main part of the pulse behind the thin layer $\delta\xi_{\rm swept}$ and set $\xi=0$ immediately behind this layer.


\section{Stimulated scattering in the pulse}

\begin{figure*}
\centering
\includegraphics[width=0.97\textwidth]{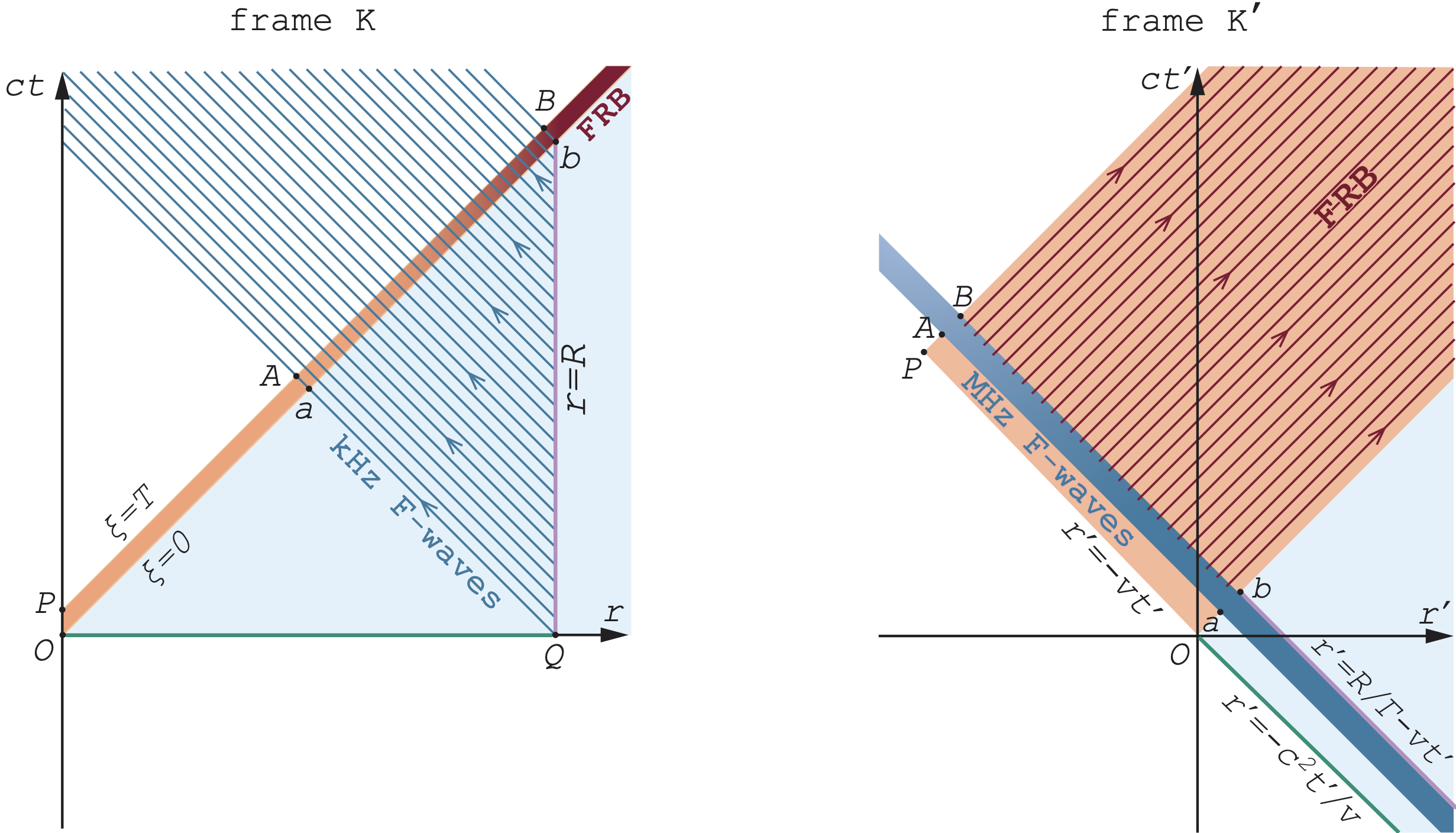}
\caption{{\em Left:} spacetime diagram of FRB emission in the lab frame. For clarity of the figure, a source of kHz waves is placed at a fixed radius $R$ (purple worldline) and sending  waves in the $-\hat r$ direction (dark blue worldlines). The FRB is generated where the kHz waves cross the explosion pulse (orange) of width $cT\ll R$. The generated GHz waves (dark brown) propagate outward with the pulse. FRB production peaks when the explosion pulse approaches radius $R$ where the kHz wave-pulse interaction ends.   {\em Right:} same process viewed in frame $\Kf$ moving outward with a high Lorentz factor $\Gamma=(1-v^2/c^2)^{-1/2}$ (fluid frame). Spacetime coordinates in frame $\Kf$ are $t'=\Gamma(t-v r/c^2)$ and $r'=\Gamma(r-vt)$. In this frame, the entire kHz wave train is a small bullet (dark blue) traveling through a wide region occupied by the pulse (orange): the width of the kHz wave train is compressed from $\sim R$ in the lab frame to $R/2\Gamma$ in frame $\Kf$ while the pulse width is increased from $cT$ to $2\Gamma cT\gg R/2\Gamma$. Events $O$, $P$, $a$, $b$, $A$, $B$ are indicated in both plots for easier comparison. Event $Q$ is not seen in the right plot, because it is located at $ct'=-\Gamma R$ (at the intersection of the green and purple lines) far from the origin $O$.}
\label{fig:spacetime}
\end{figure*}

\subsection{Global picture of ambient wave--pulse interaction}
\label{global}

Our final equations describing FRB production will be stated and solved in the fixed lab frame ${\cal K}$; however, the scattering rate will be given first in the moving fluid frame $\Kf$. It is useful to begin with a global picture, demonstrated with a spacetime diagram in frames ${\cal K}$ and $\Kf$ (Figure~\ref{fig:spacetime}). It shows a curious feature. In the lab frame, the kHz waves cross the pulse width $cT$ on the timescale $\sim T\ll r/c$, so the wave-pulse interaction has a quasi-steady pattern. By contrast, when viewed in frame $\Kf$, the entire kHz wave train of length $\sim r$ becomes a small ''bullet'' --- a short packet of MHz radiation. The bullet travels though a much wider region occupied by the explosion pulse ($T'=2\Gamma T$) and leaves behind a wake of exponentially amplified F-waves. This wake is the FRB (in the lab frame it has a GHz frequency and propagates in the radial direction). The ratio of bullet width $c\,\Delta\xi'_{\rm kHz}=r/2\Gamma$ to pulse width $cT'$ is 
\beq
  \frac{\Delta\xi'_{\rm kHz}}{T'}  = \frac{r}{4\Gamma^2 cT} \sim \frac{1}{4\Gamma},
\eeq
where we used $\Gamma\approx r/r_0$ and $cT\sim r_0$. Note also that the kHz waves are Doppler boosted in frame $\Kf$ to frequency $\nu' \sim 2\Gamma\nu_0$, and the number of wave oscillations in the train is invariant: $\nu\Delta\xi_{\rm kHz}=\nu'\Delta\xi'_{\rm kHz}$.

Our calculations will treat waves as wave packets with well-defined locations, traveling with the group speed. This usual picture of macroscopic radiative transfer (geometric optics) holds if the wave intensities vary on scales $\delta\xi'$ exceeding wavelength $c/\nu'$. Note that this condition is stated in the fluid frame $\Kf$, where wave interaction processes are defined and all participating waves have similar (MHz) frequencies $\sim\nu'$. Transforming the condition $\delta\xi'>c/\nu'$ to the lab frame, one obtains
\beq
    \delta\xi \! = \! \frac{\delta\xi'}{2\Gamma}  > \frac{1}{\Gamma\nu'}
    \sim \frac{1}{\Gamma^2\nu_0} \! \sim  \nu_{\rm FRB}^{-1},
\eeq
which defines a minimum timescale $\sim 1$ nanosecond. The actual smallest scales appearing in our calculations are $\delta\xi > 10^{-5}\,{\rm ms}\gg \nu_{\rm FRB}^{-1}$, and the corresponding smallest $\delta\xi'$ are large, comparable to the entire kHz wave train in frame $\Kf$. Therefore, geometric optics is a good approximation, and one can think of the interacting waves as particles with well defined worldlines.

We now describe the local interaction rate in the fluid frame $\Kf$, and then will shift the view to the lab frame where global calculations can be performed.

\subsection{Local scattering rate in fluid frame}
\label{scattering}

The pulse would be transparent to the ambient F-waves if they are treated as infinitesimal linear waves, since they propagate with no interaction. However, waves with any finite amplitudes experience nonlinear processes and become scattered inside the pulse.

All waves, F or A, which may exist or appear inside the pulse, have small amplitudes compared to the pulse field $\Bpf$. Such waves are in the realm of weak turbulence physics. They are well described in the leading order of  weak nonlinear processes called three-wave interactions: one wave transforms into two new waves. When viewed in the local fluid frame, this interaction obeys energy and momentum conservation, which is expressed as ``resonant conditions'' for the frequencies and wavevectors of the participating waves (similar conditions for Raman scattering in laser plasma are discussed e.g. in \cite{Kruer2003}). The F-waves can experience interactions of two types:\footnote{The process $F\rightarrow F+F$ vanishes when the fluid rest mass energy is strongly dominated by its magnetic field energy.}  (1) $F\rightarrow F+A$ and (2) $F\rightarrow A+A$.
Hereafter, we consider the situation where seeds of A-waves necessary for $F\rightarrow A+A$ are weaker than seeds of F-waves that drive the scattering process $F\rightarrow F+A$. Then, one can consider only
$F\rightarrow F+A$, since its coupling to process $F\rightarrow A+A$ is weak (Appendix~\ref{appB}).

\begin{figure}
\centering
\includegraphics[width=0.52\columnwidth]{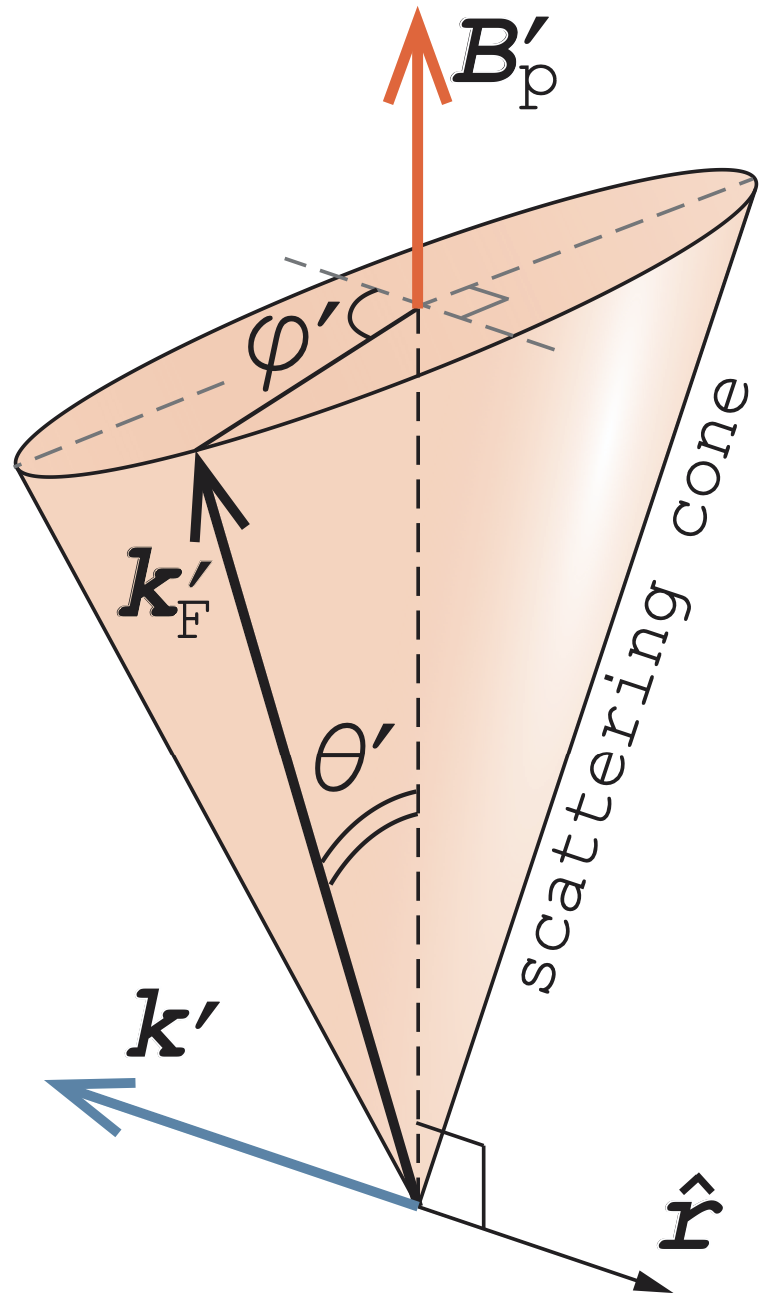}
 \caption{Wavevectors of F-waves participating in the process $F\rightarrow F+A$, viewed in the fluid frame $\Kf$. The pump wave has wavevector $\bkk$ in the $-\hat r$ direction, and amplified F-waves have wavevectors $\bkkF$ described by angles $\thf$ and $\phif$. Amplification peaks for $(\thf,\phif)$ near $(0,\pi/2)$ [and the symmetric direction $(\pi,-\pi/2)$]. The solid angle of strong amplification (the shaded cone) is around the direction of the pulse magnetic field $\bBp'$ (orange arrow).}
\label{Fig4}
\end{figure}

An ambient kHz F-wave with some random wavevector $\bk$ and angular frequency $\omega=ck$ is Doppler boosted in frame $\Kf$ by the factor $\omf/\omega\sim\Gamma$, and its wavevector $\bkk$ becomes directed along $-\hat{r}$ (Figure~\ref{Fig4}). The process $F\rightarrow F+A$ trades the pump wave $\bkk$ to a new F-wave with wavevector $\bkkF$ and frequency $\omFf=c\kkF$ plus an A-wave with wavevector $\bkkA$ and frequency $\omAf=c\kk_{\rm A\parallel}$, where subscript $\parallel$ indicates the component parallel to $\bBpf$. The resonance conditions for the three-wave interaction $F\rightarrow F+A$ are
\beq
\label{eq:res_om}
  \bkkF+\bkkA=\bkk, \qquad \omFf+\omAf=\omf.
\eeq
They imply 
\beq
\label{eq:omF_res}
  \omFf=\frac{\omf}{1+|\cos\thf|},
\eeq
where $\thf$ is the angle between $\bkkF$ and $\bBpf$. 

Like all three-wave processes, the rate of $F\rightarrow F+A$ is given by the standard weak-turbulence theory, applied here to the magnetically dominated fluid \citep{Thompson1998, Lyubarsky2019}, as summarized in Appendix~\ref{appA:KinEq}. The accuracy of the three-wave interaction formalism is verified by fully nonlinear simulations \citep{Solanki2026a}. The scattering rate is formulated using Lorentz-invariant ``occupation numbers.'' For instance, occupation numbers for the pump F-waves are defined as 
\beq
\label{eq:occup}
  n_{\bk}\equiv \frac{U_{\bk}}{\omega},
\eeq
where $U_{\bk}$ is the energy density distribution in Fourier space.\footnote{This definition gives $n_{\bk}$ with dimension of $\hbar$. The usual dimensionless occupation number in the quantum picture would be $n_{\bk}/\hbar$; however, the Planck constant $\hbar$ anyway cancels in the final result since the calculated process is classical.}
Occupation numbers $n_{\bkF}$ and $n_{\bkA}$ for the generated F and A waves have similar definitions.

The process $F\rightarrow F+A$ generates $n_{\bkF}$ and $n_{\bkA}$ with rate
\beq
\label{eq:rate}
   \dot n'_{\bkF}=\dot n'_{\bkA}=\WFA\frac{\Uf_{\omf}}{\omf}\,(n_{\bkF}+n_{\bkA}),
\eeq
where $\Uf_\omf=\dd\Uf/\dd\omf$ is the spectral energy density of the primary beam,
\beq
\label{eq:W}
   \WFA=\frac{\pi^2 \omf^3}{\Bpf^2}\,\fFA(\bOmFf), 
\eeq
and $\fFA(\bOmFf)$ is a dimensionless factor that depends on the direction vector $\bOmFf=\bkkF/\kkF$ and has a peak value of 1 (see Appendix~\ref{appA:KinEq}). While the occupation numbers are Lorentz invariant, their time derivatives are not; the prime in $\dot n'_{\bkF}=\dd n_{\bkF}/\dd t'$ highlights that the rate is measured in the local fluid frame $\Kf$.

Equation~(\ref{eq:rate}) assumes $n_{\bkF},n_{\bkA}\ll n_{\bk}$ and neglects the inverse process $F+A\rightarrow F$. This approximation holds even when most of the pump beam is consumed by $F\rightarrow F+A$. This is easy to see in frame $\Kf$: while $\kkF$ and $\kkA$ are comparable with $\kk$ in magnitude, the pump beam occupies a much smaller phase-space volume (since it is collimated within the angle $\sim\Gamma^{-1}$) and hence has a much higher occupation number. The condition $n_{\bkF},n_{\bkA}\ll n_{\bk}$ breaks when the scattering suppresses the pump beam by the factor of $\sim\Gamma^{-2}$.

The general expression for $\WFA$ \citep{Lyubarsky2019, GolbraikhLyubarsky2023} simplifies in our case because the primary F-waves are strongly beamed in the fluid frame $\Kf$ in the direction perpendicular to $\bBpf$. The angular-dependent factor $\fFA$ then becomes 
\beq
\label{eq:f1}
   \fFA \equiv \frac{2(1-\sin\thf\cos\phif)}{1+|\cos\thf|} |\cos\thf|\sin^2\!\phif,
\eeq
where $\thf$ and $\phif$ are defined by
\beq 
   \bOmFf=(\sin\thf\cos\phif,\sin\thf\sin\phif,\cos\thf),
\eeq
with the $x$ axis along $-\hat r$ (the direction of ambient waves in frame $\Kf$) and the $z$ axis along $\bBpf$.

Note that it is sufficient to have only F seeds to initiate the process $F\rightarrow F+A$, as the rates $\dot n'_{\bkF}=\dot n'_{\bkA}$ are proportional to $(n_{\bkF}+n_{\bkA})$. After the generated F and A waves have grown significantly above the seed level, we have $n_{\bkF}\approx n_{\bkA}$. The scattered F-waves then continue to grow with rate
 \beq
\label{eq:qf}
 \qf\equiv \frac{\dd\ln n_{\bkF}}{\dd\tf}=2\WFA \frac{\Uf_{\omf}}{\omf}=\frac{2\pi^2 \omf^2\Uf_{\omf}}{\Bpf^2} \, \fFA(\bOmFf).
\eeq
One can see that the growth rate is largest for $\bkkF$ with $|\cos\thf|=1$ and $\sin^2\!\phif=1$.

\subsection{FRB production rate in the lab frame}
\label{FRB_lab}

In the lab frame, the generated F-waves are Doppler shifted to frequency 
\beq
\label{eq:omF}
  \omF=\Gamma(1-\beta\sin\thf\cos\phif)\,\omFf\sim \Gamma^2\omega,
\eeq 
and become strongly beamed along $\hat r$, within angle $\Gamma^{-1}$. Therefore, they continue to ride with the explosion pulse at nearly constant $\xi=t-r/c$. We wish to know how $n_{\bkF}$ grows in the lab frame.

Note that $\kF^{\mu}=(\omF/c,\bkF)$ is a four-vector, which obeys Lorentz transformation, and $\kF^{\mu}/\kF=(1,\bOmF)$ describes the wave direction in spacetime. The rate $\dot{n}'_{\bkF}$ given in \Eq~(\ref{eq:rate}) is the growth of occupation number along the wave propagation direction in frame $\Kf$:
\beq
   \dot{n}'_{\bkF}=\frac{\kF^{\mu'}}{\kF'}\,\partial_{\mu'} n_{\bkF},
\eeq 
where $\partial_\mu\equiv c\,\partial/\partial x^{\mu}$ in spacetime coordinates $x^\mu$. The same expression but without primes gives $\dot{n}_{\bkF}$ in the lab frame. Using $\kF^{\mu}\,\partial_{\mu} =\kF^{\mu'}\partial_{\mu'}$ one finds:
\beq
\label{eq:dnF_lab}
   \dot{n}_{\bkF} = \frac{\kF^{\mu}}{\kF}\,\partial_{\mu} n_{\bkF} =\frac{\omFf}{\omF}\dot{n}'_{\bkF}.
\eeq
This gives the growth rate $q=\dd\ln n_{\bkF}/\dd t$ in the lab frame:
\beq
   q=\frac{\omFf}{\omF}\,\qf.
\eeq

The growth rate of a given wave $\bkF$ can be expressed in terms of quantities measured in the lab frame using transformation of frequency (\Eq~\ref{eq:omF}) and pulse field $\Bp=\Gamma\Bpf$. One can also relate the pump beam density $\omf\Uf_\omf\sim \Uf$ to the corresponding density $U=U_0$ in the lab frame. Note that $\omf$ scales as $\Gamma$, and $\omf\Uf_\omf$ scales as $\Gamma^2$. It is convenient to define the following dimensionless quantities
\beq
\label{eq:bY}
   b\equiv \frac{\omf}{2\Gamma\omega_0}, \qquad Y\equiv \frac{\omf\Uf_\omf}{4\Gamma^2 U_0},
   \qquad Z\equiv b\,Y,
\eeq
where $\omega_0$ is the characteristic frequency of the ambient F-waves; it is in the kHz band. Note that both $b$ and $Y$ are independent of $\Gamma\gg 1$. Thus, the function $Y(b)$ [or $Z(b)$] describes the spectrum of the pump beam in a way that is independent of the local fluid motion $\Gamma(\xi)$. $Y(b)$ and $Z(b)$ are uniform across the pulse if the depletion of the pump wave is negligible. 

We can now express the growth rate $q$ as follows:
\beq
\label{eq:q}
   q = \frac{\pi}{2}\,\eF\,\omega_0 Z(b)\,f(\bOmFf),  \qquad f \equiv  \frac{2|\cos\thf|\sin^2\!\phif}{1+|\cos\thf|}.
\eeq
For the angular dependence of $q$ we keep the angles $\thf$ and $\phif$ that describe the scattered wave direction in the fluid frame; this gives a convenient view of the angular structure. The corresponding direction of $\bkF$ in the lab frame can be found from the Lorentz transformation of the four-wavevector $\kF^\mu$. Recall also that for a given $\bkF$, $\omf$ is determined by the resonance condition: $\omf(\omF,\bOmFf)=\omFf(1+|\cos\thf|)$ (\Eq~\ref{eq:omF_res}) with $\omFf$ related to $\omF$ by the Doppler transformation (\Eq~\ref{eq:omF}). Thus,
\beq
\label{eq:b}
   b(\omF,\thf,\phif) = \frac{\omF(1+|\cos\thf|)}{2\Gamma^2\omega_0 (1-\beta\sin\thf\cos\phif)}.
\eeq
The value of $b(\bkF)$ shows which part of the pump wave spectrum resonantly amplifies a given wave $\bkF$.

The dimensionless function $Z(b)$ depends on the spectrum of the background waves $U_\omega$. For example, an isotropic kHz background with spectral density $U_\omega$ given by \Eq~(\ref{eq:bg}) yields $Z=Z_0$ where
\beq
\label{eq:Z0}
Z_0(b) \! = \! \frac{1}{K}\times
    \left\{\begin{array}{lr}
\vspace*{2mm}
 \displaystyle{ b^4\left(\frac{1-b^{\alpha_1-2}}{\alpha_1-2}+\frac{1}{\alpha_2+2}\right)} & b<1 \\
 \displaystyle{ \frac{b^{2-\alpha_2}}{\alpha_2+2} } & \;b>1
                             \end{array}\right.
\eeq
This expression is found by integrating over all directions of the background waves $\bOm$ and taking into account that waves with different $\bOm$ contribute to $\omf\Uf_\omf$ with different Doppler shifts $\D=\Gamma(1-\bOm\cdot\bb)$. Subscript ``0'' in $Z_0$ highlights that it was calculated for the ambient kHz waves neglecting their consumption by stimulated scattering inside the pulse. For $\alpha_1=\alpha_2=3$ (used in our sample models), $Z_0(b)$ peaks at $b_{\rm max}=24/25$ and its maximum value is $Z_0^{\max}=b_{\rm max}^5/3\approx 0.272$. Note that $Z_0(b)$ is smooth even when $U_\omega$ has a sharp kink at $\omega_0$. Thus, our simple prescription for $U_\omega$ gives the FRB growth rate $q$ with a smooth spectral peak at $b_{\max}$, similar to models with elaborated smooth shapes of $U_\omega$.

\subsection{Results in the unsaturated regime}

In this section, we calculate FRB production in the unsaturated regime, when the occupation numbers of the ambient kHz waves $n_{\bk}$ are not depleted by scattering. This corresponds to a linear phase of parametric instability, which proceeds exponentially. In this regime, $Z(b)$ is unchanged from $Z_0(b)$, a fixed given function. Then, \Eqs~(\ref{eq:q}) and (\ref{eq:b}) form a complete set describing the emission process. They determine FRB production by the explosion pulse as it expands through the sea of magnetospheric kHz waves to the exit radius $\Rout \sim \RLC$.

Our sample model will assume a magnetar with $\RLC=10^{10}$\,cm, so we set $\Rout=10^{10}$\,cm. The explosion power $\Lp=4\pi r^2\Up c$ determines the fluid Lorentz factor $\Gamma$ in the pulse: 
\beq
\label{eq:gam_Lp}
  \Gamma=\left(\frac{\Lp}{32\pi r^2\Ubg c}\right)^{1/4}
  = \left(\frac{\Lp}{4 c \mu^2}\right)^{1/4} r.
\eeq
The pulse of duration $T$ resides in the interval $0<\xi<T$ with a profile $\Lp(\xi)$, which determines $\Gout(\xi)$ reached at $r=\Rout$. As an example, we will use the maximum $\Gout=500$, which corresponds to a peak power $\Lp^{\max}=7.5\times 10^{47}\mu_{33}^2$\,erg\,s$^{-1}$.

For the ambient waves, it is reasonable to  assume a spectrum peaking at $\nu_0=10$\,kHz (section~\ref{ambient}). The scattering process is controlled by two dimensionless parameters: $\nu_0\Rout/c\approx 3\times 10^3$ and $\eF(r)=U_0/\Ubg<1$. Our model assumes
$\eF\propto r^3$, so 
\beq
\label{eq:eF}
   \eF(r)=\eF^{\rm out}\left(\frac{r}{\Rout}\right)^3.
\eeq 
For typical parameters of the problem, we find that the unsaturated regime occurs for $\eF^{\rm out}\! \lesssim \! 0.02$ (depending logarithmically on the level of GHz seeds).

Consider now any location $\xi$ inside the pulse, with a given $\Gout(\xi)$. We wish to find the observed spectrum of the generated FRB luminosity (isotropic equivalent) at this $\xi$. It is given by
\beq
   \frac{\dd L}{\dd \omF}=4\pi r^2 c\,\frac{\dd\UF}{\dd\omF}, 
\eeq
where $\UF(r,\xi)$ is the energy density of the GHz F waves generated in the pulse. Since the generated waves are strongly beamed in the lab frame, they stay nearly static in $\xi$, moving with the pulse. Then, the production of GHz radio waves occurs separately at different $\xi$: the growth of $\UF(r)$ at a given $\xi$ is independent from what happens at other $\xi$ (in contrast to the saturated regime discussed in the next section). The growth of  a wave with a given $\bkF$ is described by 
\beq
   n_{\bkF}=A(\bkF,r)\,\nseed_{\bkF},
\eeq
where $\nseed_{\bkF}$ is the initial (seed) occupation number, and
\beq
\label{eq:A}
    A(\bkF,r)=e^\tau, \qquad  \tau \equiv \int_{\Rin}^{\Rout} q\left[\bkF,\Gamma(r)\right] \frac{\dd r}{c}.
\eeq
Here, we used $r/c=t-\xi\approx t$ as a time coordinate.\footnote{The pulse is a thin shell (width $cT\ll r$) located at $r\approx ct$. The coordinate $\xi\ll r/c$ is used to describe the pulse structure across its small width.}
FRB production sharply peaks at $\Rout$ and the result is insensitive to $\Rin\ll\Rout$; it is sufficient to choose e.g. $\Rin=0.2\Rout$.

The integral in \Eq~(\ref{eq:A}) is taken along the straight ray of each seed wave $\bkF$. It simplifies due to a remarkable feature of radiative transfer in relativistic outflows with $\Gamma\propto r$: the wave direction vector $\bOmFf$ measured in the local fluid frame stays constant along the ray (see section~5 in \cite{Beloborodov2011}). This occurs because two effects that could change $\bOmFf$ cancel each other: the wave angle $\psi$ with respect to the local radial direction decreases along the ray as $\sin\psi\propto r^{-1}$, and its transformation to the fluid frame $\sin\psi'=\Gamma\sin\psi$ keeps $\sin\psi'=\cnst$ since $\Gamma\propto r$. This implies $f(\bOmFf)=\cnst$ along the ray. Therefore, the wave growth rate $q\propto Z(b)$ (\Eq~\ref{eq:q}) changes along the ray only because $Z(b)=Z_0(b)$ changes as $b$ evolves: $b\propto \Gamma^{-2}\propto r^{-2}$ at $\bOmFf=\cnst$ and $\beta\approx 1$ (\Eq~\ref{eq:b}). Then, we find
\beq
\label{eq:tau}
 \tau(\bkF)=
  \frac{\pi}{2}\,\eF^{\rm out}\,\frac{\omega_0\Rout}{c} \,f(\bOmFf) \int_{x_{\rm in}}^1 \!\! Z(b)\, x^3 \dd x,
\eeq 
where $x=r/\Rout$ and $b(x)$ is given by \Eq~(\ref{eq:b}) with $\Gamma=x \Gout$ and $\beta\approx 1$.

The obtained $\tau({\bkF})$  peaks at $\bkF^{\max}$ that has $\cos\thf=1$, $\phif=\pi/2$, and $\omF\approx 0.776\,\Gout^2\omega_0$. The peak value of $\tau$ is
\beq
\label{eq:tau_max}
  \tau_{\rm max}\approx 0.505\, \eF^{\rm out}\,\frac{\nu_0\Rout}{c}.
\eeq
Our sample model has $\eFout=0.02$ and $\tau_{\rm max}\approx 34$. Since $A(\bkF)=e^\tau$ is exponentially sensitive to deviations of $\bkF$ from $\bkF^{\max}$, the peak of $A(\bkF)$ has a needle shape. A small deviation reducing $\tau$ from $\tau_{\max}$ e.g. by $\Delta\tau=0.1\tau_{\max}$ suppresses $A$ by $e^{-3.4}$. Thus, the produced FRB strongly peaks in a narrow range of frequencies and directions. The preferred direction of amplified waves $\bkkF\parallel \bBp'$ simplifies \Eq~(\ref{eq:omF}) to $\omF/\omFf\approx\Gamma$ --- the amplified waves have a narrow distribution of Doppler factors around $\Gamma$.

\begin{figure}[t]
\includegraphics[width=0.47\textwidth]{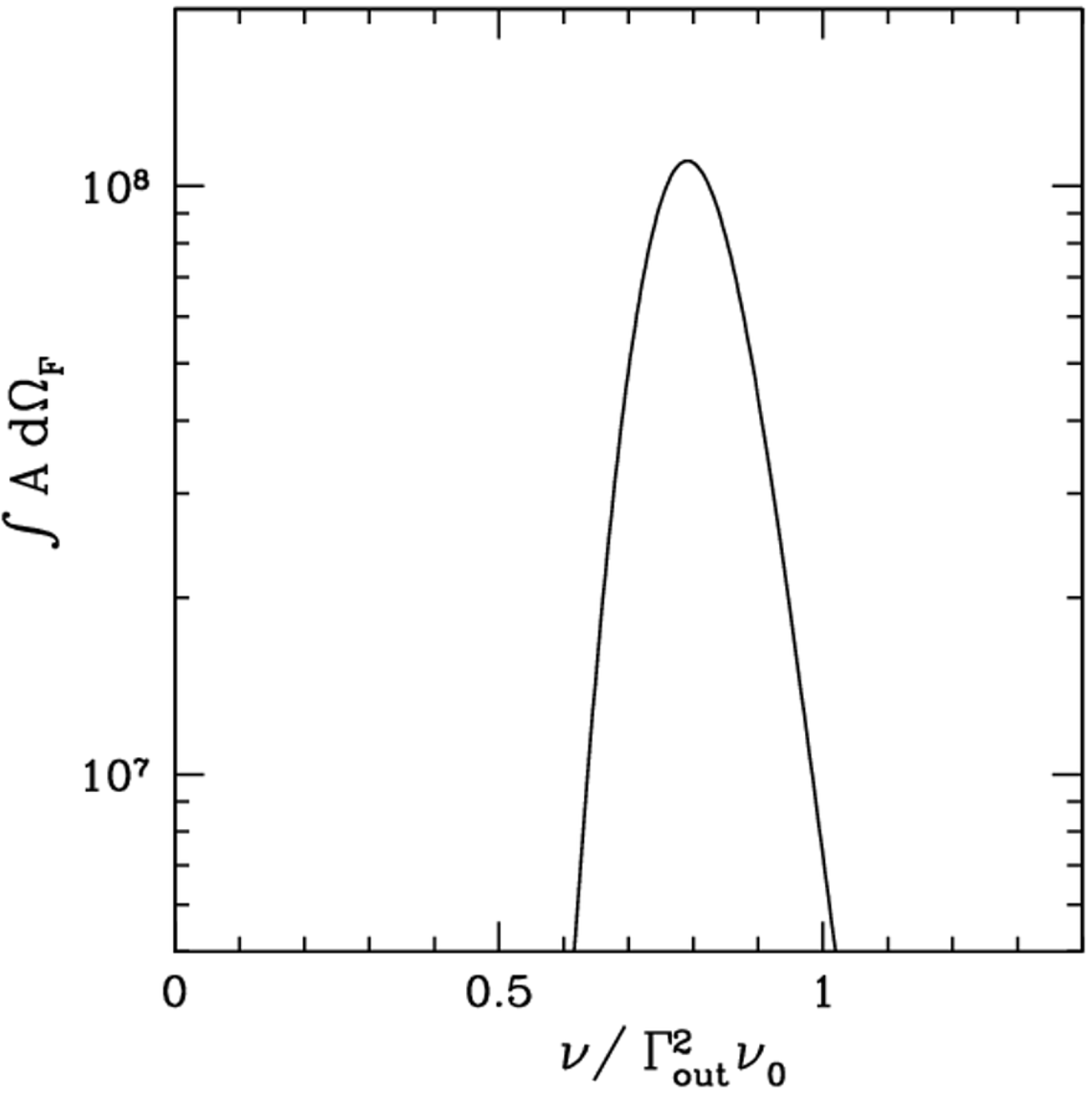} 
\caption{FRB spectrum in the unsaturated regime, found in the model with $\eF^{\rm out}=0.02$ and $\nu_0\Rout/c=3.33\times 10^3$. Frequency of the generated waves $\nu=\omF/2\pi$ is normalized to $\Gout^2\nu_0$, where $\nu_0=\omega_0/2\pi$ is the peak frequency of the ambient wave spectrum and $\Gout$ is the fluid Lorentz factor in the pulse when it reaches $\Rout$. When presented in this form, the result is independent of $\Gout\gg 1$. The spectrum peaks at $\nupeak=0.79\,\Gout^2\nu_0$; its full width at half-maximum is $\Delta\nu/\nupeak\approx 0.18$.}
\label{fig:unsat}
\end{figure}

After calculating $A(\bkF)$ on a grid in $\bkF$ space, one can find $\dd\UF/\dd\omF$ by integrating the amplified $\bkF$-waves over their directions $\bOmF$ at fixed $\omF=c\kF$:
\beq
\label{eq:UF}
  \frac{\dd\UF}{\dd\omF} \approx \frac{\omF^3}{c^3} \, \nseed_{\bkF}  {\cal A}(\omF),  
  \quad\;   {\cal A}(\omF) \! \equiv \! \int  \! A(\bkF) \, \dd\OmF.
\eeq 
Here, we used $\int A\, \nseed_{\bkF}\,\dd\OmF\approx \nseed_{\bkF} \int A \,\dd\OmF$; this approximation holds if the seed intensity of GHz waves is approximately uniform across the small solid angle $\delta\OmF\sim\pi/\Gout^2$ where the wave amplification occurs. Since $\tau\propto f(\bOmFf)$ and $\bOmFf=\cnst$ along the ray, it is convenient to rewrite the angular integral using $\dd\OmF/\dd\OmFf=(\omFf/\omF)^2$:
\beq
\label{eq:Atot}
  {\cal A}(\omF)  \! = \! \! \int  \! \frac{A(\bkF)\,\dd\OmFf}{\Gout^2(1-\sin\thf\cos\phif)^2}.
\eeq
Figure~\ref{fig:unsat} shows ${\cal A}(\omF)$ calculated for the sample model with $\eF^{\rm out}=0.02$. The integral accumulates in a small solid angle $\delta\OmFf\sim 0.1$ around $\bkF^{\max}$ where $\tau$ is close to $\tau_{\max}$. The peak value of ${\cal A}(\omF)$ is 
\beq
  {\cal A}_{\rm peak} \!\approx 0.06\,\frac{e^{\tau_{\max}}}{\Gout^2} \! \approx 10^8 \;\;\; {\rm at} \; \omF^{\rm peak} \! \approx 0.79\,\Gout^2\omega_0.
\eeq

If the source of seed GHz radiation has a radius $R_{\rm s}\ll\Rout$, the seed radiation is expected to occupy a solid angle $\Omega_{\rm s}\approx \pi R_{\rm s}^2/\Rout^2$. The model shown in Figure~\ref{fig:unsat} assumes $\Omega_{\rm s}\gg \Omega_{\rm ampl}\sim \pi/\Gout^2$ with $\nseed_{\bkF}$ approximately uniform across the amplification solid angle.\footnote{The model can be easily extended to seed radiation with any  beaming profile $w(\bOmF)$, which is equivalent to changing the initial condition $A=1$ to $A=w(\bOmF)$ and keeping isotropic $\nseed_{\bkF}$.} 
Then, the obtained FRB luminosity is related to the seed luminosity by 
\beq
\label{eq:Lnu}
    L_\nu \approx \frac{{\cal A}(\omF)}{\Omega_{\rm s}} L^{\rm seed}_\nu, \qquad \nu\equiv\frac{\omF}{2\pi}.
\eeq
The FRB spectrum is shaped by ${\cal A}(\omF)$, assuming that the seed spectrum $\dd L_{\rm seed}/\dd \omF$ weakly varies across the narrow peak of ${\cal A}(\omF)$. Its full width at half-maximum (FWHM) in the example shown in Figure~\ref{fig:unsat} is $\Delta\nu/\nu\approx 0.18$. The bolometric luminosity of the FRB is $L\sim 0.1\,\nu L_\nu$.


\section{Saturated regime}
\label{saturated}

Amplification of GHz radiation $n_{\bkF}=e^\tau\nseed_{\bkF}$ cannot continue indefinitely with increasing $\tau$. When $\tau$ exceeds some $\tau_{\rm sat}$ the amplification process saturates by depleting the ambient kHz waves available for scattering inside the pulse. The value of $\tau_{\rm sat}$ weakly depends on $\nseed_{\bkF}$: if $A=e^{30}$ is not enough for saturation then $e^{50}$ will certainly suffice; so, conservatively,  $\tau_{\rm sat}=40\pm 10$. Using $\tau_{\max}$ from \Eq~(\ref{eq:tau_max}), one can see that the transition to saturation, $\tau_{\max}>\tau_{\rm sat}$, occurs when
\beq
\label{eq:sat}
   \eF^{\rm out}\! > \! \frac{ 2c\tau_{\rm sat}}{\nu_0\Rout} \!
     =\! 0.024\! \left(\frac{\tau_{\rm sat}}{40}\right) \! \left(\frac{\nu_0}{10\,\rm kHz}\right)^{-1}
     \!\left(\frac{\Rout}{10^{10}\,\rm cm}\right)^{-1}.
\eeq
This condition corresponds to a moderate level of magnetospheric turbulence, so it can be satisfied in magnetars.

The FRB energy $\EFRB$ in the saturated regime was estimated in section~\ref{picture}. One can compare it with the cumulative energy of the explosion pulse,
\beq
  \Enp(\xi)\approx \Bp^2 r^2 c\xi \approx 4\Gamma^4\Bbg^2 r^2 c\xi,  
\eeq  
where $\xi$ equals the observer time $\tobs$ measured from the leading edge of the pulse, $0<\xi<T$. This gives 
\beq
\label{eq:eta}
    \eta\equiv \frac{\EFRB}{\Enp}\sim \frac{\eF }{4\Gamma^2 \xi q} \sim \frac{1}{4\nu\xi},
\eeq
where $\nu\equiv \omF/2\pi$. Although the numerical coefficient in this estimate is not accurate, one can anticipate efficiencies $\eta\sim 10^{-6}$ for FRBs with durations $\tobs\sim\xi\sim 1$\,ms, and higher $\eta$ for bursts of shorter durations. This is confirmed with detailed calculations below. Our next main goal is to find the spectrum of the produced FRB.

\subsection{Depletion of ambient kHz waves inside the pulse}

FRB production at each wavevector $\bkF$ is given by the growth rate of GHz waves (\Eq~\ref{eq:q}). It is controlled by the function $Z(b)$ that describes the spectrum of the pump beam. In the saturated regime, one can no longer set $Z(b)$ equal to a given, fixed $Z_0(b)$, as the depletion of kHz waves inside the pulse implies a local reduction of $Z(b)$ from $Z_0(b)$. So, the calculation of FRB production now involves tracking the diminishing $Z(b)$ at all locations $\xi$ across the explosion pulse, $0<\xi<T$. At any moment of time, the state of the GHz and kHz waves inside the pulse is described by $A(\bkF,\xi)$ and $Z(b,\xi)$, and we now derive the equation for the self-consistent (depleted) $Z(b,\xi)$.

The scattering process consuming kHz waves should be first viewed in the fluid frame $\Kf$, where all waves have MHz frequencies and the scattering $F\rightarrow F+A$ satisfies conservation of energy and momentum. It is convenient to use the wave occupation numbers $n_{\bkF}$ (amplified wave) and $n_{\bk}$ (pump wave), as defined in section~\ref{scattering}. The consumption rate of $n_{\bk}$ equals the rate of process $F\rightarrow F+A$ integrated over the final F + A  states. This integration yields (see Appendix~\ref{appA:KinEq}): 
\beq
\label{eq:dn_fluid}
 \dot{n}'_{\bk} =- 2 n_{\bk}\frac{\omf^2}{c^3} \int \frac{\WFA n_{\bkF}  \, \dd\OmFf}{(1+|\cos\thf|)^3}.
\eeq

Now  we switch the view to the lab frame. Similar to \Eq~(\ref{eq:dnF_lab}), $\dot n_{\bk}$ in the lab frame is related to $\dot{n}'_{\bk}$ in the fluid frame by
\beq
\label{eq:dn_trans}
   k\, \dot{n}_{\bk} = k^\mu\,\partial_{\mu} n_{\bk} = k' \dot{n}'_{\bk}.
\eeq
The four-vector $k^\mu$ in the lab frame has $k^t=k$ and $k^r=k\cos\psi$, where $\psi$ is the propagation angle relative to the radial direction. It is convenient to use $t,\xi$ as spacetime coordinates instead of $t,r$. The relations $\left.\partial_t\right|_r= \left.\partial_t\right|_{\xi} + \left.\partial_\xi\right|_{t}$ and $\left.c\,\partial_r\right|_t = -\left.\partial_\xi\right|_t$ imply
\beq
\label{eq:t_xi}
  k^{\mu}\partial_\mu=k\left.\partial_t\right|_{\xi}+(k-k^r)\left.\partial_\xi \right|_t,
\eeq
and \Eq~(\ref{eq:dn_trans}) becomes
\beq
   k\,(\partial_t n_{\bk})_{\xi}+(k-k^r)(\partial_\xi n_{\bk})_t=\Gamma(k-\beta k^r) \dot{n}'_{\bk},
\eeq
where we used $k'=\Gamma(k-\beta k^r)$. In this relation, $\beta=\sqrt{1-\Gamma^{-2}}$ can be replaced with $\beta=1$ with high accuracy, since $\Gamma^{-2}\lesssim 10^{-5}$. Note also that the ambient waves cross the explosion pulse $0<\xi<T$ on a short timescale,
\beq
  t_{\rm cross}=\frac{T}{1-k^r/k}\ll \frac{r}{c},
\eeq
and hence the $\xi$-profile of $n_{\bk}$ across the pulse is quasi-steady: $(\partial_t n_{\bk})_\xi\ll (\partial_\xi n_{\bk})_t$. Thus, the equation describing $n_{\bk}(\xi)$ simplifies to
\beq
   \partial_\xi n_{\bk}=\Gamma \dot{n}'_{\bk}.
\eeq
Substituting \Eq~(\ref{eq:dn_fluid}) for $\dot{n}'_{\bk}$, we find
\beq
\label{eq:dn_dxi}
   \partial_\xi n_{\bk}
   = - \frac{2\pi^2}{\Bpf^2 c^3 }\, \Gamma \omf^5n_{\bk} \int   \frac{n_{\bkF}\fFA(\bOmFf)\, \dd\OmFf}{(1+|\cos\thf|)^3},
\eeq
where $\fFA(\bOmFf)$ is given in \Eq~(\ref{eq:f1}).

Note that $\Uf_{\omf}$  and $Z$ are related to $n_{\bk}$ as follows
\beq
  \Uf_\omf = \frac{\omf^3}{c^3}\!\int  n_{\bk}\, \dd\Omf, \quad\;\; 
  Z= \frac{4\omega_0^4}{c^3U_0}\, b^5  \Gamma^2\! \! \int  n_{\bk}\, \dd\Omf,
\eeq
where we used the definitions of $b$ and $Z$ (\Eq~\ref{eq:bY}). One can find $\partial_\xi Z$ by integrating \Eq~(\ref{eq:dn_dxi}) over $\dd\Omf$. The pump waves are strongly beamed in the fluid frame, so the integral peaks in a narrow solid angle $\delta\Omf\propto \Gamma^{-2}\ll 1$ around $\bkk= -(\omf/c)\hat{r}$.\footnote{Furthermore, $\Gamma^2 d\Omf$ is independent of the local $\Gamma(\xi) \gg 1$. This implies $\partial_\xi (\Gamma^2 \! \int n_{\bk}\, d\Omf)= \Gamma^2\! \int \partial_\xi n_{\bk}\, d\Omf$.}
Therefore, when integrating \Eq~(\ref{eq:dn_dxi}) over $\dd\Omf$, we simply integrate $n_{\bk}$ on both sides (while all other terms remain constant) and obtain
\beq
\label{eq:Z_ev}
   \partial_\xi \ln Z
   = - \frac{64\pi^2 \omega_0^5}{\Bp^2 c^3 }\, \Gamma^8 b^5 \int   \frac{n_{\bkF}  \fFA(\bOmFf)\, \dd\OmFf}{(1+|\cos\thf|)^3},
\eeq
where we substituted $\Bpf=\Bp/\Gamma$.

Onset of the saturation regime occurs when the pump wave spectrum $Z(b)$ is significantly depleted below $Z_0(b)$, which requires 
\beq
\label{eq:onset}
  \xi \,\partial_\xi \ln Z\sim 1 \qquad ({\rm depletion~onset}).
\eeq 
Since $\partial_\xi \ln Z$ scales with $n_{\bkF}\propto A=e^\tau$, this condition quickly becomes satisfied at large $\tau$. Then, $Z(b)$ has two distinct parts: $Z\ll Z_0$ in a certain frequency band calculated below, 
and $Z\approx Z_0$ outside this band.

\subsection{Growth of GHz waves inside the pulse}

\Eq~(\ref{eq:Z_ev}) for $Z$ is coupled to the evolution of $n_{\bkF}$, which is described by \Eq~(\ref{eq:dnF_lab}). We rewrite it in coordinates $t,\xi$ using the identity stated in \Eq~(\ref{eq:t_xi}):
\beq
\label{eq:nkF_ev_}
 \frac{\kF^{\mu}}{\kF}\,\partial_{\mu} n_{\bkF} 
 = (\partial_t n_{\bkF})_{\xi} +(1-\cos\psi)(\partial_\xi n_{\bkF})_t = q n_{\bkF},
\eeq
where $q\propto Z$ (\Eq~\ref{eq:q}) and $\cos\psi\equiv \kF^r/\kF$. Lorentz transformation of $\kF^\mu$ gives angle $\psi$ in terms of $\thf$ and $\phif$ that describe wave $\bkF$ in the fluid frame $\Kf$: 
\beq
  1-\cos\psi=\frac{(1-\beta)(1+\sin\thf\cos\phif)}{1-\beta\sin\thf\cos\phif}.
\eeq
The equation for $\n_{\bkF}$ then becomes (using $\Gamma\gg 1$):
\beq
\label{eq:nkF_ev}
   \partial_t \ln n_{\bkF} +\frac{(1+\sin\thf\cos\phif)}{2\Gamma^2(1-\sin\thf\cos\phif)}\, \partial_\xi \ln n_{\bkF} = q.
\eeq

The term $(1-\cos\psi)(\partial_\xi \ln n_{\bkF})_t$ describes advection of the amplified waves in $\xi$. This term is normally negligible because $1-\cos\psi\sim \Gamma^{-2}\lesssim 10^{-5}$ for $\Gamma$ of main interest. However, advection is not negligible everywhere. We will show below that in the saturation regime $A$ and $Z$ develop sharp gradients in $\xi$ at the leading edge of the pulse, and here advection in $\xi$ becomes important. Therefore, our numerical simulations  will use the full \Eq~(\ref{eq:nkF_ev}). The advection term will be neglected in the simplified analytical model discussed in the next section; then, \Eq~(\ref{eq:nkF_ev}) simplifies to $(\partial_t \ln n_{\bkF})_{\xi}=q$.

\subsection{A simplified analytical model}
\label{toy}

Before presenting the complete numerical solution for the coupled evolution of $n_{\bk}$ and $n_{\bkF}$, we demonstrate its basic features with a simplified analytical model which helps develop an intuitive picture.

Consider the explosion pulse. It  occupies a shell of thickness $cT$ and remains static in coordinate $\xi=t-r/c$. Suppose we are given the fluid Lorentz factor inside the pulse $\Gamma\gg 1$, which corresponds to magnetic field $\Bp=2\Gamma^2\Bbg$. The pulse contains tiny GHz seeds $\nseed_{\bkF}$. Our simplified model will assume that (1) the ambient kHz waves $\bk$ are beamed in the $-\hat r$ direction, and (2) the amplified GHz waves $\bkF$ at each $\xi$ are beamed along $\bBpf$ in the local fluid frame $\Kf$, so that the direction of $\bkkF$ has the angles $|\cos\thf|=1$ and $\sin^2\!\phif=1$ (such waves experience strongest amplification, see section~\ref{scattering}).

In this ``two-beam'' model, induced scattering occurs from beam~1 ($\bk$) to beam~2 ($\bkF$). Then, $\WFA$ (\Eq~\ref{eq:W}) becomes 
\beq
   \WFA = \frac{\pi^2\omf^3}{\Bpf^2}.
\eeq 
At each $\xi$, frequencies $\omega=ck$ and $\omF=c\kF$ of the waves participating in induced scattering  are related by
\beq
\label{eq:om_beam}
  \omF=\Gamma\omFf=\Gamma\frac{\omf}{2}=\Gamma^2\omega.
\eeq
\Eqs~(\ref{eq:dn_dxi})  and (\ref{eq:nkF_ev}) for $n_{\bk}$ and $n_{\bkF}$ then simplify to
\begin{align}
\label{eq:toy1}
      \partial_\xi n_{\bk} &=-2\Gamma n_{\bk} \frac{\pi^2\omf^2}{\Bpf^2} \,\Uf_{\omFf}, 
 \\ 
\label{eq:toy2}
      \partial_t n_{\bkF}  &= \frac{2}{\Gamma} n_{\bkF} \frac{\pi^2\omf^2}{\Bpf^2} \Uf_{\omf}, 
\end{align}
where we neglected the advection term $\propto  \partial_\xi \ln n_{\bkF}$. Spectral energy density $U_\omega$ of any F-waves in any frame is related to their occupation number by 
\beq
  U_\omega = \omega n_\omega= \frac{\omega}{c} \int n_{\bk}\, k^2 \dd\Omega 
  =  k^3 \! \int \! n_{\bk}\,\dd\Omega.
\eeq
$U_\omega$ may be thought of as number density (multiplied by $\hbar$) of waves with frequencies around a given $\omega$: $n\equiv \omega n_\omega$. Integration of \Eqs~(\ref{eq:toy1}) and (\ref{eq:toy2}) over the solid angles of the two beams gives two coupled equations for $\n'\equiv \Uf_{\omf}$ and $\nF'\equiv \Uf_{\omFf}$.

Transformation of the beams from fluid frame to lab frame leaves $n_\omega$ invariant: $n_\omega=n_{\omf}$ and $n_{\omF}=n_{\omFf}$. This allows one to rewrite the equations in the form
\begin{align}
\label{eq:n_ev}
      2c\,\partial_s \n &=- \R\, \n\,\nF,
       \\
\label{eq:nF_ev}
      \partial_t \nF  &=  \R\, \n\, \nF,    
\end{align}
where $\n \equiv  \omega n_{\omega}$, $\nF \equiv \omF n_{\omF}$, $s\equiv c\,\xi$ is the distance measured from the leading edge of the pulse, and 
\beq
  \R=\frac{4\pi^2\omf^2}{\Bpf^2}=\frac{16\pi^2\omF^2}{\Bp^2}=\frac{4\pi^2\omega^2}{\Bbg^2}.
\eeq
Note that the evolution of $\nF$ at a given frequency $\omF$ is coupled only to $\n$ at frequency $\omega=\omF/\Gamma^2$. The system breaks up into independent evolution problems for each pair $(\omega,\omF)$, with no coupling between different pairs.

\begin{figure}[t]
\vspace*{1mm}
\includegraphics[width=0.47\textwidth]{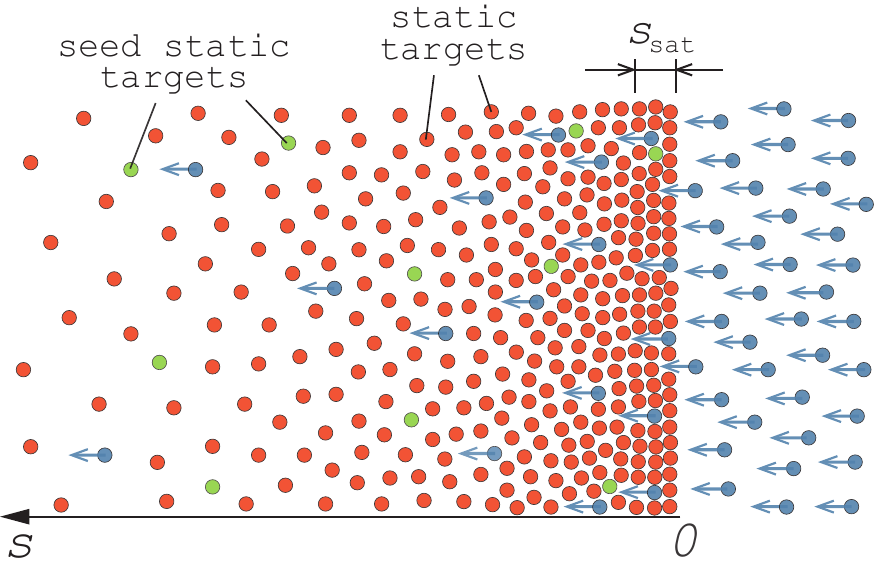} 
\caption{Stimulated scattering described by \Eqs~(\ref{eq:n_ev}) and (\ref{eq:nF_ev}) at each pair of frequencies $(\omega,\omF)$ is equivalent to a particle collision problem: kHz waves correspond to a particle beam (blue) and GHz waves correspond to static targets (green and orange) residing at $s>0$. At $t=0$, sparse seed targets (green) weakly attenuate the beam. A beam particle colliding with a target turns into a new static target (orange). This leads to an exponential growth of the target population and strong attenuation of the beam. Then, the target density keeps growing only at $s<s_{\rm sat}$, where most of the incident beam is absorbed, while thickness $s_{\rm sat}$ of this shielding front layer  decreases exponentially with time. Density at $s>s_{\rm sat}$ saturates at marginal attenuation (attenuation length $=\!s$) which sets the density profile $\propto s^{-1}$.}
\label{fig:beam}
\end{figure}

Note also that the GHz beam does not move in $s$ (its advection with small speed $\dd s/\dd t\approx c/\Gamma^2$ is neglected in our simplified model). The GHz waves may be thought of as static particles (targets) with density $\nF(s)$ occupying the slab $0<s<cT$ while the kHz beam may be thought of as particles streaming through the slab with speed $\dd s/\dd t=2c$. Thus, for each pair of frequencies $(\omega,\omF)$, our simplified model is equivalent to static targets interacting with an incident particle beam  (Figure~\ref{fig:beam}), with the interaction rate coefficient $\R(\omega)$.

The initial target density $\nF=\nseedF$ is tiny, so it weakly attenuates the incident beam. At $t>0$, the targets begin to ``catch'' particles from the beam and convert them to new static targets. This initiates an exponential growth of the target density $\nF$. From $\partial_t\ln\nF=\R \n$ (\Eq~\ref{eq:nF_ev}) one can see that $\nF$ grows exponentially until it strongly attenuates the incident beam from its original density $\n=\n_0$. This occurs where $s\,\partial_s\ln\n \approx -1$, which corresponds to $\R\nF\xi\approx 2$ (\Eq~\ref{eq:n_ev}). Then, the beam-targets interaction proceeds with $\nF$ saturating at $\nF^{\rm sat}=2(\R\xi)^{-1}$.

This evolution is demonstrated by the explicit solution of \Eqs~(\ref{eq:n_ev}) and (\ref{eq:nF_ev}) derived in Appendix~\ref{solution}:
\beq
\label{eq:solution}
   \nF = \frac{ \nseedF e^{\tau_0(t)}}{1 +\tauseed(s) e^{\tau_0(t)}}, 
   \quad \n = \frac{\n_0}{1 +\tauseed(s) e^{\tau_0(t)}},
\eeq
where $\tauseed=\R \nseedF s/2c\ll 1$, and $\tau_0=\R\n_0 t$ grows linearly with time for a steady incident beam of density $\n_0$. The solution holds at each $\omega$ and the corresponding $\omF=\Gamma^2\omega$. Note that the optical depth $\tauseed\propto s$ describes attenuation of the kHz beam in space at fixed $t=0$ while $\tau_0\propto t$ describes the growth of GHz waves in time at fixed $s$ near the edge of the pulse (where the beam is not attenuated yet). For a beam with energy density $U_0$ and some spectral shape $U_\omega^0$ around a characteristic frequency $\omega_0$, one can express $\tau_0$ as\footnote{Note that $\tau_0(t)$ in the simplified  analytical model differs from that in the full model (\Eq~\ref{eq:tau_max}) by a factor of $\approx 2\pi^2 ct/\Rout$.}
\beq
\label{eq:tau0_}
  \tau_0(t,\omega)=\frac{\pi}{2} \eF\, Z_0(b)\,\omega_0 t,
\eeq
where $b=\omega/\omega_0$, $\eF=U_0/\Ubg$, and $Z_0(b)=b\,\omega U_\omega^0/U_0$.

The solution~(\ref{eq:solution}) shows that the exponential growth of $\nF$ saturates when
\beq
\label{eq:sat_cond}
  \tau_0> \tausat=\ln \tauseed^{-1}(\omega,s)
   \quad\; ({\rm saturated~regime}).
\eeq
The onset of saturation first occurs at frequency $\omega=\omega_\star$ where $Z_0(\omega)$ is maximum. Then, the condition~(\ref{eq:sat_cond}) becomes satisfied in a growing frequency band $\omsat<\omega<\omega_{\rm sat,1}$. Inside this band, $\n/\n_0=U_\omega/U_\omega^0$ is exponentially suppressed while $\nF$ saturates at
\beq
\label{eq:Usat}
     U_{\omF}^{\rm sat}=\nF^{\rm sat} = \frac{ \nseedF}{\tauseed(s)} = \frac{4\Ubg}{\pi\omega^2 \xi }  
    = \frac{\Up}{2\pi \omF^2 \xi}.  
\eeq
This result is the expected relation $\nF^{\rm sat}=2(\R\xi)^{-1}$, which may be more intuitive in the equivalent problem of particle beam collision with static targets (Figure~\ref{fig:beam}). The corresponding solution for $\n(\xi)$ may be expressed as $\n(s)=\n_0\ssat/s$ in the saturation domain $s>\ssat$, which is exponentially shrinking with time: $\ssat\propto\exp(-\tau_0)$. The shrinking $s_{\rm sat}(t)$ implies the progressing suppression of $\n$ at all $s>s_{\rm sat}$ while $\nF$ settles at $\nF^{\rm sat}(s)$. Note that the saturated level of GHz waves $\nF^{\rm sat}$ is independent of both $\nseedF$ and $\n_0$; these parameters only determine where saturation sets in.

\begin{figure}[t]
\vspace*{1mm}
\includegraphics[width=0.47\textwidth]{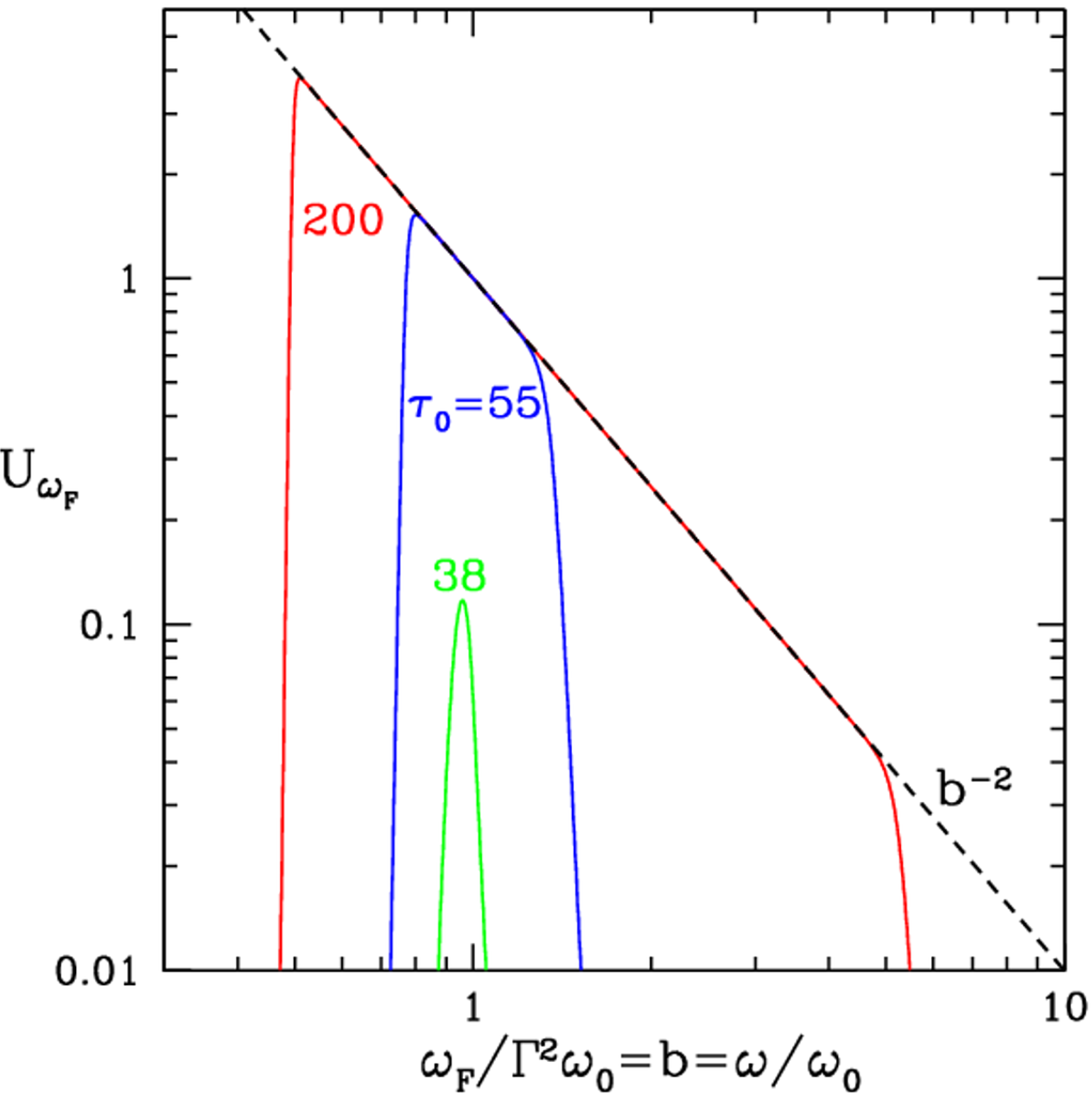} 
\caption{Spectrum of amplified GHz waves found in the analytical model ($U_{\omF}$ is normalized to its saturated value $U_{\omF}^{\rm sat}$ at $b=1$). The spectrum is determined by $\tausat$ and $\tau_0(t,b)$  (\Eq~\ref{eq:tau0_}). In this plot, $\tausat=40$ and we chose three moments when $\tau_0$ at $b=1$ equals 38, 55, and 200. Saturation occurs at $\tau_0>\tausat$: then a frequency band appears where $U_{\omF}$ reaches the ceiling of $U_{\omF}^{\rm sat}\propto b^{-2}$ (\Eq~\ref{eq:Usat}). The saturated band widens with the growing $\tau_0$. The sharp cutoff of $U_{\omF}$ at the boundary of this band determines the position of the spectral peak $\omF^{\rm peak}$. The full-width at half-maximum of the peak is $\Delta\omF/\omF^{\rm peak}\approx 0.4$.}
\label{fig:toy}
\end{figure}

As one can see from \Eq~(\ref{eq:Usat}), the saturated GHz spectrum declines with frequency as $U_{\omF}\propto \omF^{-2}$, and its overall amplitude decreases with distance from the leading edge of the pulse as $s^{-1}$. Outside the saturation band, one finds $\nF\propto e^{\tau_0}$ with a smaller $\tau_0\propto Z(b)$. The smaller $\tau_0$ implies an exponentially sharp drop of the GHz spectrum at $\omF<\Gamma^2\omsat$, which corresponds to $b<\bsat$.\footnote{The cutoff is shaped by $\tau_0\propto Z_0(b)\propto b^4$ at low $b$ (\Eq~\ref{eq:Z0}). Descending from $\bsat$ by $\delta b$ reduces $\nF\propto e^{\tau_0}$ by the factor of $e^{-\delta\tau_0}$ with $\delta \tau_0\approx (4\delta b/\bsat)\tausat\sim 10^2 \delta b/\bsat$.}

The resulting FRB spectrum is shown in Figure~\ref{fig:toy}. It is shaped by the decline of $U_{\omF}\propto \omF^{-2}$ inside the saturated band and the sharp cutoff at lower frequencies. The spectrum has a pronounced peak at 
\beq
  \omF^{\rm peak}\approx \Gamma^2\omsat.
\eeq
The width of the peak is controlled by the $\omF^{-2}$ wing. The full width $\Delta \omF$ at half-maximum (FWHM) is found from $[(\ompeak+2\Delta\omF)/\ompeak]^{-2}\approx 1/2$, which gives
\beq
\label{eq:width}
   \frac{\Delta\nu}{\nupeak}\approx \sqrt{2}-1\approx 0.4,
   \qquad \nu\equiv\frac{\omF}{2\pi}.
\eeq

The analytical model demonstrates two distinct zones in the pulse. At sufficiently small $s<\ssat^\star$, the condition~(\ref{eq:sat_cond}) is not met at any $\omega$, so here $\nF$ grows exponentially at all $\omF$. At $s>\ssat^\star$, the condition~(\ref{eq:sat_cond}) is satisfied in a finite frequency band, and this saturation band widens with increasing $s$. The boundary $\ssat^\star$ is set by the condition $\tauseed e^{\tau_0}\sim 1$ evaluated with the maximum $Z=Z_{\max}$ that gives the maximum $\tau_0$:
\beq
  \frac{\ssat^\star(t)}{c}=\xisat^\star \! \sim \! \frac{\Up}{2\pi\omF^\star U_{\rm seed}} 
  \exp\left[-\frac{\pi}{2} \eF\, Z_{\max}\,\omega_0 t\right],
\eeq
where $\omF^\star=\Gamma^2\omega_\star$ and $Z(\omega_\star/\omega_0)=Z_{\max}$. The evolution starts at $t=0$ with $\xisat^\star\gg T$ (no saturation). When the saturation regime is reached, $\xisat^\star$ exponentially quickly moves to the leading front of the pulse, $\xisat^\star\ll T$.

In essence, the depletion of the kHz beam develops in the layer of $\xi\sim\xisat^\star$. The depletion occurs in the frequency band where stimulated scattering drives so much growth of GHz waves that it consumes the kHz waves. The explosion pulse then resembles a car with a wind shield. The ``wind shield'' at $\xi\sim\xisat^\star$ collects most of the ambient kHz waves, converting them to GHz radiation; it produces the brightest GHz luminosity. The ``cabin'' at $\xi\gg\xisat^\star$ emits GHz waves in the self-regulated saturated state (\Eq~\ref{eq:Usat}), an attractor for the system, with a universal spectral peak at $\omF^{\rm peak}$. The total energy emitted by the shield is comparable to that emitted by the cabin. The simplified model thus gives a rough estimate for the FRB energy
\beq
  \EFRB\sim 8\pi r^2 c \int_{\xisat^\star}^T \frac{\Up}{2\pi\omF \xi}\, \dd\xi 
  \sim \frac{\ln(T/\xisat^\star)}{\pi \omF T} \, \Enp.
\eeq
Full numerical simulations presented below give $\EFRB$ that is larger by a numerical factor. 

The shield at $\xi\sim\xisat^\star$ becomes exponentially thin with time while its GHz luminosity becomes exponentially large. The simplified model then breaks for two reasons. (i) Throughout the paper we use the standard radiative transfer equations where waves have well defined locations --- they are treated as wave packets moving with group velocity. This description fails if the solution has gradients on scales $\delta\xi<2\pi/\omF$ (section~\ref{global}), so $\xisat^\star$ certainly cannot shrink below $2\pi/\omF$. (ii) A more limiting condition is related to the neglected drift of GHz wave packets in $\xi$, which occurs on scale $\xiadv\sim t/\Gamma^2$. The shield cannot become thinner than $\xiadv$. 

Advection of GHz waves can be included by keeping the term $\partial_\xi \ln n_{\bkF}$ in \Eq~(\ref{eq:nkF_ev}), as done in numerical simulations presented below. The simulations will also have other features of realistic explosions. In particular, the explosion develops in time: the pulse radius $r\approx ct$ grows and $\n_0(r)$, $\Bbg\propto r^{-3}$, and $\Gamma\propto r$ all vary with time. Realistic ambient kHz waves are not collimated along $-\hat{r}$ in the lab frame. However, only their spectrum measured in the fluid frame matters for stimulated scattering, and it is well described as a beam spectrum with appropriately defined $Z(b)$ (section~\ref{scattering}). The amplified GHz radiation is not strictly confined to angles $\thf$ and $\phif$ that satisfy $|\cos\thf|=1$ and $\sin^2\!\phif=1$; the full angular distribution of waves will be tracked by the simulations. The numerical model will, however, show same basic features as the analytical model described above.


\subsection{Setup of the full numerical model}
\label{unsaturated}

FRB production is described by \Eqs~(\ref{eq:Z_ev}) and (\ref{eq:nkF_ev}). They form a closed system for $Z(b,t,\xi)$ (the kHz pump wave) and $n_{\bkF}(\kF,\thf,\phif,t,\xi)$ (the amplified GHz wave). The system can be integrated numerically in $t,\xi$ as the explosion pulse propagates in $r\approx ct$. The integration starts at $r\ll \Rout$ with a given initial condition $n_{\bkF}=\nseed_{\bkF}(\kF)$. At each timestep, $A=n_{\bkF}/\nseed_{\bkF}$ and $Z$ are calculated with the boundary conditions $A=1$ and $Z=Z_0(b)$  at the leading edge of the pulse $\xi=0$. 

The ambient kHz waves prior to scattering are described by the function $Z_0(b)$ (\Eq~\ref{eq:Z0}) with slopes $\alpha_1=\alpha_2=3$ and the peak at frequency $\nu_0$. The parameter $\Rout\nu_0/c$ in our models is set to $3.33\times 10^3$. The amplitude of ambient kHz waves is described by \Eq~(\ref{eq:eF}) with $\eF^{\rm out}$ left as a free parameter. The model with $\eF^{\rm out}=0.1$ will be presented in detail below.

The calculation can be done for explosions with any profile of Lorentz factor $\Gamma$, which is related to the pulse power $\Lp(\xi)$ by \Eq~(\ref{eq:gam_Lp}). For illustration, we use the following pulse profile:
\beq
\label{eq:Lp}
  \frac{\Lp(\xi)}{\Lp^{\max}}=\left\{\begin{array}{lr}
  \vspace*{2mm}
  0.01+0.99(\xi/\xip) & 0<\xi<\xip \\
  (T-\xi)/(T-\xip) & \;\; \xip<\xi<T    
  \end{array} \right.
\eeq 
The profile rises to its peak at a chosen $\xip$ and gradually decreases toward the end of the pulse at $\xi=T$. In our sample models, we set $T=1$\,ms and assume that the rise of the pulse is fast, $\xip\ll T$.

The fluid Lorentz factor reaches its maximum $\Gamma_{\max}$ at $r=\Rout$ and $\xi=\xip$, so $ \Gamma(r,\xi)$ can be expressed as
\beq
  \Gamma(r,\xi) = \Gamma_{\max} \frac{r}{\Rout} \left(\frac{\Lp}{\Lp^{\max}}\right)^{1/4}. 
\eeq
We set $\Gamma_{\max}=500$, which corresponds to a peak power $\Lp^{\max}=7.5\times 10^{47}\mu_{33}^2$\,erg\,s$^{-1}$. Only $\Gamma\gg 1$ is of interest for FRB production, and our prescription ensures $\Gamma\gg 1$ in the simulation domain. $\Gamma\gg 1$ at $\xi=0$ corresponds either to a shock at the leading edge (where $\Gamma$ jumps to a high value), or simply to a shift of the reference point $\xi=0$ to exclude from our simulation the thin leading layer where $\Gamma$ may be small. The excluded thin layer also contains the swept-up outer magnetosphere with its Alfv\'enic turbulence (see Figure~\ref{fig:pulse}).  

The results weakly depend on the initial level of GHz seeds; however, it must be specified in the simulations. The spectral luminosity of seeds is related to $\nseed_{\bkF}$ by 
\beq\label{eq:Fseed}
  \frac{\Lseed_{\omF}}{4\pi r^2 c}\approx \Useed_{\omF} \approx \Omseed\kF^3\nseed_{\bkF},
\eeq 
where $\Omseed(r)\approx \pi \Rseed^2/r^2$ is the solid angle occupied by seed radiation flowing from a source of size $\Rseed$. In our simulations, the FRB radiation is amplified in a solid angle $\Omega_{\rm ampl}\sim \pi/\Gamma^2<\Omseed$, with $\nseed_{\bkF}\approx {\rm const}$ across $\Omega_{\rm ampl}$. Entering the opposite regime of $\Omega_{\rm ampl}>\Omseed$ would involve $\nseed_{\bkF}$ that steeply drops outside $\Omseed$; however, the results in the deeply saturated regime $\tau\gg\tausat$ would remain nearly the same: saturation would still be reached, since $\tausat$ depends slowly (logarithmically) on the seed level. The results would change for marginally saturated FRBs, as amplification would be suppressed in the parts of the explosion pulse with low $\Gamma$ that correspond to $\Omega_{\rm ampl}>\Omseed$.

We also need to choose a seed spectrum. Seed waves are strongly amplified at frequency $\omF\approx\Gamma^2\omega_0$, and a characteristic $\omF$ can be defined with $\Gamma=\Gamma_{\max}$. We will assume seeds with a power-law spectrum with slope $\als$ in the GHz band, $\Lseed_{\omF}\propto \omF^{\als}$, which gives
\beq
   \omF\Lseed_{\omF}=\left(\frac{\omF}{\Gamma_{\max}^2\omega_0}\right)^{1+\als} L_{\rm seed}.
\eeq
Here, $L_{\rm seed}$ is a parameter with the dimension of power. It is convenient to define a dimensionless parameter
\beq
  \Qseed \equiv \frac{L_{\rm seed}}{\Lp^{\max}}\,\frac{\Rout^2}{\Rseed^2}.
\eeq
The seed intensity of GHz waves then can
be expressed as
\beq
\label{eq:seeds}
   \nseed_{\bkF} = \frac{ c^2\Lp^{\max}\Qseed}{4\pi^2 \Rout^2\omF^4} 
   \left(\frac{\omF}{\Gmax^2\omega_0}\right)^{1+\als}.
\eeq
Note that $\nseed_{\bkF}$ is independent of $r$: occupation number (or intensity) or any unscattered radiation remains constant along the ray. 

Our numerical models will use $\Qseed=10^{-16}$ and $\als=0$. This choice is rather conservative, as it assumes seeds a few orders of magnitude weaker than could be supplied by persistent radio pulsations detected in some magnetars. The observed pulsations produce luminosity $L_{\rm seed}\sim 10^{30}$\,erg\,s$^{-1}$ \citep{Camilo2006}, possibly emitted at $r\sim 3\times 10^7$\,cm \citep{Zeng2026}. Any reasonable $\nseed_{\bkF}$ works for our proposed model, since the amplification of GHz waves is a fast exponential process with many $e$-foldings. We have verified that the results weakly depend on $\Qseed$ and $\als$ if FRB production occurs in the saturated regime.

\subsection{Numerical results}

As a first simple test  we have calculated models with small $\eFout<0.02$, which do not reach saturation. These models  showed the unsaturated exponential growth at each location $\xi$ inside the pulse, with $Z\approx Z_0(b)$ everywhere. At each $\xi$, we found the spectrum shown in Figure~\ref{fig:unsat}, with the narrow peak at $\nupeak(\xi)\approx 0.8\,\Gout^2(\xi)\,\nu_0$. 

Then, we increased $\eFout$ above 0.02 and observed the transition to the saturated regime (in agreement with \Eq~\ref{eq:sat}), with a changed FRB spectrum. As expected from the analytical model, a saturated frequency band appears in the spectrum, and the produced FRB has a well defined spectral peak, with a cutoff below $\nupeak$ and a  wing at $\nu>\nupeak$. We have calculated models with $\eFout=0.03-0.3$ and all of them show a similar spectral structure --- the result of self-regulation by the depletion of the kHz pump wave.

\begin{figure*}
\includegraphics[width=0.997\textwidth]{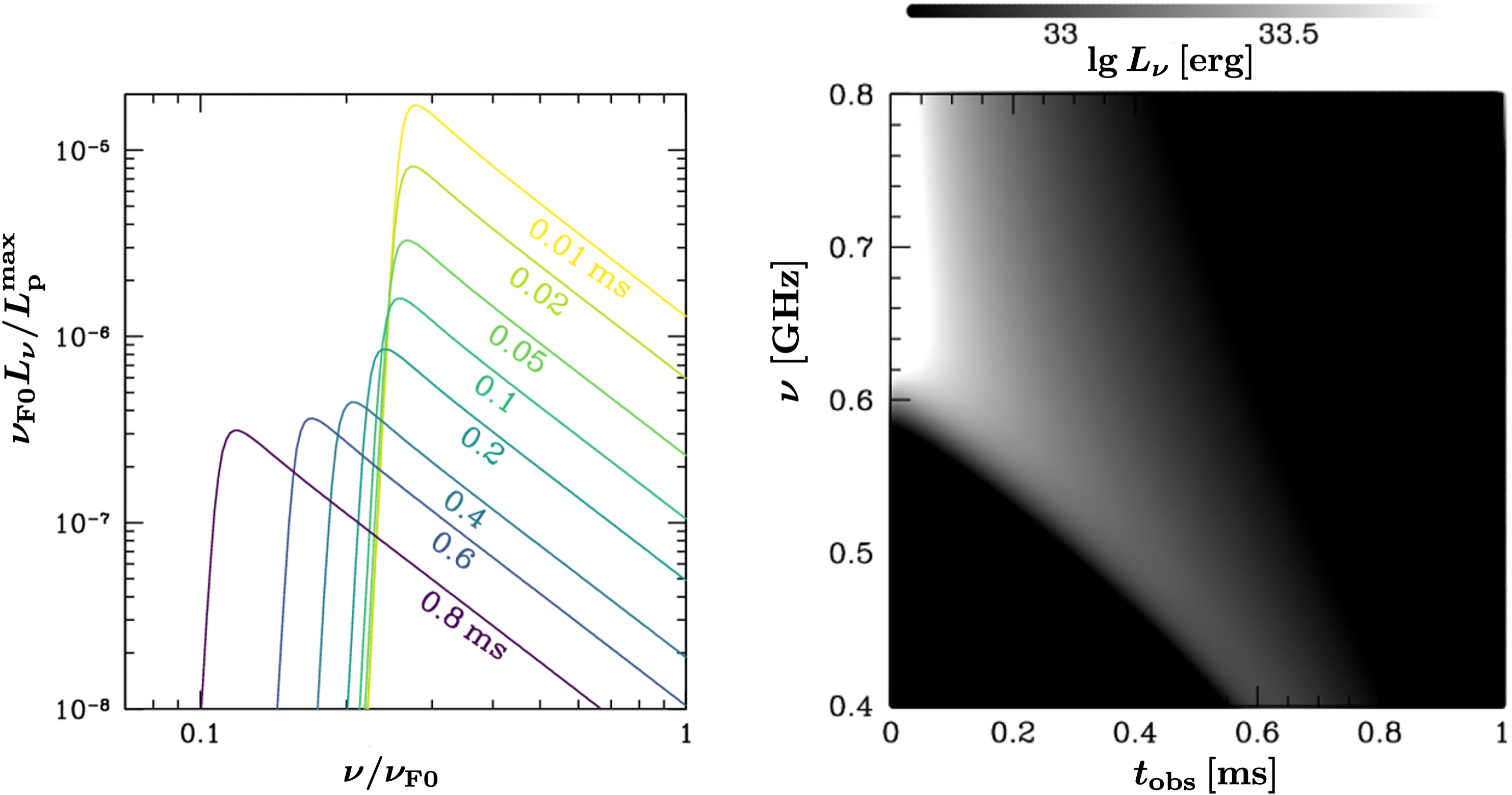} 
\caption{FRB emission calculated in the model with $\eF^{\rm out}=0.1$ and $\nu_0\Rout/c=3.33\times 10^3$. {\em Left:} Evolution of the emitted spectrum with $\tobs$ ($\tobs\approx \xi$ is indicated next to each curve). Frequency of the generated waves $\nu=\omF/2\pi$ is normalized to $\nu_{\rm F0}\equiv \Gmax^2\nu_0$, where $\nu_0=\omega_0/2\pi$ is the peak frequency of the ambient kHz wave spectrum, and $\Gmax=500$ is the peak Lorentz factor of the MHD fluid in the pulse at $\Rout$. The emitted spectrum $L_\nu$ is normalized to $\Lp^{\max}/\nu_{\rm F0}$, where $\Lp^{\max}$ is the peak power of the explosion pulse. {\em Right:} The spectral and temporal structure of the produced emission is shown on the $\nu$--$\tobs$ plane in the CHIME spectral band $400{\rm \, MHz}<\nu<800$\,MHz.}
\label{fig:spectrum}
\end{figure*}

\begin{figure}
\vspace*{2mm}
\includegraphics[width=0.47\textwidth]{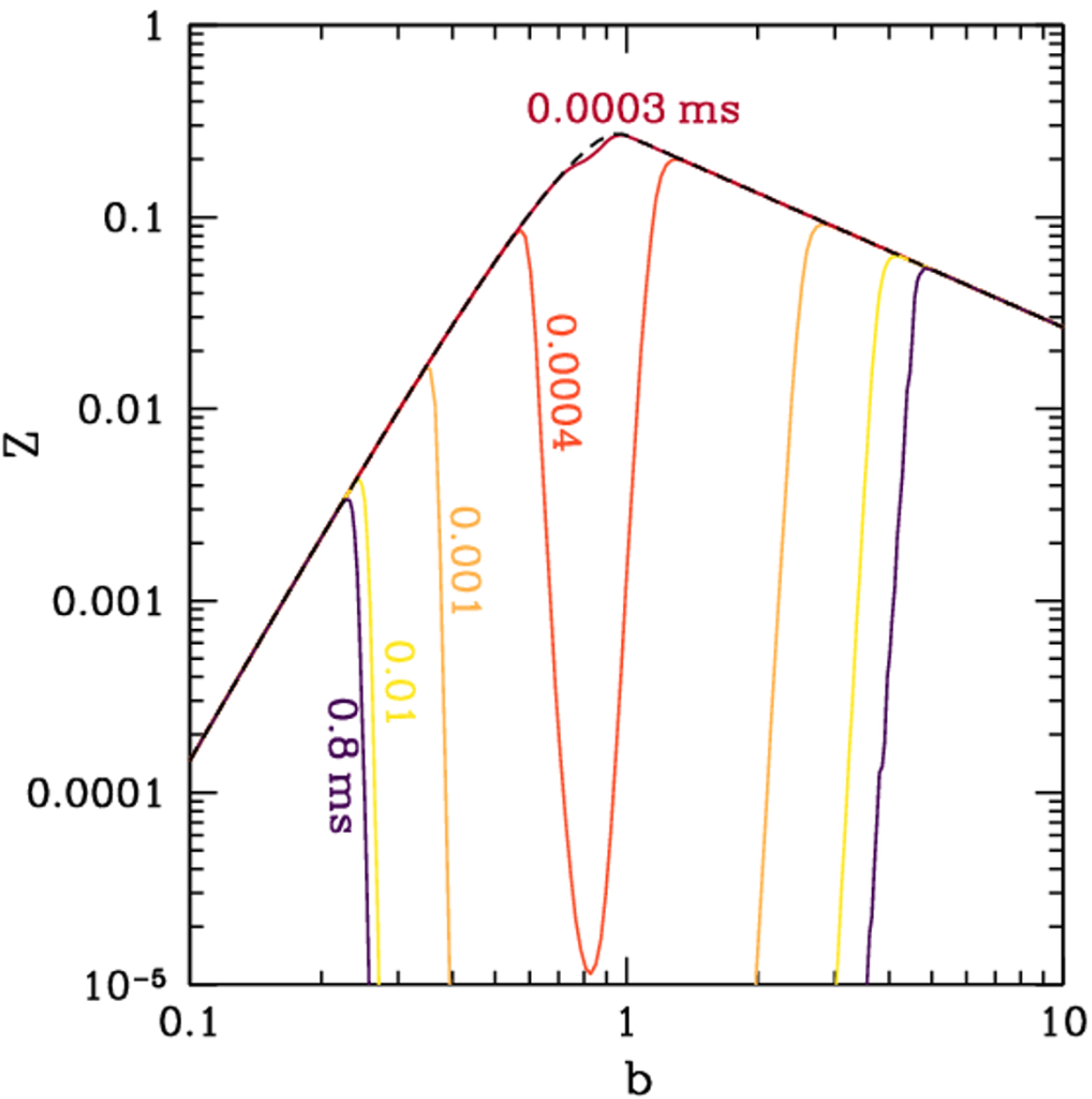}
\caption{Function $Z(b)$ (defined in (\Eq~\ref{eq:bY})) describing the spectrum of ambient kHz waves  boosted to the local fluid frame $\Kf$. We use the normalized frequency $b\equiv \nu'/2\Gamma\nu_0$, which is independent of $\Gamma\gg 1$ (as $\nu'\propto\Gamma$). $Z(b)$ is shown at different locations $\xi$ inside the explosion pulse; $\xi$ is indicated next to each colored curve. The black dashed curve shows the original (unattenuated) $Z=Z_0(b)$. The drop of $Z$ below $Z_0$ appears at $\xi\approx 3\times 10^{-4}$\,ms. At larger $\xi$, the depletion band widens to $\bsat<b<b_{\rm sat,1}$ with sharp boundaries at $\bsat$ (where $Z$ drops below $Z_0$) and $b_{\rm sat,1}$ (where $Z$ returns to $Z_0$). FRB production peaks at frequency $\nupeak(\xi)\approx\Gout^2(\xi)\,\bsat(\xi)\,\nu_0$, so $\bsat(\xi)$ controls the position of the FRB spectral peak, as one can see in Figure~\ref{fig:peak}.}
\label{fig:Z}
\end{figure}

\begin{figure}
\vspace*{2mm}
\includegraphics[width=0.47\textwidth]{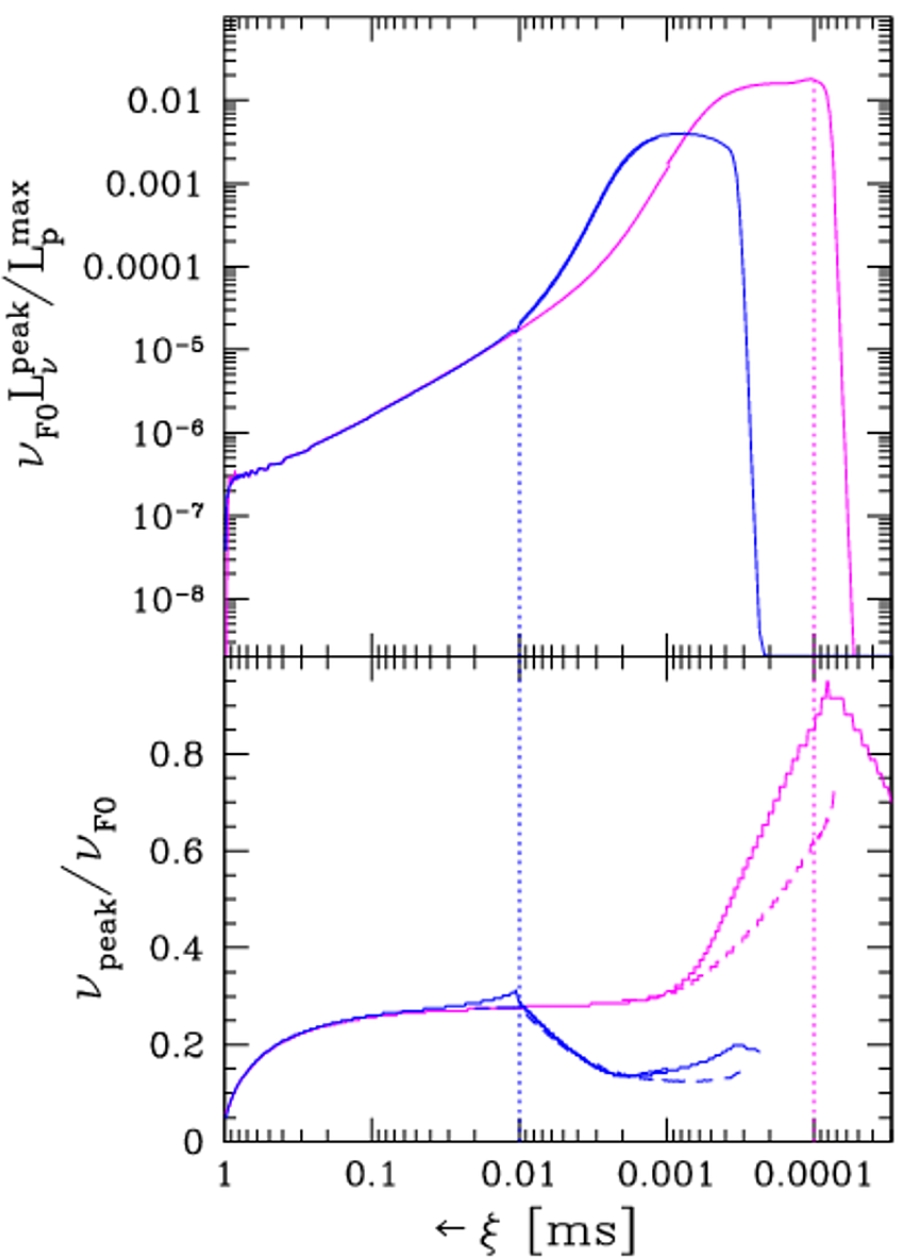}
\caption{Evolution of frequency $\nupeak$ and luminosity $L_\nu^{\rm peak}$ of the produced FRB with $\xi$; observer time $\tobs$ is related to $\xi$ by \Eq~(\ref{eq:tobs}). The evolution is shown for Model~A ($\xip=10^{-4}$\,ms, magenta) and Model~B ($\xip=10^{-2}$\,ms, blue). The parameter $\xip$ (the rise time of the explosion power to its peak $\Lp^{\max}$) is indicated by the dotted vertical lines. Both models have $\eFout=0.1$. The dashed curves in the lower panel show the approximation $\nupeak=1.1\,\Gout^2\bsat\nu_0$.}
\label{fig:peak}
\end{figure}

\begin{figure}
\vspace*{2mm}
\includegraphics[width=0.47\textwidth]{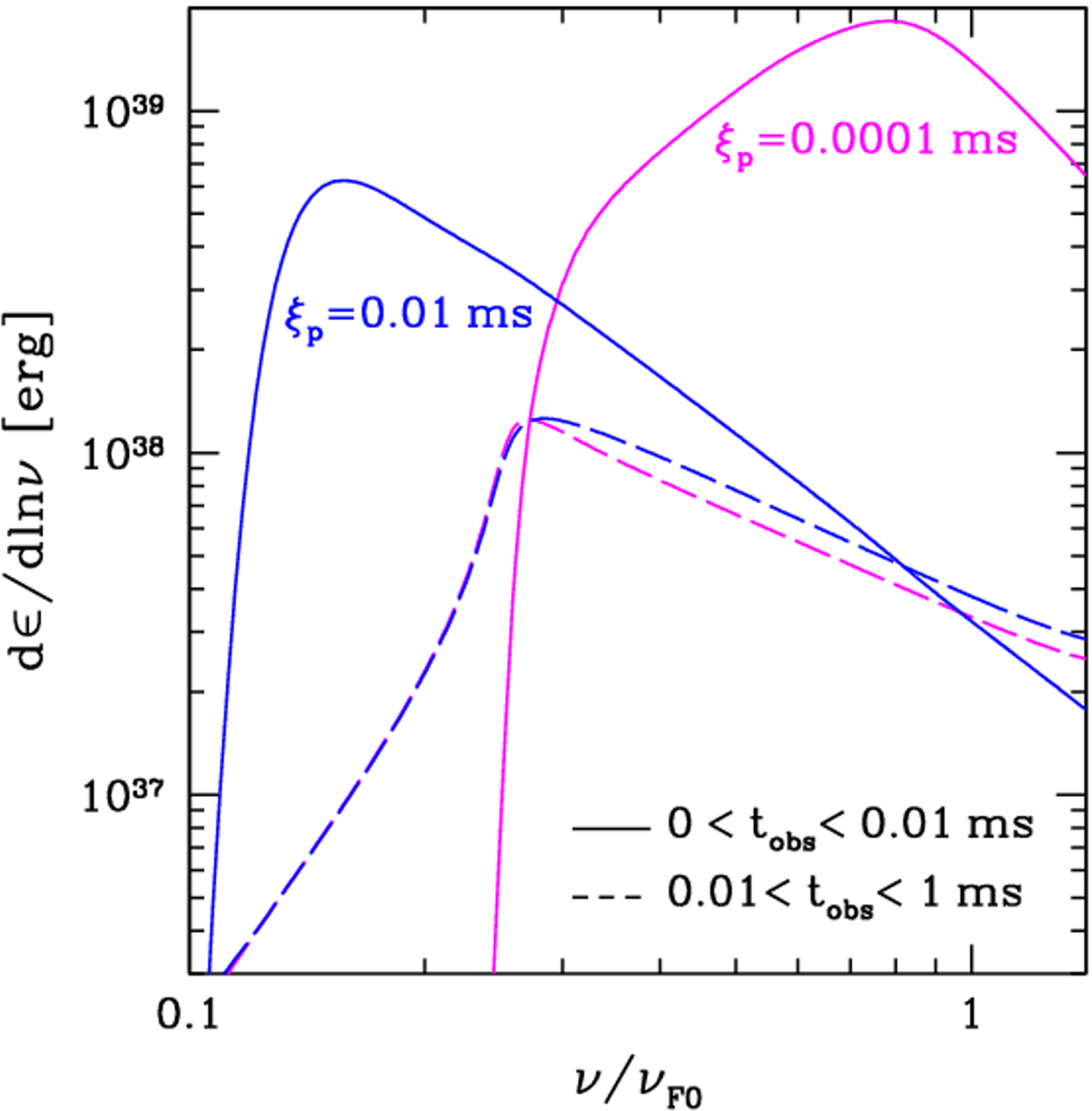}
\caption{Comparison of the time-integrated FRB spectra in the intervals $0<\tobs<0.01$\,ms (solid) and $0.01\,{\rm ms}<\tobs<1$\,ms (dashed), calculated for Model~A ($\xip=10^{-4}$\,ms, magenta) and Model~B ($\xip=10^{-2}$\,ms, blue). Frequency $\nu$ is normalized to $\nu_{\rm F0}=\Gmax^2\nu_0$.}
\label{fig:shield}
\end{figure}

Figures~\ref{fig:spectrum}--\ref{fig:shield} present the model with $\eFout=0.1$. Figure~\ref{fig:spectrum} shows the GHz emission accumulated in the explosion pulse by the end of FRB production, when the pulse exits the magnetosphere at radius $\Rout$. The emission is generated at each $\xi$, and the figure shows how the resulting GHz spectrum varies with $\xi$. The spectrum is normalized using the peak explosion power $\Lp^{\max}$ as indicated in the figure.\footnote{In the unsaturated regime, Figure~\ref{fig:unsat} presented the amplification factor $\A(\nu)$---the result of parametric instability in its linear phase, which can be multiplied with any $L_\nu^{\rm seed}$ to get the FRB spectrum $L_\nu$ (\Eq~\ref{eq:Lnu}). Saturation places a ceiling on the produced GHz radiation, which depends on the explosion power, and it is useful to see the FRB spectrum normalized to $\Lp^{\max}$.}
We have calculated two versions of the model with $\xip=0.01$\,ms and $\xip=10^{-4}$\,ms ($\xip$ is the onset time of $\Lp^{\max}$, see \Eq~(\ref{eq:Lp})). The two versions gave the same results at $\xi>0.01$\,ms.

Figure~\ref{fig:Z} demonstrates the key feature of the saturation regime --- the depletion of kHz radiation in a frequency band of a certain width. The spectrum of kHz waves (Doppler boosted to the fluid frame) is described by the function $Y(b)$ or the corresponding $Z(b)=b\,Y(b)$ (section~\ref{FRB_lab}). This function is shown in Figure~\ref{fig:Z} at several $\xi$ for the model with $\eFout=0.1$ and $\xip=0.01$\,ms. One can see how the saturation band $\bsat<b<b_{\rm sat,1}$ develops at $\xi>\xi_{\rm sat}^\star\approx 3\times 10^{-4}$\,ms. It first appears near the peak of $Z_0(b)$, as here the kHz waves experience the fastest stimulated scattering into GHz waves, with rate $q\propto Z$. The saturation band then widens at $\xi>\xisat^\star$: $\bsat(\xi)$ decreases and $b_{\rm sat,1}(\xi)$ grows. The widening is very fast at the onset of saturation at $\xi\lesssim 0.001$\,ms, and practically stalls at $\xi>0.01$\,ms.  $Z(b)$ is strongly depleted between $\bsat$ and $b_{\rm sat,1}$.

This depletion enforces a ceiling for FRB production $L_\nu^{\rm sat}\propto \nu^{-2}$ (cf. Figures~\ref{fig:toy} and \ref{fig:spectrum}). As expected, Figure~\ref{fig:spectrum} confirms that the generated GHz spectrum exponentially cuts off at $\nu<\nupeak\approx \Gamma^2\bsat$, which corresponds to $b<\bsat$. Similarly to the analytical model in section~\ref{toy}, the obtained spectra are shaped by the ceiling $L_\nu^{\rm sat}$ inside the saturation band and the cutoff at the lower boundary of this band. The resulting spectral peak at each $\xi$ has the width $\Delta\nu/\nu_{\rm peak}\approx 0.5$ (FWHM). It is slightly larger than 0.4 found in the simplified model (\Eq~\ref{eq:width}). The increase of $\Delta\nu$ is related to the angular structure of the amplified GHz radiation. Unlike the analytical two-beam model, the simulation tracks all amplified GHz waves, with $\bkF$ varying around the most favorable directions that maximize amplification. This variation is, however, modest (amplification is exponentially suppressed for unfavorable directions), resulting in the modest increase of $\Delta\nu$.

The obtained evolution of the FRB spectrum with $\xi$ translates to the evolution in observer time $\tobs$. The arrival time of the amplified GHz radiation to a distant observer is
\beq
\label{eq:tobs}
  \tobs=\xi+(1-\cos\psi)\frac{\Rout}{c},
\eeq 
where $\psi$ is the emission angle relative to the radial direction at radius $\Rout$. The typical $\psi=\Gout^{-1}$ gives $(1-\cos\psi)\Rout/c\sim 10^{-3}$\,ms, and hence $\tobs\approx \xi$ at $\tobs>1\,\mu{\rm s}$. As seen in Figure~\ref{fig:spectrum}, $\nupeak$ drifts downward with $\tobs\approx \xi$, giving the ``sad trombone'' evolution of FRB emission, while the spectral peak $L_{\nu}^{\rm peak}(\tobs)$ decreases. 

Figure~\ref{fig:peak} shows $\nupeak(\xi)$ and $L_{\nu}^{\rm peak}(\xi)$ in the entire range of $\xi$, including $\xi\lesssim  1\,\mu{\rm s}$. The results are shown for the models with $\xip=10^{-4}$\,ms (Model~A) and $\xip=0.01$\,ms (Model~B). One can draw two main conclusions from Figure~\ref{fig:peak}:

(1) At $\xi>0.01$\,ms, both choices of $\xip$ give nearly the same $\nupeak(\xi)$ and $\Lpeak(\xi)$. This demonstrates that the saturated GHz emission at $\tobs\approx\xi>0.01$\,ms is insensitive to the details of the explosion pulse near its leading edge where the pulse power $\Lp$ rises to its peak $\Lp^{\max}$.  We also find that $\nupeak$ is well described by 
\beq\label{eq:nupeak}
  \nupeak = 1.1\,\bsat\Gout^2\nu_0.
\eeq
Since $\bsat(\xi)$ varies very little at $\xi\gtrsim 0.01$\,ms (Figure~\ref{fig:Z}), the sad-trombone drift of $\nupeak$ is mainly shaped by the decreasing $\Gout^2(\xi)\propto \Lp^{1/2}(\xi)$. We have verified that if $\Lp=\cnst$ at $\xip<\xi<T$, the peak stays at nearly constant position $\nupeak$ until the end of the pulse. 

(2) At $\xi<0.01$\,ms, the produced FRB has an extremely high luminosity, and its details  depend on $\xip$. This dependence enters via the ratio $\xip/\xiadv$, where 
\beq
   \xiadv\sim (1-\cos\psi)\,\frac{r}{c}\sim \frac{r}{c \Gamma^2} \sim 1\,r_{10} \left(\frac{\Gamma}{500}\right)^{-2} \! \mu{\rm s}
\eeq
is a characteristic scale for advection of GHz radiation in $\xi$ due to its finite angle $\psi\sim\Gout^{-1}$ relative to the radial direction. Model~A has $\xip/\xiadv<1$ while Model~B has $\xip/\xiadv>1$. 

In Model~B, advection delays the onset of FRB emission to $\xi_{\rm onset}\approx 3\times 10^{-4}$\,ms. Here, amplification occurs in the zone of rising $\Gamma(\xi)$, and hence the advected GHz wave packets have a reduced time to stay in resonance with the peak of $Z(b)$, which delays $\xi_{\rm onset}$. The onset in Model~A is faster, $\xi_{\rm onset}\approx 7\times 10^{-5}$\,ms. It occurs with $\Gamma(\xi)$ approaching $\Gmax=\cnst$, so that advected GHz wave packets can stay in resonance with the peak of $Z(b)$ for stronger amplification. As one can see from Figure~\ref{fig:peak}, $\nupeak$ at $\xi<0.01$\,ms is significantly higher in Model~A, because the maximum FRB luminosity is generated where $\Gamma\approx\Gamma_{\max}$. By contrast, in Model~B, strong emission develops at $\xi\ll\xip$ where $\Gamma<\Gamma_{\max}$, which gives a lower $\nupeak$. 

These details of FRB emission on small (microsecond) timescales are important because of the huge GHz luminosity generated at small $\xi$. The origin of this effect was illustrated in Figure~\ref{fig:beam}: in the deeply saturated regime, the density of created GHz radiation increases toward the front edge $\xi=0$ as $\xi^{-1}$ for $\xi>\xisat$, and $\xisat$ shrinks exponentially until it approaches the advection scale $\xiadv$. Advection redistributes the amplified GHz radiation in $\xi$ on the scale $\xiadv$, shaping the FRB peak. The resulting maximum of GHz luminosity occurs on a microsecond timescale and reaches huge values comparable to $1\%$ of the explosion power (Figure~\ref{fig:peak}). This feature is a predicted hallmark of FRBs generated in the deeply saturated regime. A small leading part of the explosion pulse carries a large fraction of the total FRB energy $\E$. At $\xi\gtrsim\xiadv$, the cumulative energy $\E(\xi)$ is close to the estimate found in the simplified analytical model, $\dd\E/\dd\xi=L\propto \xi^{-1}$, which  gives a flat distribution of $\E$ over $\ln\xi$. 

It may also be useful to compare the spectra generated in the shield and the main body of the pulse. Figure~\ref{fig:shield} shows the time-integrated spectra in the intervals of $0<\xi<0.01$\,ms and $0.01\,{\rm ms}<\xi<1$\,ms, calculated for  Models~A and B. In both models, the front layer $\xi<0.01$\,ms dominates $\E$ for saturated FRB production because the layer collects most of the ambient kHz waves as the explosion pulse plows through the magnetosphere (cf. the analogous  problem of particle beam conversion into static targets in Figure~\ref{fig:beam}).

It is worth emphasizing that the presented models assume a fast rise of the explosion power $\Lp(\xi)$, on a timescale $\xip\leq 0.01$\,ms. A slower rise, e.g. $\xip\sim 1$\,ms, would place the ``shield'' emission at lower frequencies and make it less concentrated in time. The main body of the pulse (where $\Gamma$ exceeds a few hundred) would still generate the saturated GHz emission similar to that shown in Figure~\ref{fig:spectrum}.


\section{Conclusions}

This paper proposes a novel mechanism of FRB emission which naturally accompanies magnetospheric explosions. In contrast to most discussed scenarios of FRBs, it involves no kinetic plasma phenomena. A range of previous proposals, from shock synchrotron maser \citep{Lyubarsky2014,Beloborodov2017,Sironi2021} to curvature emission by charge bunches \citep{Kumar2017} to boosted magnetic reconnection \citep{Lyubarsky2020} involved physics on kinetic scales. The mechanism presented here is a pure MHD process. Although it can be described using the quantum picture of stimulated scattering, it is in fact a classical parametric instability that exponentially amplifies seed GHz waves to extreme luminosities. 

FRBs produced by this mechanism have remarkable features: they can have narrow spectra (as narrow as $\Delta\nu/\nu\sim 0.2$) and contain an ultrastrong microsecond component. The narrow spectrum is a simple consequence of the exponential amplification from tiny seeds: the enormous amplification $\exp[{\tau(\bkF)}]$ forms a narrow peak near wavevector $\bkF$ that maximizes $\tau$. The microsecond component appears in the saturated regime described in section~\ref{saturated}; the FRB then reaches a huge luminosity ($\sim 1$\% of the explosion power) concentrated within $\delta \tobs\sim 1\,\mu{\rm s}$. 

The produced GHz radio wave $E_{\rm wave}=B_{\rm wave}$ is embedded in the explosion pulse that has field $\Bp\gg B_{\rm wave}$. The FRB propagates inside the pulse to large radii far beyond the magnetosphere. It does not encounter the condition $E^2>B^2$ in the dangerous zone $r\lesssim 10^{11}$\,cm, where $E^2>B^2$ would cause quick damping of the FRB \citep{Beloborodov2024,Beloborodov2026}. Additional nonlinear effects influencing GHz waves embedded in the strong electromagnetic pulse are also inefficient \citep{Lyubarsky2020} and should not prevent the FRB escape to distant observers. 
 
The spectrum and temporal structure of the predicted FRB can be directly calculated from first principles for any given explosion pulse, as demonstrated with sample models presented in this paper. Main observational properties of predicted FRBs are as follows:

(1) Emission occurs in the GHz band. This is a consequence of the fact that the magnetosphere is filled with kHz MHD waves excited by magnetar quakes. Note also that the proposed mechanism requires explosions with power $\Lp\sim 10^{45}-10^{48}$\,erg/s, which develop Lorentz factors $\Gamma\propto\Lp^{1/4}$ exceeding a few hundred, sufficient to boost the kHz waves to the GHz band.

(2) The GHz luminosity can reach $L\sim 10^{45}$\,erg\,s$^{-1}$ for explosions with power $\Lp\gtrsim 10^{47}$\,erg\,s$^{-1}$. Such ultra-luminous peaks of emission are predicted to have short durations $\sim 1\,\mu{\rm s}$. On longer timescales $\tobs$, the predicted luminosity varies roughly as $L\sim 10^{-3}\Lp/\tobs$.

(3) The luminosity distribution of FRBs is expected to have a tail of exponentially low $L$, which corresponds to the unsaturated regime of FRB production.  

(4) The instantaneous FRB spectrum has a pronounced peak. In the unsaturated regime, its FWHM is $\Delta\nu/\nupeak\approx 0.18$. In the saturated regime, the spectrum has a practically universal shape at times $\tobs>0.01$\,ms, with FWHM $\Delta\nu/\nupeak\approx 0.5$.

(5) The model predicts a ``sad trombone'' pattern where the explosion power $\Lp$ decreases.

(6) The FRB temporal structure is controlled by the temporal structure of the explosion. It may be simple or complicated, with multiple peaks. The spectral evolution and the time-integrated spectrum depend on the profile of the explosion pulse $\Lp(\xi)$.

(7) The GHz emission is nearly radial, within the beaming angle $\psi\sim\Gamma^{-1}\sim 10^{-3}$. It will appear to distant observers as if the source is located at $\sim 10^7$\,cm instead of its actual radius $\sim 10^{10}$\,cm. 

(8) The predicted FRB has a 100\% linear polarization. Its position angle is controlled by the orientation of the magnetic field in the explosion pulse $\bBp$. It can have any orientation in the plane transverse to the line of sight, depending on how the explosion was produced. Furthermore, it may vary with $\xi\approx\tobs$, perhaps explaining the rare polarization swing detected in FRB~20221022A \citep{Mckinven2025}. An additional change in polarization can occur at much larger radii where the FRB escapes the explosion pulse \citep{Lyubarsky2020}.

Search of the predicted features in existing observational data, including spectral shapes and early microsecond peaks, is left for future work. The narrowest spectra found in our models, $\Delta\nu/\nu\approx 0.18$ are still broader than the narrowest spectra reported in a few FRBs, $\Delta\nu/\nu\sim 0.1$ \citep{Kumar2024}. A remaining open question is how much the spectral and temporal structure of FRBs can be changed by propagation effects on larger scales before they reach the distant observer.

The proposed FRB mechanism can be investigated further. It would be useful to study the expected profiles of the explosion pulses $\Lp(\xi)$ using MHD simulations of concrete explosion scenarios. Calculations in the present paper show that the shape of $\Lp(\xi)$ has a significant effect on the produced FRB: the rise of $\Lp$ to $\Lp^{\max}$ influences the brightest component of the burst, and the subsequent fall of $\Lp(\xi)$ shapes the sad trombone effect. 

Our emission calculations can be extended to include the process $F\rightarrow A+A$ (Appendix~\ref{appB}), which may become competitive in the presence of suitable Alfv\'enic seeds inside the pulse. Note also that both processes, $F\rightarrow F+A$ and $F\rightarrow A+A$, rely on the MHD description of A-waves. It may break in parts of the explosion pulse that carry a low plasma density corresponding to density parameter $\N\lesssim 10^{38}$ (see section~\ref{pulse}). This would have imprints on the produced FRB, which may be interesting to investigate.

\medskip

A.M.B acknowledges support by NASA grants 21-ATP21-0056 and 80NSSC24K1229, NSF grant AST-2408199, and Simons Foundation grant 446228. This work was facilitated by MPPC grant PHY-2206609. I.D. acknowledges support from the Israel Science Foundation under grant No. 2126/22.


\newpage 

\appendix

\vspace*{-5mm}

\section{Process $F\rightarrow F+A$}
\label{appA:KinEq}

In Appendices~\ref{appA:KinEq} and \ref{appB} we use only the local fluid frame $\Kf$. In contrast to main text, we omit primes for quantities measured in frame $\Kf$ for simpler notation. This should cause no confusion, as no other frames are used here.

We describe waves by their Lorentz-invariant occupation numbers (\Eq~\ref{eq:occup}) and use the standard weak-turbulence kinetic approximation (e.g. \citealt{Zakharov1992}). It assumes that the waves have random phases and small amplitudes $\delta B$ compared to the  background magnetic field inside the pulse, $\delta B\ll \Bp$.

The three-wave process $F\rightarrow F+A$ involves waves $\bk,\bkF,\bkA$. Resonance conditions stated below imply that the magnitudes of the three wavevectors are comparable in the fluid frame. The pump wave $\bk$ is beamed within the solid angle $\Delta\Omega\sim \pi/\Gamma^2$, so it is concentrated in a phase space volume reduced by the factor of $\sim \Gamma^{-2}$ compared to that of the generated waves $\bkF$ and $\bkA$ (which are spread over a solid angle $\sim 1$). Therefore, $n_{\bkF}+n_{\bkA}\ll n_{\bk}$ remains satisfied until the pump wave is very strongly depleted, by the factor of $\sim \Gamma^{-2}$. The inverse process $F+A\rightarrow F$ may be neglected as long as $n_{\bkF}+n_{\bkA}\ll n_{\bk}$, and the kinetic equation for the amplified F-waves $\bkF$ takes the form (e.g. \citealt{GolbraikhLyubarsky2023})
\begin{equation}\label{kFMS}
    \frac{\dd n_{\bkF}}{\text{d}t}\equiv\frac{\partial n_{\bkF}}{\partial t'}+\boldsymbol{v}_{{\rm F}}\cdot\frac{\partial n_{\bkF}}{\partial \boldsymbol{r}}\simeq \int \dd\bk\,\dd\bkA \WFA n_{\bk}(n_{\bkF}+n_{\bkA})\delta^{(3)}(\boldsymbol{k}-\bkF-\bkA)\delta(\omega-\omF-\omA),
\end{equation}
where $\boldsymbol{v}_{{\rm F}}=\partial \omF/\partial\bkF$ is the group velocity of amplified F-waves. Here, the three-wave interaction rate is integrated over all $\bk$ and $\bkA$ that participate in the amplification of wave $\bkF$. The factor $n_{\bkF}+n_{\bkA}$ describes stimulated emission, and the small spontaneous emission term is neglected. The delta-functions impose resonance conditions that express conservation of momentum and energy in the $F\rightarrow F+A$ process: $\bk=\bkF+\bkA$ and $\omega=\omF+\omA$. We choose local Cartesian coordinates $x,y,z$ with the $z$ axis along the local mean magnetic field $\boldsymbol{B}_{\rm p}$ and the $x$ axis along the direction of the pump beam $\bk$. The resonance conditions then require
\begin{equation}
  k=k_{{\rm A}x}+k_{{\rm F}x}, \qquad
  k_{{\rm A}y}=-k_{{\rm F}y},
\qquad
k_{{\rm A}z}=-k_{{\rm F}z}, \qquad k=\kF+|k_{{\rm A}z}|.
\label{eq:resonance_components}
\end{equation}

The general expression for the 3-wave interaction coefficient $\WFA$ for the process $F\rightarrow F+A$ is given by \citep{Lyubarsky2019}
\begin{equation}
    \WFA=\frac{\pi^2c^4\omA}{2\omega\omF}\frac{k^2_{A\perp}}{\Bp^2k^2_\perp k^2_{{\rm F}\perp}}\left[\kF-\frac{k_{{\rm A}z}}{|k_{{\rm A}z}|}k_{{\rm F}z}\right]^2([\bkA\times\bkF]\cdot\hat{\boldsymbol{z}})^2,
    \label{eq:matrix_element}
\end{equation}
where subscript $\perp$ denotes the component perpendicular to $\bBp$. It can be simplified for the relevant case of a pump wave beamed perpendicular to $\bBp$. Using \Eq~(\ref{eq:resonance_components}) and $[\bkA\times\bkF]\cdot\hat{\boldsymbol{z}}=k k_{{\rm F}y}$, we obtain
\begin{equation}
 \WFA=\frac{\pi^2c^2\omega}{\Bp^2}\frac{|k_{{\rm F}z}|k_{{\rm F}y}^2}{\kF}\frac{(k-k_{{\rm F}x})^2+k_{{\rm F}y}^2}{k_{{\rm F}x}^2+k_{{\rm F}y}^2}.
\label{eq:W_cartesian}
\end{equation}
The resonance conditions allow one to express all wavevector components appearing in \Eq~(\ref{eq:W_cartesian}) in terms of $\omega$ and the direction $\bOmF$ of the amplified F-wave $\bk_{\rm F}$. We parameterize this direction by the polar angle $\theta$, measured from $\bBp$, and the azimuthal angle $\varphi$, measured from the direction of the pump beam: $\bkF=\kF\left(\sin\theta\cos\varphi,\sin\theta\sin\varphi,\cos\theta\right)$. Energy conservation gives $\kF=k/(1+|\cos\theta|)$. Equation~(\ref{eq:W_cartesian}) then reduces to
\begin{equation}
\WFA(\omega,\theta,\varphi) = \frac{\pi^2\omega^{3}}{\Bp^2}
\frac{2|\cos\theta|\sin^2\!\varphi}{1+|\cos\theta|}
\left(1-\sin\theta\cos\varphi\right)\equiv \frac{\pi^2\omega^3}{\Bp^2}\fFA(\bOmF).
\label{eq:Wfa}
\end{equation}
Note that $\WFA$ vanishes when $\cos\theta=0$ (then the resonant A-wave has zero frequency) and when $\sin\varphi=0$ (then all three wavevectors lie in the plane formed by $\bk$ and $\bBp$). $W$ is also proportional to $k_{\rm A\perp}\propto 1-\sin\theta\cos\varphi$.

\begin{figure*}
\vspace*{2mm}
\includegraphics[width=0.95\textwidth]{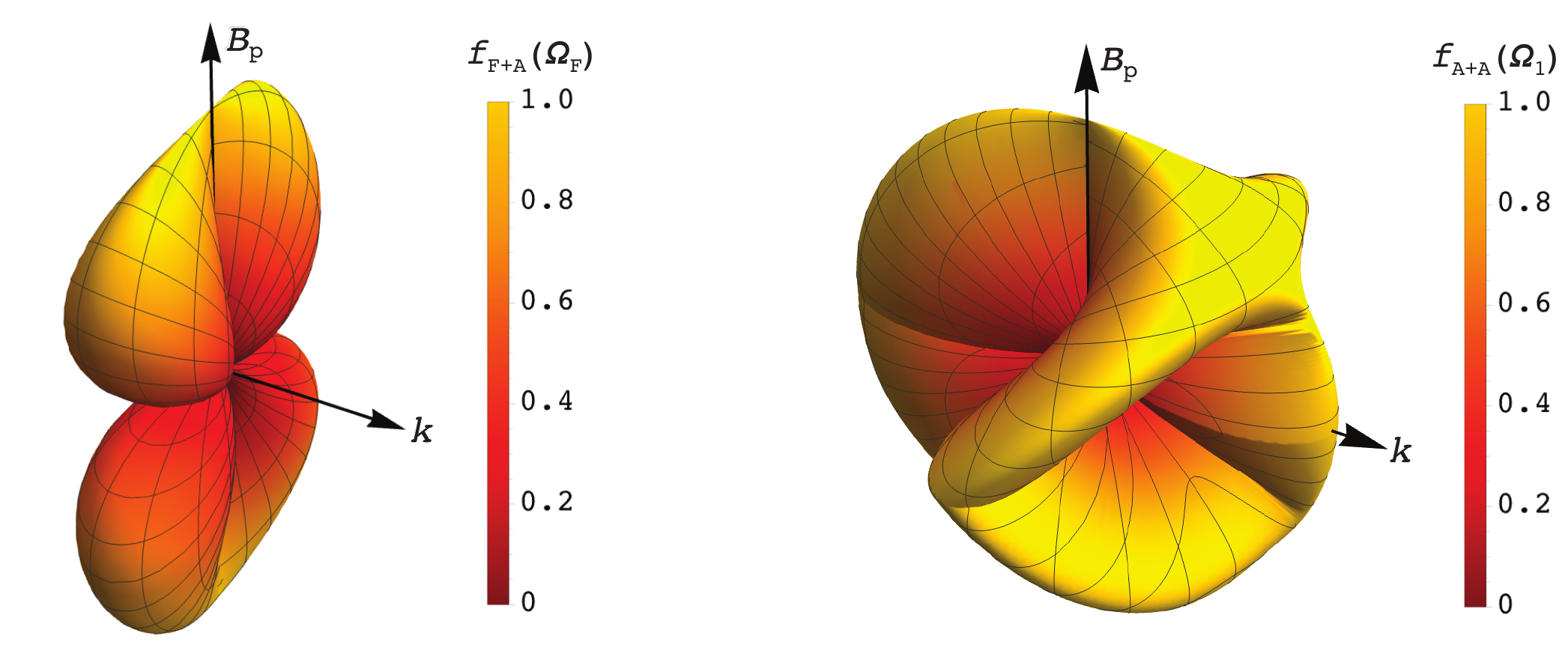}
\caption{Left: angular factor $\fFA(\bOmF)$ in the interaction coefficient $\WFA$ for the 3-wave process $F\rightarrow F+A$, viewed in the local rest frame of the fluid $\Kf$ (in Appendices~\ref{appA:KinEq} and \ref{appB} we use only frame $\Kf$ and omit primes for simpler notation). The plot is constructed using a 3D space of vectors $\boldsymbol{f}$; we show the surface defined by $\boldsymbol{f}=\fFA(\bOmF)\,\bOmF$ with directions $\bOmF$ spanning the full $4\pi$ solid angle. In this way, $\fFA(\bOmF)$ is visualized as the distance from the origin $\boldsymbol{f}=0$ [$\fFA(\bOmF)={\rm const}$ would give a sphere in this representation]. In addition, the value of $\fFA(\bOmF)$ is shown by color. Right: angular factor $\fFA(\bOm_1)$ for the process $F\rightarrow A+A$.}
\label{fig:angular}
\end{figure*}

Equation~(\ref{kFMS}) can be simplified by integrating the first delta-function with respect to $\bkA$ (this gives $\bkA=\bk-\bkF$) and then writing $\text{d}\boldsymbol{k}=\omega^2\text{d}\omega\text{d}\Omega/c^3$. Since the pump wave is strongly beamed in the $x$-direction, we replace its angular distribution by $\delta^{(2)}(\bOm-\hat{\boldsymbol{x}})$. This delta-function is removed by the integration over $\dd\Omega$, and $\delta(\omega-\omF-\omA)$ is removed by the integration over $\dd\omega$. As a result, we obtain
\begin{equation}
    \frac{\text{d} n_{\bkF}}{\text{d} t} =
    \WFA\frac{U_{\omega}}{\omega}(n_{\bkF}+n_{\bkA}).
\label{eq:fineq}
\end{equation}
Similarly, one finds the same expression for the evolution of $n_{\bkA}$: $\dd n_{\bkA}/\dd t=\dd n_{\bkF}/\dd t$.

The generation of $n_{\bkF}$ and $n_{\bkA}$ implies some depletion of the pump wave $n_{\bk}$. It is described by the kinetic equation
\begin{equation}
  \frac{\dd n_{\bk}}{\dd t} \equiv\frac{\partial n_{\bk}}{\partial t}+\boldsymbol{v}\cdot\frac{\partial n_{\bk}}{\partial \boldsymbol{r}}
  = - \int\dd \bkF\dd \bkA \WFA n_{\bk} \left(n_{\bkF}+n_{\bkA}\right) 
  \delta^{(3)}(\bk-\bkF-\bkA) \delta(\omega-\omF-\omA).
\label{eq:kHz_depletion_general}
\end{equation}
Here $\boldsymbol{v}=\partial\omega/\partial\bk$, and the interaction rate is integrated over all pairs of amplified A and F waves that can resonantly interact with the pump wave $\bk$. We first use integration over $\bkA$ to remove $\delta^{(3)}(\bk-\bkF-\bkA)$, which enforces $\bkA=\bk-\bkF$. We then write $\dd\bkF=\kF^2\dd \omF\dd\OmF/c$. Integration over $\omF$ now removes $\delta(\omega-\omF-\omA)$ while enforcing $\kF=k/(1+|\cos\theta|)$. Thus, we find
\beq
  \frac{\dd n_{\bk}}{\dd t}
   = -\frac{\omega^2}{c^3} n_{\bk}
\int\frac{\WFA\left(n_{\bkF}+n_{\bkA}\right)}{(1+|\cos\theta|)^3} \dd\OmF.
\label{eq:kHz_depletion_angular}
\eeq
The equal rates $\dd n_{\bkF}/\dd t=\dd n_{\bkA}/\dd t$ imply $n_{\bkF}\approx n_{\bkA}$, once $n_{\bkF}$ and $ n_{\bkA}$ have grown well above the initial seed level. Then, \Eq~(\ref{eq:kHz_depletion_angular}) becomes \Eq~(\ref{eq:dn_fluid}) stated in the main text.


\section{Process $F\rightarrow A+A$}
\label{appB}

\subsection{Interaction rate}

We continue to use here the fluid frame $\Kf$ and omit primes for all quantities for simpler notation. Kinetic equations for the process $F\rightarrow A+A$ are similar to those for $F\rightarrow F+A$. However, the difference between the dispersion relations for F and A waves enters the resonance conditions and leads to a different interaction coefficient $\WAA$. Let $\bk_1$ and $\bk_2$ be the wavevectors of the two amplified A-waves. Their frequencies are $\omega_1=ck_{1z}$ and $\omega_2=ck_{2z}$. We continue to use the coordinates with the $z$-axis along $\bBp$ and the $x$-axis along the pump F-beam. The resonance conditions require
\begin{equation}\label{eq:resAA}
    k=k_{1x}+k_{2x}, \qquad
    k_{1y}=-k_{2y}, \qquad
    k_{1z}=-k_{2z}, \qquad
    k=|k_{1z}|+|k_{2z}|.
\end{equation}
These relations imply $|k_{1z}|=|k_{2z}|=k/2$ and $\omega_1=\omega_2=\omega/2$, so the two A-waves share equally the energy received from the pump F-wave. The received transverse momentum $\bk=\bk_{1\perp}+\bk_{2\perp}$ can be partitioned between the two A-waves in various ways. This freedom increases the phase space of the $F\rightarrow A+A$ process. 

The general 3-wave interaction probability for $F\rightarrow A+A$ is \citep{Lyubarsky2019,GolbraikhLyubarsky2023}
\begin{equation}\label{eq:VAA}
  \WAA=4\pi^2c^2\frac{\omega_1\omega_2}{\omega}\left(\frac{k_{1\perp}(\boldsymbol{k}_{2\perp}\cdot\boldsymbol{k}_{\perp})}{k_{2\perp}}+\frac{k_{2\perp}(\boldsymbol{k}_{1\perp}\cdot\boldsymbol{k}_\perp)}{k_{1\perp}}\right)^2\frac{H(-k_{1z}k_{2z})}{\Bp^2k^2_\perp},
\end{equation}
where $H$ is the Heaviside function (only A-waves with opposite propagation directions along $\bBp$ participate in the interaction). Using \Eq~(\ref{eq:resAA}), one finds
\begin{equation}
  \WAA=\frac{\pi^2\omega^3}{\Bp^2}\frac{(k_{1x}k_{2x}-k_{1y}k_{2y})^2}{k^2_{1\perp}k^2_{2\perp}}.
\end{equation}
The resonance conditions allow all wavevector components to be expressed in terms of $\omega=c\bk$ and the direction of $\bk_1$. This direction $\bOm_1$ is described by two angles: the polar angle $\theta_1$ measured from $\bBp$ and the azimuthal angle $\varphi_1$ measured from the pump wave direction; then, $\boldsymbol{k}_1=k_1(\sin\theta_1\cos\varphi_1, \sin\theta_1\sin\varphi_1, \cos\theta_1)$ and one can express $\WAA$ as follows
\begin{equation}
    \WAA(\omega,\theta_1,\varphi_1)=\frac{\pi^2\omega^3}{\Bp^2} \fAA(\bOm_1),
    \qquad 
    \fAA(\bOm_1)\equiv\frac{(2|\cos\theta_1|\cos\varphi_1-\sin\theta_1\cos2\varphi_1)^2}{1+3\cos^2\theta_1-4|\cos\theta_1|\sin\theta_1\cos\varphi_1}.
\label{eq:Waa}
\end{equation}
The denominator in the expression for $\fAA$ exceeds the numerator by $4\sin^2\!\varphi_1\,(\sin\theta_1\cos\varphi_1-|\cos\theta_1|)^2\geq 0$ and hence $\fAA\leq 1$. The denominator equals numerator (and $\fAA$ reaches its maximum value $\fAA=1$) in two cases: $\sin\varphi_1=0$ or $\sin\theta_1\cos\varphi_1=|\cos\theta_1|$.

The processes $F\rightarrow F+A$ and $F\rightarrow A+A$ have the same maximum rates, $\WFA^{\max}=\WAA^{\max}=\pi^2\omega^3/\Bp^2$. However, their angular dependencies, described by $\fFA(\bOmF)$ and $\fAA(\bOm_1)$, are quite different (cf. \Eqs~(\ref{eq:Wfa}) and (\ref{eq:Waa}), and see Figure~\ref{fig:angular}). For $F\rightarrow F+A$, the maximum is reached for isolated directions $\bOmF$, whereas for $F\rightarrow A+A$ it is reached along two branches of $\bOm_1$. The first branch $\sin\varphi_1=0$ has $k_{1y}=k_{2y}=0$, so both amplified A-waves lie in the $\boldsymbol{k}$-$\boldsymbol{B}$ plane (this branch forms a circular ridge in Figure~\ref{fig:angular}b). The second branch,
\beq
  \sin\theta_1\cos\varphi_1=|\cos\theta_1|, \qquad \cos^2\!\theta_1\leq\frac{1}{2},
\eeq
has $k_{1x}=k_{2x}=k/2$ and $k_{1y}=-k_{2y}$, so the two A-waves equally share the momentum received from the pump wave, and their wavevectors $\bk_1$ and $\bk_2$ are symmetric about the $\boldsymbol{k}$-$\boldsymbol{B}$ plane (this branch appears as two symmetric curved ridges in Fig.~\ref{fig:angular}b).

\subsection{Coupling of process $F\rightarrow A+A$ to the primary process $F\rightarrow F+A$}

Consider A-waves generated by the process $F\rightarrow F+A$ that trades the pump wave $\bk$ to waves $\bkF$ and $\bkA$. The A-waves stimulate the process $F\rightarrow A+A$, which converts another pump wave $\tilde \bk$ into two A-waves $\bk_1$ and $\bk_2$ with $\bk_1=\bkA=\bk-\bkF$. The partner wave $\bk_2$ is determined by the resonant conditions $\tilde\bk=\bk_1+\bk_2$ and $\tilde k=|k_{1z}|+|k_{2z}|$. For a given $\bkF=\kF\bOmF$, the resonance conditions for $F\rightarrow F+A$ and $F\rightarrow A+A$ uniquely determine both $\bk_1$ and $\bk_2$:
\begin{align}
\label{eq:res1}
  & k_{1x} = \kF(|\nz|+1-\nx), \quad & k_{1y} = -\kF\ny, \qquad & k_{1z} = -\kF\nz, \qquad  & k_{1\perp}^2 = 2\kF^2(1-\nx)(1+|\nz|), \\
\label{eq:res2}
  & k_{2x} = \kF(|\nz|-1+\nx), \quad & k_{2y} = \kF\ny, \qquad & k_{2z} = \kF\nz, \qquad  & k_{2\perp}^2 = 2\kF^2(1-\nx)(1-|\nz|).
\end{align}
This also yields 
\beq
  \tilde k=k_{1x}+k_{2x}=2|\nz|\kF=\frac{2|\nz| k}{1+|\nz|}\leq k,
\eeq
which corresponds to $\tilde{b}<b$. The reaction $F\rightarrow A+A$ consumes the pump beam at slightly lower frequencies compared to the primary reaction $F\rightarrow F+A$. For instance, $\nz=\cos\theta=0.8$ gives $\tilde b\approx 0.89\, b$.

The relations between $\bk_1$, $\bk_2$, and $\bkF$ imply that $F\rightarrow F+A$ and $F\rightarrow A+A$ work together to amplify the coupled triad $\{\bkF,\bk_1,\bk_2\}$. The original driver is the process $F\rightarrow F+A$ (since we assume that A-seeds are small compared to F-seeds). It controls the directions of strongly amplified waves $\bkF$: they grow fastest around $\bn$ with $|\cos\theta|=1$ and $\sin^2\!\varphi=1$, which in turn determines the directions of the fastest growing $\bk_1$ and $\bk_2$. Note that both F and A waves stay at approximately constant $\xi$ inside the pulse, and their directions $\bOmF=\bkF/\kF$, $\bOm_1=\bk_1/k_1$, and $\bOm_2=\bk_2/k_2$ all stay constant with time when measured in the fluid frame, which accelerates as $\Gamma\propto r$. 

\Eqs~(\ref{eq:res1}) and (\ref{eq:res2}) have an interesting implication: the generated waves $\bk_1$ and $\bk_2$ form two distinct classes of A-waves, with no overlap. This is seen from the fact that $k_{1x}-|k_{1z}|=\kF(1-\nx)>0$ for all $\bk_1$ while $k_{2x}-|k_{2z}|=-\kF(1-\nx)<0$ for all $\bk_2$ (taking into account that no amplified triads involve $\bkF$ with $\nx=1$, as it implies $\fFA=0$). Since each wave $\bk_2$ is outside the class of waves that  participate in $F\rightarrow F+A$, it is amplified only by $F\rightarrow A+A$. Then, the evolution equations for a resonant triad $\{\bkF,\bk_1,\bk_2\}$ can be stated as follows
\begin{align}
\label{eq:triad1}
  \dot{n}_{\bkF} & =\frac{\pi^2 \omega^2U_\omega}{\Bp^2}(n_{\bkF}+n_{\bk_1}) \fFA,   \\
\label{eq:triad2}
 \dot{n}_{\bk_1} & =\frac{\pi^2 \omega^2U_\omega}{\Bp^2}(n_{\bkF}+n_{\bk_1}) \fFA
  +\frac{\pi^2 \tilde{\omega}^2U_{\tilde{\omega}}}{\Bp^2}(n_{\bk_1}+n_{\bk_2}) \fAA, 
                       \\
\label{eq:triad3}
  \dot{n}_{\bk_2} & =\frac{\pi^2 \tilde{\omega}^2U_{\tilde{\omega}}}{\Bp^2}(n_{\bk_1}+n_{\bk_2}) \fAA, 
\end{align}
where $\tilde\omega=c\tilde k$ and $\tilde\bk=\bk_1+\bk_2$; $\fFA(\bOmF)$ is known, and $\fAA(\bOm_1)$ is determined by $\bk_1=\bk-\bkF$. 

Since $\dot{n}_{\bk_1}=\dot{n}_{\bkF}+\dot{n}_{\bk_2}>\dot{n}_{\bkF}$, one always finds $n_{\bk_1}>n_{\bkF}$. The presence of process $F\rightarrow A+A$ increases $n_{\bkF}+n_{\bk_1}$ and speeds up FRB production $\dot{n}_{\bkF}$. This speed-up is accompanied by an enhanced energy deposition into A-waves and a quicker depletion of the pump beam. In the saturated regime, this effect somewhat reduces the efficiency of FRB production, as the consumed waves from the pump beam become unequally partitioned between the amplified F and A waves, with a higher proportion of A-waves. Compared to $F\rightarrow F+A$ operating in isolation (which would give $n_{\bk_1}=n_{\bkF}$ and $n_{\bk_2}=0$), the presence of $F\rightarrow A+A$ produces an excess of A-waves 
\beq
   \Delta n\equiv n_{\bk_1}+n_{\bk_2}-n_{\bkF}>0. 
\eeq

We will now show that $\Delta n< n_{\bkF}$, and hence the presence of $F\rightarrow A+A$ weakly affects FRB production. Taking the sum of \Eqs~(\ref{eq:triad1})+(\ref{eq:triad2}), and the sum of \Eqs~(\ref{eq:triad2})+(\ref{eq:triad3}), we obtain two equations for $u\equiv n_{\bkF}+n_{\bk_1}$ and  $v\equiv n_{\bk_1}+n_{\bk_2}$, which we state as a vector equation
\beq
   \dot{\boldsymbol{Q}}=A \boldsymbol{Q}, \qquad 
  A=\left(\begin{array}{cc}
    2\aFA & \aAA \\
    \aFA & 2\aAA
            \end{array}\right), 
     \qquad \boldsymbol{Q}=\left(\begin{array}{c}
    u \\
    v
            \end{array}\right)
           \equiv\left(\begin{array}{c}
    n_{\bkF}+n_{\bk_1} \\
   n_{\bk_1}+n_{\bk_2}
            \end{array}\right),
\eeq
where 
\beq
  \aFA\equiv \frac{\pi^2 \omega^2U_\omega}{\Bp^2}\fFA, \qquad \aAA=\frac{\pi^2 \tilde{\omega}^2U_{\tilde{\omega}}}{\Bp^2}\fAA.
\eeq
Diagonalizing this system, we find $\dot{\boldsymbol{Q}}_\pm=\lambda_\pm \boldsymbol{Q}_\pm$, where 
\beq
   \lambda_\pm=\aFA+\aAA\pm \sqrt{\aFA^2+\aAA^2-\aFA \aAA}, \qquad
   \boldsymbol{Q}_\pm \propto
            \left(\begin{array}{c}
            \aAA \\
        \lambda_\pm-2\aFA
                  \end{array} \right).
\eeq
The eigen vector $\boldsymbol{Q}_+$ has the highest growth rate $\lambda_+$, so the fastest growing combination of $u$ and $v$ has $v/u=(\lambda_+-2\aFA)/\aAA$. 

The importance of $F\rightarrow A+A$ relative to $F\rightarrow F+A$ is described by the ratio
\beq
   \frac{\dot{n}_{\bk_1}+\dot{n}_{\bk_2}-\dot{n}_{\bkF}}{\dot{n}_{\bkF}}
   = 2\,\frac{\dot{n}_{\bk_2}}{\dot{n}_{\bkF}} = \frac{2\aAA v}{\aFA u}
  =\frac{2(\lambda_+-2\aFA)}{\aFA},
\eeq
which gives
\beq
\label{eq:Delta_n}
   \frac{\Delta n}{n_{\bkF}}=\frac{\lambda_+-2\aFA}{\aFA} = 2\left[\sqrt{(1-w)^2+w}-1+w\right],  
   \qquad w\equiv\frac{\aAA}{\aFA}=\frac{\fAA}{\fFA}\,\frac{Z(\tilde b)}{Z(b)}.
\eeq
Since $Z(b)$ is not far from $Z_0(b)$ where strong FRB amplification occurs (including at $b\approx\bsat$ in the saturated regime) and $\tilde b$ is close to $b$, one generally expects $Z(\tilde b)/Z(b)\sim 1$. The parameter $w$ is then mainly controlled by $\fAA/\fFA$. The expression for $\fAA$ evaluated for a resonant triad $\{\bkF,\bk_1,\bk_2\}$ gives a remarkably simple result:
\beq\label{eq:f2}
    \fAA=\frac{(k_{1x}k_{2x}-k_{1y}k_{2y})^2}{k^2_{1\perp}k^2_{2\perp}}=\cos^2\!\varphi.
\eeq
The strongest amplification of triads occurs with $\bkF$ that gives $\fFA$ near $\fFA^{\max}=1$, which requires $\sin^2\!\varphi\approx 1$ and implies $\cos^2\!\varphi\ll 1$. Then, we find $\fAA/\fFA\ll 1$ and $w\ll 1$, and expansion of \Eq~(\ref{eq:Delta_n}) in $w$ gives
\beq\label{eq:nw}
    \frac{\Delta n}{n_{\bkF}}\approx w< 1.
\eeq
We conclude that the process $F\rightarrow A+A$ remains subdominant, wasting only a fraction of the pump wave energy to amplify the additional A-waves $\bk_2$ that do not participate in FRB production.

This conclusion can be made more quantitative by estimating $w$ for the triads $\{\bkF,\bk_1,\bk_2\}$ that dominate FRB production, i.e. involve $\bkF$-waves with strongest amplification. Consider mode $\bkF^{\max}$ with $\cos\theta=1$ and $\varphi=\pi/2$ that has $\fFA=1$ and gets the maximum amplification factor $A_{\max}=e^{\tau_{\max}}$. A different mode $\bkF$ whose direction $\bOmF$ differs from that of $\bkF^{\max}$ (by some $\delta\theta$ and $\delta\phi$) has $\fFA<1$. Its amplification $A=e^{\fFA\tau_{\max}}$ is smaller than $A_{\max}$ by the factor $\exp[-\tau_{\max}(1-\fFA)]$. Expanding $\fFA$ near $\bkF^{\max}$, one finds
\beq
1-\fFA\simeq
\left(\delta\varphi-\frac{\delta\theta}{2}\right)^2
+\frac{5}{16}\delta\theta^4.
\eeq
Modes $\bkF$ of main interest have $A\sim A_{\max}$, which corresponds to $\tau_{\max}(1-\fFA)\lesssim 1$. This requires $|\delta\varphi-\delta\theta/2|\lesssim\tau_{\max}^{-1/2}$ and $\delta\theta\lesssim \left[16/(5\tau_{\max})\right]^{1/4}$. Then, using \Eq~(\ref{eq:f2}), we find that the resonant triads with the strongly amplified F-waves satisfy
\beq
\fAA=\cos^2\!\varphi
\approx \delta\varphi^2
\approx \frac{\delta\theta^2}{4}
 \lesssim \frac{1}{5\tau_{\max}},
\qquad
\fFA\approx 1.
\eeq
Thus, the characteristic $w$ may be estimated using $\fAA/\fFA\sim (5\tau_{\max})^{-1/2}$:
\beq
  w\sim \frac{Z(\tilde b)}{Z(b)}\frac{1}{\sqrt{5\tau_{\max}}}
  \approx 0.07\,\frac{Z(\tilde b)}{Z(b)} \left(\frac{\tau_{\max}}{40}\right)^{-1/2}.
\eeq


\section{Solution in two-beam model}
\label{solution}

The simplified two-beam model is described by \Eqs~(\ref{eq:n_ev}) and (\ref{eq:nF_ev}) for each pair of frequencies $\omega$ and $\omF=\Gamma^2\omega$. We rewrite these equations as 
\begin{align}
\label{eq:n_ev_}
      2c\,\partial_s \ln\n & =- \R\, \nF,  \\
\label{eq:nF_ev_}
      \partial_t \ln\nF  & =  \R\, \n,  
\end{align}
where $\R=4\pi^2\omega^2/\Bbg^2$.
Let $\n_0$ be the density of beam $\n$ prior to  scattering, and $\nseedF$ be the initial $\nF$, which fills the slab $0<s<cT$. It is instructive to examine the simple case where $\n_0$, $\Bbg$, and $\Gamma(s)$ are all independent of time $t$. 

Without scattering, $\nF=\nseedF$ and $n=\n_0$ would stay unchanged. Scattering attenuates beam $\n$ and generates $\nF$. The spatial attenuation pattern of $\n$ at any given time $t$ is determined by \Eq~(\ref{eq:n_ev_}),
\beq
      \n\! = \! \n_0 e^{-\tauatt}, \qquad \tauatt(\omega,t,s) = \frac{1}{2} \int_0^{s/c} \!\! \R \, \nF\,d\xi,
\eeq  
and the growth of $\nF$ is determined by \Eq~(\ref{eq:nF_ev_}),
\beq
   \nF=\nseedF e^{\tau}, \qquad \tau(\omF,t,s)=\int_0^t \R \,n\, d\tilde{t}.
\eeq
The initial $\nF=\nseedF$ is tiny and gives a weak initial attenuation of beam $\n$:
\begin{align}
  \tauseed \equiv \tauatt(\omega,0,s) =  \frac{2\pi^2\omega^2}{\Bbg^2} \nseedF \xi \ll 1.
\end{align}
At each location $s$, density $\nF$ grows exponentially with time as long as $n\approx \n_0=const$, which requires $\tauatt(s)\ll 1$.

The condition $\tauatt\ll 1$ is always satisfied near $s=0$, so here $\nF$ continues to grow indefinitely as $e^{\tau_0}$, where
\beq
\label{eq:tau0}   
  \tau_0\equiv \tau(\omF,t,0) = \R\, \n_0\,t = \frac{4\pi^2\omega^2}{\Bbg^2} \n_0\,t.
\eeq
Inside the slab, $\tauatt$ eventually reaches unity (this occurs where $s\,\partial_s\ln n\sim 1$) and $n$ drops below $\n_0$. This depletion develops exponentially fast, with $\n$ dropping first at the end of the slab $s=cT$, and then quickly dropping at progressively smaller $s$. The depletion of beam $\n$ establishes a new, saturated, phase of the scattering process. 

The exact solution describing this evolution can be obtained as follows. Applying $\partial_t$ to \Eq~(\ref{eq:n_ev_}) and $\partial_s$ to \Eq~(\ref{eq:nF_ev_}), one finds
\beq
   \partial_t\partial_s\ln\n=\partial_s\partial_t\ln\nF.
\eeq 
Hence, $\partial_t\partial_s \ln(\n/\nF)=0$ and $\ln(n/\nF)$ has the form 
\beq
  \ln \frac{\n}{\nF}=a(t)+b(s).
\eeq
The unknown $a(t)$ and $b(s)$ are determined by the initial and boundary conditions:
\begin{align}
\label{eq:t_0}
   t=0: & \quad  \ln \frac{\n_0\, e^{-\tauseed(s)}}{\nseedF} = a(0)+b(s), \\
\label{eq:s_0}
   s=0: & \quad  \ln \frac{\n_0}{\nseedF e^{\tau_0(t)}} = a(t)+b(0),
\end{align}
which implies $a(0)+b(0)=\ln(\n_0/\nseedF)$. Taking the sum of \Eqs~(\ref{eq:t_0}) and (\ref{eq:s_0}), we find $a(t)+b(s)$ and obtain:
\beq
\label{eq:relation}
   \frac{\n}{\nF}=\frac{\n_0}{\nseedF}\,e^{-\tauseed -\tau_0}.
\eeq
This relation between $\nF$ and $\n$ reduces the problem to one unknown function, $\nF$. \Eq~(\ref{eq:nF_ev_}) then becomes 
\beq
    \partial_t \nF =  \R\, \nF^2\, \frac{\n_0}{\nseedF} \, e^{-\tauseed(s) -\tau_0(t)}.
\eeq
Its solution for $\nF$ is given by
\beq
   \frac{\nseedF}{\nF} = 1-e^{-\tauseed(s)} \left[1- e^{-\tau_0(t)}\right],
\eeq
which in turn gives $\n$, from \Eq~(\ref{eq:relation}). Expanding in $\tauseed\ll 1$, one can simplify the final result to \Eq~(\ref{eq:solution}) stated in the main text.


\bibliography{References.bib}

@ARTICLE{Zhang2022,
       author = {{Zhang}, Bing},
        title = "{Coherent Inverse Compton Scattering by Bunches in Fast Radio Bursts}",
      journal = {\apj},
         year = 2022,
        month = jan,
       volume = {925},
       number = {1},
          eid = {53},
        pages = {53},
          doi = {10.3847/1538-4357/ac3979},
archivePrefix = {arXiv},
       eprint = {2111.06571},
 primaryClass = {astro-ph.HE},
       adsurl = {https://ui.adsabs.harvard.edu/abs/2022ApJ...925...53Z}
}

@ARTICLE{Mckinven2025,
       author = {{Mckinven}, Ryan and {Bhardwaj}, Mohit and {Eftekhari}, Tarraneh and {Kilpatrick}, Charles D. and {Kirichenko}, Aida and {Pal}, Arpan and {Cook}, Amanda M. and {Gaensler}, B.~M. and {Giri}, Utkarsh and {Kaspi}, Victoria M. and {Michilli}, Daniele and {Nimmo}, Kenzie and {Pearlman}, Aaron B. and {Pleunis}, Ziggy and {Sand}, Ketan R. and {Stairs}, Ingrid and {Andersen}, Bridget C. and {Andrew}, Shion and {Bandura}, Kevin and {Brar}, Charanjot and {Cassanelli}, Tomas and {Chatterjee}, Shami and {Curtin}, Alice P. and {Dong}, Fengqiu Adam and {Eadie}, Gwendolyn and {Fonseca}, Emmanuel and {Ibik}, Adaeze L. and {Kaczmarek}, Jane F. and {Kharel}, Bikash and {Lazda}, Mattias and {Leung}, Calvin and {Li}, Dongzi and {Main}, Robert and {Masui}, Kiyoshi W. and {Mena-Parra}, Juan and {Ng}, Cherry and {Pandhi}, Ayush and {Patil}, Swarali Shivraj and {Prochaska}, J. Xavier and {Rafiei-Ravandi}, Masoud and {Scholz}, Paul and {Shah}, Vishwangi and {Shin}, Kaitlyn and {Smith}, Kendrick},
        title = "{A pulsar-like polarization angle swing from a nearby fast radio burst}",
      journal = {\nat},
         year = 2025,
        month = jan,
       volume = {637},
       number = {8044},
        pages = {43-47},
          doi = {10.1038/s41586-024-08184-4},
       adsurl = {https://ui.adsabs.harvard.edu/abs/2025Natur.637...43M}
}

@ARTICLE{Sironi2021,
       author = {{Sironi}, Lorenzo and {Plotnikov}, Illya and {N{\"a}ttil{\"a}}, Joonas and {Beloborodov}, Andrei M.},
        title = "{Coherent Electromagnetic Emission from Relativistic Magnetized Shocks}",
      journal = {\prl},
         year = 2021,
        month = jul,
       volume = {127},
       number = {3},
          eid = {035101},
        pages = {035101},
          doi = {10.1103/PhysRevLett.127.035101},
archivePrefix = {arXiv},
       eprint = {2107.01211},
 primaryClass = {astro-ph.HE},
       adsurl = {https://ui.adsabs.harvard.edu/abs/2021PhRvL.127c5101S}
}

@ARTICLE{Kumar2017,
       author = {{Kumar}, Pawan and {Lu}, Wenbin and {Bhattacharya}, Mukul},
        title = "{Fast radio burst source properties and curvature radiation model}",
      journal = {\mnras},
         year = 2017,
        month = jul,
       volume = {468},
       number = {3},
        pages = {2726-2739},
          doi = {10.1093/mnras/stx665},
archivePrefix = {arXiv},
       eprint = {1703.06139},
 primaryClass = {astro-ph.HE},
       adsurl = {https://ui.adsabs.harvard.edu/abs/2017MNRAS.468.2726K}
}

@ARTICLE{Beloborodov2011,
       author = {{Beloborodov}, Andrei M.},
        title = "{Radiative Transfer in Ultrarelativistic Outflows}",
      journal = {\apj},
         year = 2011,
        month = aug,
       volume = {737},
       number = {2},
          eid = {68},
        pages = {68},
          doi = {10.1088/0004-637X/737/2/68},
archivePrefix = {arXiv},
       eprint = {1011.6005},
 primaryClass = {astro-ph.HE},
       adsurl = {https://ui.adsabs.harvard.edu/abs/2011ApJ...737...68B}
}

@ARTICLE{Kaspi2017,
       author = {{Kaspi}, Victoria M. and {Beloborodov}, Andrei M.},
        title = "{Magnetars}",
      journal = {\araa},
         year = 2017,
        month = aug,
       volume = {55},
       number = {1},
        pages = {261-301},
          doi = {10.1146/annurev-astro-081915-023329},
archivePrefix = {arXiv},
       eprint = {1703.00068},
 primaryClass = {astro-ph.HE},
       adsurl = {https://ui.adsabs.harvard.edu/abs/2017ARA&A..55..261K}
}

@ARTICLE{Qu2026,
       author = {{Qu}, Yuanhong and {Bransgrove}, Ashley},
        title = "{3D Numerical Simulations of Magnetar Crustquakes}",
      journal = {\apj},
         year = 2026,
        month = feb,
       volume = {998},
       number = {2},
          eid = {190},
        pages = {190},
          doi = {10.3847/1538-4357/ae3a9d},
archivePrefix = {arXiv},
       eprint = {2508.12567},
 primaryClass = {astro-ph.HE},
       adsurl = {https://ui.adsabs.harvard.edu/abs/2026ApJ...998..190Q}
}

@ARTICLE{Chen2022a,
       author = {{Chen}, Alexander Y. and {Yuan}, Yajie and {Li}, Xinyu and {Mahlmann}, Jens F.},
        title = "{Propagation of a Strong Fast Magnetosonic Wave in the Magnetosphere of a Neutron Star}",
      journal = {arXiv e-prints},
         year = 2022,
        month = oct,
          eid = {arXiv:2210.13506},
        pages = {arXiv:2210.13506},
          doi = {10.48550/arXiv.2210.13506},
archivePrefix = {arXiv},
       eprint = {2210.13506},
 primaryClass = {astro-ph.HE},
       adsurl = {https://ui.adsabs.harvard.edu/abs/2022arXiv221013506C}
}

@ARTICLE{Chen2022,
       author = {{Chen}, Alexander Y. and {Yuan}, Yajie and {Beloborodov}, Andrei M. and {Li}, Xinyu},
        title = "{Relativistic Alfv{\'e}n Waves Entering Charge-starvation in the Magnetospheres of Neutron Stars}",
      journal = {\apj},
         year = 2022,
        month = apr,
       volume = {929},
       number = {1},
          eid = {31},
        pages = {31},
          doi = {10.3847/1538-4357/ac59b1},
archivePrefix = {arXiv},
       eprint = {2010.15619},
 primaryClass = {astro-ph.HE},
       adsurl = {https://ui.adsabs.harvard.edu/abs/2022ApJ...929...31C}
}

@ARTICLE{Bransgrove2020,
       author = {{Bransgrove}, Ashley and {Beloborodov}, Andrei M. and {Levin}, Yuri},
        title = "{A Quake Quenching the Vela Pulsar}",
      journal = {\apj},
         year = 2020,
        month = jul,
       volume = {897},
       number = {2},
          eid = {173},
        pages = {173},
          doi = {10.3847/1538-4357/ab93b7},
archivePrefix = {arXiv},
       eprint = {2001.08658},
 primaryClass = {astro-ph.HE},
       adsurl = {https://ui.adsabs.harvard.edu/abs/2020ApJ...897..173B}
}

@ARTICLE{Blaes1989,
       author = {{Blaes}, O. and {Blandford}, R. and {Goldreich}, P. and {Madau}, P.},
        title = "{Neutron Starquake Models for Gamma-Ray Bursts}",
      journal = {\apj},
         year = 1989,
        month = aug,
       volume = {343},
        pages = {839},
          doi = {10.1086/167754},
       adsurl = {https://ui.adsabs.harvard.edu/abs/1989ApJ...343..839B}
}

@book{Kruer2003,
  title={The Physics of Laser Plasma Interactions},
  author={Kruer, William L},
  year={2003},
  publisher={CRC Press},
  series={Frontiers in Physics},
  isbn={9780813340838}
}

@ARTICLE{Camilo2006,
       author = {{Camilo}, Fernando and {Ransom}, Scott M. and {Halpern}, Jules P. and {Reynolds}, John and {Helfand}, David J. and {Zimmerman}, Neil and {Sarkissian}, John},
        title = "{Transient pulsed radio emission from a magnetar}",
      journal = {\nat},
         year = 2006,
        month = aug,
       volume = {442},
       number = {7105},
        pages = {892-895},
          doi = {10.1038/nature04986},
archivePrefix = {arXiv},
       eprint = {astro-ph/0605429},
 primaryClass = {astro-ph},
       adsurl = {https://ui.adsabs.harvard.edu/abs/2006Natur.442..892C}
}

@ARTICLE{Kumar2024,
       author = {{Kumar}, Pawan and {Qu}, Yuanhong and {Zhang}, Bing},
        title = "{The Origins of Narrow Spectra of Fast Radio Bursts}",
      journal = {\apj},
         year = 2024,
        month = oct,
       volume = {974},
       number = {2},
          eid = {160},
        pages = {160},
          doi = {10.3847/1538-4357/ad6cda},
archivePrefix = {arXiv},
       eprint = {2406.01266},
 primaryClass = {astro-ph.HE},
       adsurl = {https://ui.adsabs.harvard.edu/abs/2024ApJ...974..160K}
}

@ARTICLE{Beloborodov2020,
       author = {{Beloborodov}, Andrei M.},
        title = "{Blast Waves from Magnetar Flares and Fast Radio Bursts}",
      journal = {\apj},
         year = 2020,
        month = jun,
       volume = {896},
       number = {2},
          eid = {142},
        pages = {142},
          doi = {10.3847/1538-4357/ab83eb},
archivePrefix = {arXiv},
       eprint = {1908.07743},
 primaryClass = {astro-ph.HE},
       adsurl = {https://ui.adsabs.harvard.edu/abs/2020ApJ...896..142B}
}

@ARTICLE{Beloborodov2017,
       author = {{Beloborodov}, Andrei M.},
        title = "{A Flaring Magnetar in FRB 121102?}",
      journal = {\apjl},
         year = 2017,
        month = jul,
       volume = {843},
       number = {2},
          eid = {L26},
        pages = {L26},
          doi = {10.3847/2041-8213/aa78f3},
archivePrefix = {arXiv},
       eprint = {1702.08644},
 primaryClass = {astro-ph.HE},
       adsurl = {https://ui.adsabs.harvard.edu/abs/2017ApJ...843L..26B}
}

@ARTICLE{Lyubarsky2014,
       author = {{Lyubarsky}, Yu.},
        title = "{A model for fast extragalactic radio bursts.}",
      journal = {\mnras},
         year = 2014,
        month = jul,
       volume = {442},
        pages = {L9-L13},
          doi = {10.1093/mnrasl/slu046},
archivePrefix = {arXiv},
       eprint = {1401.6674},
 primaryClass = {astro-ph.HE},
       adsurl = {https://ui.adsabs.harvard.edu/abs/2014MNRAS.442L...9L}
}

@ARTICLE{Yuan2020,
       author = {{Yuan}, Yajie and {Beloborodov}, Andrei M. and {Chen}, Alexander Y. and {Levin}, Yuri},
        title = "{Plasmoid Ejection by Alfv{\'e}n Waves and the Fast Radio Bursts from SGR 1935+2154}",
      journal = {\apjl},
         year = 2020,
        month = sep,
       volume = {900},
       number = {2},
          eid = {L21},
        pages = {L21},
          doi = {10.3847/2041-8213/abafa8},
archivePrefix = {arXiv},
       eprint = {2006.04649},
 primaryClass = {astro-ph.HE},
       adsurl = {https://ui.adsabs.harvard.edu/abs/2020ApJ...900L..21Y}
}

@ARTICLE{Mahlmann2023,
       author = {{Mahlmann}, J.~F. and {Philippov}, A.~A. and {Mewes}, V. and {Ripperda}, B. and {Most}, E.~R. and {Sironi}, L.},
        title = "{Three-dimensional Dynamics of Strongly Twisted Magnetar Magnetospheres: Kinking Flux Tubes and Global Eruptions}",
      journal = {\apjl},
         year = 2023,
        month = apr,
       volume = {947},
       number = {2},
          eid = {L34},
        pages = {L34},
          doi = {10.3847/2041-8213/accada},
archivePrefix = {arXiv},
       eprint = {2302.07273},
 primaryClass = {astro-ph.HE},
       adsurl = {https://ui.adsabs.harvard.edu/abs/2023ApJ...947L..34M}
}

@ARTICLE{Beloborodov2026,
       author = {{Beloborodov}, Andrei M.},
        title = "{Compression Fronts from Fast Radio Bursts}",
      journal = {\apj},
         year = 2026,
        month = apr,
       volume = {1000},
       number = {2},
          eid = {157},
        pages = {157},
          doi = {10.3847/1538-4357/ae4692},
archivePrefix = {arXiv},
       eprint = {2503.16054},
 primaryClass = {astro-ph.HE},
       adsurl = {https://ui.adsabs.harvard.edu/abs/2026ApJ..1000..157B}
}

@ARTICLE{Beloborodov2024,
       author = {{Beloborodov}, Andrei M.},
        title = "{Damping of Strong GHz Waves near Magnetars and the Origin of Fast Radio Bursts}",
      journal = {\apj},
         year = 2024,
        month = nov,
       volume = {975},
       number = {2},
          eid = {223},
        pages = {223},
          doi = {10.3847/1538-4357/ad698c},
archivePrefix = {arXiv},
       eprint = {2307.12182},
 primaryClass = {astro-ph.HE},
       adsurl = {https://ui.adsabs.harvard.edu/abs/2024ApJ...975..223B}
}

@ARTICLE{Petroff2022,
       author = {{Petroff}, E. and {Hessels}, J.~W.~T. and {Lorimer}, D.~R.},
        title = "{Fast radio bursts at the dawn of the 2020s}",
      journal = {\aapr},
         year = 2022,
        month = dec,
       volume = {30},
       number = {1},
          eid = {2},
        pages = {2},
          doi = {10.1007/s00159-022-00139-w},
archivePrefix = {arXiv},
       eprint = {2107.10113},
 primaryClass = {astro-ph.HE},
       adsurl = {https://ui.adsabs.harvard.edu/abs/2022A&ARv..30....2P}
}

@ARTICLE{Lyubarsky2020,
       author = {{Lyubarsky}, Yuri},
        title = "{Fast Radio Bursts from Reconnection in a Magnetar Magnetosphere}",
      journal = {\apj},
         year = 2020,
        month = jul,
       volume = {897},
       number = {1},
          eid = {1},
        pages = {1},
          doi = {10.3847/1538-4357/ab97b5},
archivePrefix = {arXiv},
       eprint = {2001.02007},
 primaryClass = {astro-ph.HE},
       adsurl = {https://ui.adsabs.harvard.edu/abs/2020ApJ...897....1L}
}

@ARTICLE{Parfrey2013,
       author = {{Parfrey}, Kyle and {Beloborodov}, Andrei M. and {Hui}, Lam},
        title = "{Dynamics of Strongly Twisted Relativistic Magnetospheres}",
      journal = {\apj},
         year = 2013,
        month = sep,
       volume = {774},
       number = {2},
          eid = {92},
        pages = {92},
          doi = {10.1088/0004-637X/774/2/92},
archivePrefix = {arXiv},
       eprint = {1306.4335},
 primaryClass = {astro-ph.HE},
       adsurl = {https://ui.adsabs.harvard.edu/abs/2013ApJ...774...92P}
}

@ARTICLE{Beloborodov2023,
       author = {{Beloborodov}, Andrei M.},
        title = "{Monster Radiative Shocks in the Perturbed Magnetospheres of Neutron Stars}",
      journal = {\apj},
         year = 2023,
        month = dec,
       volume = {959},
       number = {1},
          eid = {34},
        pages = {34},
          doi = {10.3847/1538-4357/acf659},
archivePrefix = {arXiv},
       eprint = {2210.13509},
 primaryClass = {astro-ph.HE},
       adsurl = {https://ui.adsabs.harvard.edu/abs/2023ApJ...959...34B}
}

@ARTICLE{Chatterjee2026,
       author = {{Chatterjee}, Koushik and {Philippov}, Alexander and {Beloborodov}, Andrei M. and {Parfrey}, Kyle and {Ripperda}, Bart and {Most}, Elias R.},
        title = "{Relativistic Magnetohydrodynamic Simulations of Giant Magnetar Bursts}",
      journal = {arXiv e-prints},
         year = 2026,
        month = feb,
          eid = {arXiv:2602.17755},
        pages = {arXiv:2602.17755},
          doi = {10.48550/arXiv.2602.17755},
archivePrefix = {arXiv},
       eprint = {2602.17755},
 primaryClass = {astro-ph.HE},
       adsurl = {https://ui.adsabs.harvard.edu/abs/2026arXiv260217755C}
}

@ARTICLE{GolbraikhLyubarsky2023,
       author = {{Golbraikh}, Ephim and {Lyubarsky}, Yuri},
        title = "{On the Escape of Low-frequency Waves from Magnetospheres of Neutron Stars}",
      journal = {\apj},
         year = 2023,
        month = nov,
       volume = {957},
       number = {2},
          eid = {102},
        pages = {102},
          doi = {10.3847/1538-4357/acfa78},
archivePrefix = {arXiv},
       eprint = {2309.09218},
 primaryClass = {astro-ph.HE},
       adsurl = {https://ui.adsabs.harvard.edu/abs/2023ApJ...957..102G}
}

@ARTICLE{Lyubarsky2019,
       author = {{Lyubarsky}, Yuri},
        title = "{Radio emission of the Crab and Crab-like pulsars}",
      journal = {\mnras},
         year = 2019,
        month = feb,
       volume = {483},
       number = {2},
        pages = {1731-1736},
          doi = {10.1093/mnras/sty3233},
archivePrefix = {arXiv},
       eprint = {1811.11122},
 primaryClass = {astro-ph.HE},
       adsurl = {https://ui.adsabs.harvard.edu/abs/2019MNRAS.483.1731L}
}

@ARTICLE{Takamoto2016,
       author = {{Takamoto}, Makoto and {Lazarian}, Alexandre},
        title = "{Compressible Relativistic Magnetohydrodynamic Turbulence in Magnetically Dominated Plasmas and Implications for a Strong-coupling Regime}",
      journal = {\apjl},
         year = 2016,
        month = nov,
       volume = {831},
       number = {2},
          eid = {L11},
        pages = {L11},
          doi = {10.3847/2041-8205/831/2/L11},
archivePrefix = {arXiv},
       eprint = {1610.01373},
 primaryClass = {astro-ph.HE},
       adsurl = {https://ui.adsabs.harvard.edu/abs/2016ApJ...831L..11T}
}

@ARTICLE{Cho2005,
       author = {{Cho}, Jungyeon},
        title = "{Simulations of Relativistic Force-free Magnetohydrodynamic Turbulence}",
      journal = {\apj},
         year = 2005,
        month = mar,
       volume = {621},
       number = {1},
        pages = {324-327},
          doi = {10.1086/427493},
archivePrefix = {arXiv},
       eprint = {astro-ph/0408318},
 primaryClass = {astro-ph},
       adsurl = {https://ui.adsabs.harvard.edu/abs/2005ApJ...621..324C}
}

@ARTICLE{Thompson1998,
       author = {{Thompson}, Christopher and {Blaes}, Omer},
        title = "{Magnetohydrodynamics in the extreme relativistic limit}",
      journal = {\prd},
         year = 1998,
        month = mar,
       volume = {57},
       number = {6},
        pages = {3219-3234},
          doi = {10.1103/PhysRevD.57.3219},
       adsurl = {https://ui.adsabs.harvard.edu/abs/1998PhRvD..57.3219T}
}

@ARTICLE{Lyubarsky2021,
       author = {{Lyubarsky}, Yuri},
        title = "{Emission Mechanisms of Fast Radio Bursts}",
      journal = {Universe},
         year = 2021,
        month = mar,
       volume = {7},
       number = {3},
          eid = {56},
        pages = {56},
          doi = {10.3390/universe7030056},
archivePrefix = {arXiv},
       eprint = {2103.00470},
 primaryClass = {astro-ph.HE},
       adsurl = {https://ui.adsabs.harvard.edu/abs/2021Univ....7...56L}
}

@ARTICLE{Thompson2023,
       author = {{Thompson}, Christopher},
        title = "{Direct emission of strong radio pulses during magnetar flares}",
      journal = {\mnras},
         year = 2023,
        month = feb,
       volume = {519},
       number = {1},
        pages = {497-518},
          doi = {10.1093/mnras/stac3565},
archivePrefix = {arXiv},
       eprint = {2209.11136},
 primaryClass = {astro-ph.HE},
       adsurl = {https://ui.adsabs.harvard.edu/abs/2023MNRAS.519..497T}
}

@ARTICLE{Mahlmann2022,
       author = {{Mahlmann}, J.~F. and {Philippov}, A.~A. and {Levinson}, A. and {Spitkovsky}, A. and {Hakobyan}, H.},
        title = "{Electromagnetic Fireworks: Fast Radio Bursts from Rapid Reconnection in the Compressed Magnetar Wind}",
      journal = {\apjl},
         year = 2022,
        month = jun,
       volume = {932},
       number = {2},
          eid = {L20},
        pages = {L20},
          doi = {10.3847/2041-8213/ac7156},
archivePrefix = {arXiv},
       eprint = {2203.04320},
 primaryClass = {astro-ph.HE},
       adsurl = {https://ui.adsabs.harvard.edu/abs/2022ApJ...932L..20M}
}

@ARTICLE{Zeng2026,
       author = {{Zeng}, Shuzhe and {Philippov}, Alexander and {Juno}, James and {Beloborodov}, Andrei M. and {Popova}, Elena},
        title = "{Origin of Pulsed Radio Emission from Magnetars}",
      journal = {\apjl},
         year = 2026,
        month = jan,
       volume = {996},
       number = {2},
          eid = {L20},
        pages = {L20},
          doi = {10.3847/2041-8213/ae2ade},
archivePrefix = {arXiv},
       eprint = {2509.13419},
 primaryClass = {astro-ph.HE},
       adsurl = {https://ui.adsabs.harvard.edu/abs/2026ApJ...996L..20Z}
}

@ARTICLE{Solanki2026a,
       author = {{Solanki}, Siddhant and {Mahlmann}, Jens and {Philippov}, Alexander},
        title = "{Nonlinear Decay of Fast Magnetosonic Waves through Weak Turbulence: Force-Free Electrodynamics Simulations}",
      journal = {arXiv e-prints},
         year = 2026,
        month = jun,
          eid = {arXiv:2606.19450},
        pages = {arXiv:2606.19450},
          doi = {10.48550/arXiv.2606.19450},
archivePrefix = {arXiv},
       eprint = {2606.19450},
 primaryClass = {astro-ph.HE},
       adsurl = {https://ui.adsabs.harvard.edu/abs/2026arXiv260619450S}
}

@ARTICLE{Solanki2026b,
       author = {{Solanki}, Siddhant and {Mahlmann}, Jens and {Philippov}, Alexander and {Beloborodov}, Andrei},
        title = "{Damping of Fast Radio Bursts in the Inner Magnetospheres of Magnetars}",
      journal = {arXiv e-prints},
         year = 2026,
        month = jun,
          eid = {arXiv:2606.19448},
        pages = {arXiv:2606.19448},
          doi = {10.48550/arXiv.2606.19448},
archivePrefix = {arXiv},
       eprint = {2606.19448},
 primaryClass = {astro-ph.HE},
       adsurl = {https://ui.adsabs.harvard.edu/abs/2026arXiv260619448S}
}

@BOOK{Zakharov1992,
       author = {{Zakharov}, V.~E. and {L'Vov}, V.~S. and {Falkovich}, G.},
        title = "{Kolmogorov spectra of turbulence 1. Wave turbulence.}",
         year = 1992,
       adsurl = {https://ui.adsabs.harvard.edu/abs/1992kst..book.....Z}
}

\end{document}